\documentclass[11pt,a4paper,fleqn,usenatbib]{mnras}

\usepackage{newtxtext,newtxmath}

\usepackage{ulem}

\usepackage[T1]{fontenc}
\usepackage{ae,aecompl}

\usepackage{graphicx}	
\usepackage{amsmath}	

\newcommand{\Msun}{\, {\rm M_{\odot}}}

\title[Tantalizing Models of Self-Interacting Dark Matter]{TangoSIDM: Tantalizing models of Self-Interacting Dark Matter}
\author[C. A. Correa et al.]{Camila A. Correa,$^{1}$\thanks{E-mail: camila.correa@uva.nl}
Matthieu Schaller,$^{2,3}$ 
Sylvia Ploeckinger,$^{2}$ 
Noemi Anau Montel,$^{1}$
\newauthor
Christoph Weniger$^{1}$
and Shin'ichiro Ando$^{1}$ 
\\
$^{1}$GRAPPA Institute, University of Amsterdam, Science Park 904, 1098 XH Amsterdam, The Netherlands \\
$^{2}$Lorentz Institute for Theoretical Physics, Leiden University, PO Box 9506, 2300 RA Leiden, The Netherlands\\
$^{3}$Leiden Observatory, Leiden University, PO Box 9513, 2300 RA Leiden, The Netherlands
}

\date{Accepted XXX. Received YYY; in original form ZZZ}

\pubyear{2022}

\begin{document}
\label{firstpage}
\pagerange{\pageref{firstpage}--\pageref{lastpage}}
\maketitle

\begin{abstract}
We introduce the TangoSIDM project, a suite of cosmological simulations of structure formation in a $\Lambda$-Self-Interacting Dark Matter (SIDM) universe. TangoSIDM explores the impact of large dark matter (DM) scattering cross sections over dwarf galaxy scales. Motivated by DM interactions that follow a Yukawa potential, the cross section per unit mass, $\sigma/m_{\chi}$, assumes a velocity dependent form that avoids violations of current constraints on large scales. We demonstrate that our implementation accurately models not only core formation in haloes, but also gravothermal core collapse. For central haloes in cosmological volumes, frequent DM particle collisions isotropise the particles orbit, making them largely spherical. We show that the velocity-dependent $\sigma/m_{\chi}$ models produce a large diversity in the circular velocities of satellites haloes, with the spread in velocities increasing as the cross sections reach 20, 60 and 100 cm$^2$/g in $10^9~\rm{M}_{\odot}$ haloes. The large variation in the haloes internal structure is driven by DM particles interactions, causing in some haloes the formation of extended cores, whereas in others gravothermal core collapse. We conclude that the SIDM models from the Tango project offer a promising explanation for the diversity in the density and velocity profiles of observed dwarf galaxies.
\end{abstract}

\begin{keywords}
methods: numerical - galaxies: haloes - cosmology: theory - dark matter.
\end{keywords}


\section{Introduction}

Uncovering the nature of dark matter (DM) is one of the most pressing pursuits in modern physics and cosmology. The long-held cosmological paradigm of $\Lambda$ collisionless cold dark matter ($\Lambda$CDM) accurately predicts the large-scale structure of the Universe (\citealt{Planck20, eBOSS2021}), however, significant discrepancies on galactic and sub-galactic scales are constantly challenging it.  

Galactic observations targeting (i) the number of observed satellite galaxies, and (ii) the dynamical mass in the inner regions of dwarf galaxies, are key to understand the nature of DM. The number of satellite galaxies has introduced the `missing satellite problem', or problem of abundance, stating that CDM simulations overpredict the abundance of satellites around the Milky Way (hereafter MW) (\citealt{Klypin99,Moore99}). Several works have concluded that the missing satellite problem is solved when introducing baryonic effects from supernova feedback and reionisation (e.g. \citealt{Fattahi16,Sawala16,Wetzel16, GarrisonKimmel19, Applebaum21,Engler21}). However, in recent years, the discovery of several new satellites, more careful survey selection functions and new development of higher-resolution simulations, have altered the CDM+baryons predictions, showing now that CDM simulations may underpredict the abundance of luminous satellites (e.g. \citealt{Kim18,Jethwa18,Torrealba18,Kelley19,Homma19, Nadler20,Kim21}). 

The dynamical mass in some dwarf galaxies appears to be low compared to CDM predictions (e.g. \citealt{Moore94,deBlok97, Oh11,Walker11}). CDM simulations without baryons predict that dwarf galaxies reside in dark matter haloes that have dense central regions, with a density profile showing a steep slope and `cusp' shape (\citealt{NFW97}). Differently, many dwarf galaxies appear to have lower central densities, with a `cored' density profile following a flat slope (\citealt{Walker10, BoylanKolchin11, Ferrero12,Read19}). This problem originally called the `core-cusp problem', is also referred as `diversity problem' due to the large variety in shape and central densities in the local dwarf galaxies (e.g. \citealt{Read19,Hayashi20}), as well as in several galaxy rotation curves (e.g., \citealt{Oman15,Read16,Tollet16,Relatores19,Ren19,Santos20}).

An additional disagreement between the prediction of CDM simulations and observations is the so-called `too-big-to-fail problem', which states that the most massive haloes in CDM simulations are too dense in the centre to host the observed luminous MW satellites (\citealt{BoylanKolchin11,BoylanKolchin12}). The too-big-to-fail is a problem related to the internal structure of haloes, that has not only been found in the MW satellites, but also in the ones of M31 (\citealt{Tollerud14}), and in galaxies from the field (\citealt{GarrisonKimmel14,Papastergis15}).

To this day, CDM with the addition of baryons, does not seem to convincingly solve the missing satellite (e.g. \citealt{Graus18,Kelley19}), the to-big-to-fail problem (e.g. \citealt{Kaplinghat19}), nor the cusp/core/diversity problem (e.g. \citealt{Santos20}). In the latter, baryonic feedback processes from star formation and supernova explosions produce gravitational fluctuations that allow the redistribution of dark matter, and the formation of cores (\citealt{Governato12}). But this appears to be very model-dependent (e.g. \citealt{Dutton20}), with numerical simulations producing either too many cores in dwarf galaxies of $M_{\rm{DM}}\sim10^{9}-10^{10}~\rm{M}_{\odot}$ (e.g. \citealt{DiCintio14,Tollet16,Hopkins18,Lazar20}) or none at all (e.g. \citealt{Bose19}).

This motivates to question the nature of DM and to consider DM physics beyond standard models. A promising alternative to CDM is to assume non-gravitational interactions among DM particles (\citealt{Spergel00}). These types of DM models, widely known as `self-interacting dark matter' (hereafter SIDM), consider that DM particles experience collisions with each other. DM particles collisions transfer heat towards the colder central regions of DM haloes, lowering central densities and creating constant density cores (e.g. \citealt{Dave01,Colin02,Vogelsberger12,Rocha13,Dooley16,Vogelsberger19,Robles19}). 

The cross section per unit mass, $\sigma/m_{\chi}$, is the main parameter that controls the rate of DM particles interactions in numerical simulations (e.g. \citealt{Robertson17,Kahlhoefer19, Robertson19, Kummer19,Banerjee20,Shen21}, among others), as well as in semi-analytic models (e.g. \citealt{Balberg02,Ahn05, Essig19, Nishikawa19}). A low cross section ($\sigma/m_{\chi}{<}1$ cm$^{2}$/g) produces low DM collisions rates, allowing DM haloes to keep cuspy density profiles. Alternatively, high cross sections ($\sigma/m_{\chi}{>}1$ cm$^{2}$/g) lead to very frequent DM collisions that are able to produce central density cores (e.g. \citealt{Rocha13, Zavala13}).

In the regime of very large cross sections (e.g. $\sigma/m_{\chi}{>}10$ cm$^{2}$/g), DM particle interactions are so frequent that they are able to rapidly heat the central DM halo core, causing it to contract and raise in density. In this regime, known as gravothermal core collapse (\citealt{Balberg02,Elbert15}), DM haloes form a density core early on, which changes to a cuspy profile at latter times. Although it has been known for some time that it takes longer than a hubble time for a halo to enter in the gravothermal core collapse regime (\citealt{Balberg02, Koda11, Sameie20}), recent studies of satellites have changed this and showed that the gravothermal collapse is accelerated by mass loss via tidal stripping (\citealt{Nishikawa19}). In fact, not very eccentric orbits, nor excessive mass loss, are needed for satellite haloes to enter in gravothermal core collapse (\citealt{Kahlhoefer19,Turner21,Zeng21}). 

Several studies have constrained the cross section to be smaller $\sigma/m_{\chi}{<}1.25$ cm$^{2}$/g on galaxy cluster scales (see e.g. \citealt{Randall08,Dawson13,Jee14, Massey15,Wittman18,Harvey19, Sagunski21,Andrade22}). On dwarf galaxy scales, current constraints of $\sigma/m_{\chi}$ rely on predicting the DM density profile of galaxies following the isothermal Jeans modelling. In this manner, \citet{Read18} analysed the density profile of Draco, a cuspy MW dwarf spheroidal galaxy (dSph), and concluded that its high central density gives an upper bound on the SIDM cross section of $\sigma/m_{\chi}<0.57$ cm$^2$g$^{-1}$. While \citet{Valli18} derived a similar upper limit on $\sigma/m_{\chi}$ for Draco (but probed different cross sections, ranging between 0.1 and 40 cm$^2$g$^{-1}$, for the remaining dSphs, see also \citealt{Kaplinghat16}), others (e.g. \citealt{Hayashi21,Ebisu22}) analysed the cuspy profiles of some dwarfs and ultra-faint dwarfs, and concluded that zero self-interactions are favoured. Although this method provides an accurate description of simulated SIDM density profiles (\citealt{Robertson21}), it does not consider the gravothermal core collapse scenario. Therefore cuspy galaxy DM profiles can only result from low $\sigma/m_{\chi}$.

In a scenario where SIDM has a large $\sigma/m_{\chi}$ on dwarf galaxy scales, galaxies in small pericenter orbits that have lost mass from tidal stripping, can quickly enter in gravothermal core collapse and exhibit a cuspy DM density profile. Differently, satellite galaxies that have not lost mass from tidal interactions are able to keep a flat density core. This naturally gives rise to a diversity in the shape of DM density profiles in systems that are quite DM-dominated, and therefore not expected to be altered by the presence of baryons (see e.g.~\citealt{Oman15}). Interestingly, \citet{Kaplinghat19} reported an anti-correlation between the central DM densities of the bright dwarf spheroidal galaxies of the MW (dSphs) and their orbital pericenter distances, so that the dSphs that have come closer to the MW centre are more dense in DM than those that have not come so close. This anti-correlation has been proposed as a potential signature of SIDM (\citealt{Correa21}), with $\sigma/m_{\chi}$ depending on the relative velocity of DM particles, in such a way that DM behaves almost collisionless over cluster scales but as a collisional fluid on satellite galaxy scales. \citet{Correa21} derived a semi-analytic model of the local dSphs evolution, to determine the range of $\sigma/m_{\chi}$ able to explain the large central densities of the dwarfs under the gravothermal core-collapse regime. They found that the densities of dSphs, such as Carina and Fornax, can be explained with $\sigma/m_{\chi}$ ranging between 30 and 50 cm$^{2}$/g, whereas other dSphs prefer larger values ranging between 70 and 100 cm$^{2}$/g. 

A velocity-dependent SIDM cross section model, where DM behaves as a collisional fluid on small scales while it is essentially collisionless over large scales, has been suggested as early as \citet{Yoshida00}. In addition, particle physics models favour such framework for the DM particle (e.g. \citealt{Buckley10, Boddy14}), arguing that DM exists in a `hidden sector', where forces between DM particles are mediated by analogues to electroweak or strong forces (e.g. \citealt{Pospelov08, ArkaniHamed09,Buckley10,Feng10,Boddy14,Tulin18}). While studies on galaxy clusters scales have set robust upper limits on the DM self-interaction cross section, robust constraints of $\sigma/m_{\chi}$ on dwarf galaxy scales are currently missing. The possibility of gravothermal core collapse indicates that $\sigma/m_{\chi}$ could be larger than 10 cm$^{2}$/g on dwarf galaxy scales (\citealt{Correa21}). But more thorough studies with detailed modelling of SIDM, and galaxy formation in a cosmological context, are needed to prove or rule out scenarios of large $\sigma/m_{\chi}$. 

The goal of this study is to improve the current modelling of dwarf galaxies embedded in a SIDM universe, in order to derive robust constraints of large $\sigma/m_{\chi}$ on dwarf galaxy scales, and to prove (or alternatively rule out) that large $\sigma/m_{\chi}$ can explain the diversity in the density and velocity profiles of observed dwarf galaxies. To do so we introduce the `TANtalasinG mOdels of Self-Interacting Dark Matter' project (hereafter TangoSIDM). TangoSIDM consists of a suite of cosmological, hydrodynamical simulations of structure formation in a $\Lambda$-SIDM universe. The main models presented in this work are dark matter-only volumes of 25 comoving Mpc on a side and employ a resolution that allows the study of satellite haloes as small as $10^9~\rm{M}_{\odot}$. The simulations use state-of-the-art numerical techniques and new modelling of the DM-DM particles interactions. The TangoSIDM suite includes many simulations that will be presented in future works, including simulations using state-of-the-art galaxy formation models. In this study we present and describe in detail the methodology employed to model SIDM in a cosmological set-up. Additionally we analyse the dark matter-only simulations and show the first results on the DM haloes' internal structure from TangoSIDM.

This paper is organized as follows. Section 2 describes the velocity-dependent cross section assumed to model the DM particles interactions (Sec. 2.1). It also outlines the simulations (Sec. 2.2) and the details of the SIDM implementation (Sec. 2.3). Section 3 presents our results. A brief discussion is presented in Section 4. Finally, we summarise our key results in Section 5. In addition, comparison with previous works, as well as further validation and numerical convergence tests, are included in the appendix sections.

\section{Self-Interacting Dark Matter Models}

\subsection{Scattering cross-section}\label{Scattering_cross_section_Section}

\begin{figure} 
	\includegraphics[angle=0,width=0.48\textwidth]{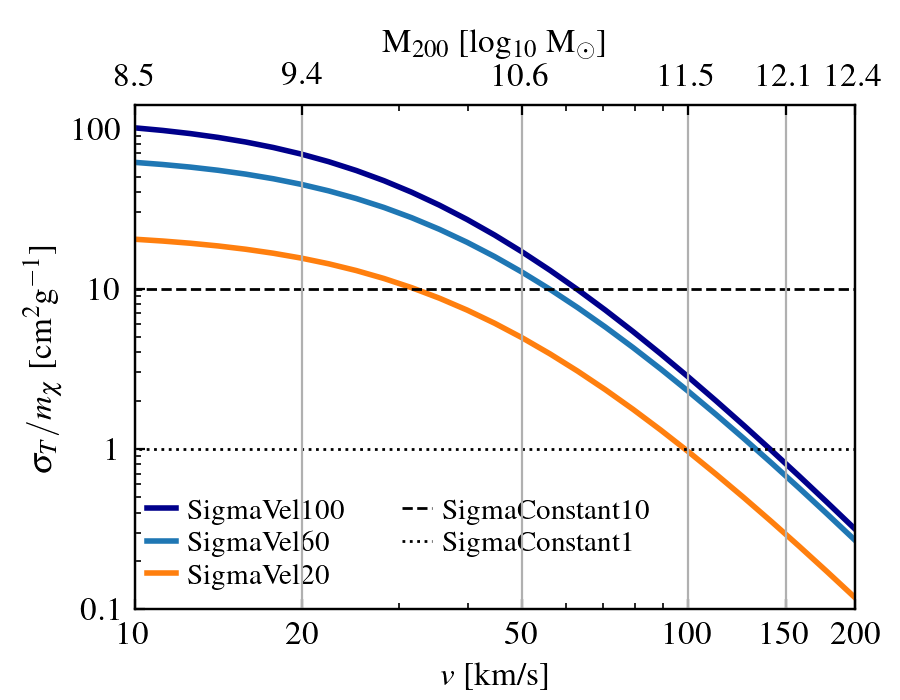}
	\vspace{-0.5cm}
	\caption{Momentum transfer cross sections as a function of relative DM particles scattering velocity of the SIDM models presented in this work (see Table~\ref{Table_models}). The figure highlights three velocity-dependent models (dark blue, light blue and orange lines) where $\sigma_{T}/m_{\chi}$ reaches 100, 60 and 20 cm$^{2}$/g on $10^{9}~\rm{M}_{\odot}$ dwarf galaxies. Additionally three constant cross section models are also studied, $\sigma_{T}/m_{\chi}=10$ and 1 cm$^{2}$/g (black dashed and dotted lines) and $\sigma_{T}/m_{\chi}=0$ (CDM). While the bottom x-axis highlights the relative velocity between DM particles, the top x-axis indicates the typical halo mass that hosts orbits of such velocities. The labels of the different curves indicate the simulations names.}
	\label{SIDM_models}
\end{figure}

To model the interaction among DM particles we assume that DM particles $\chi$ interact under the exchange of a light mediator $\phi$, with the scattering following a Yukawa potential,

\begin{equation}
V(r)=-\frac{\alpha_{\chi}e^{-m_{\phi}r}}{r},
\end{equation}

\noindent where $\alpha_{\chi}\equiv g_{\chi}^{2}/4\pi$ is the dark fine structure constant and $g_{\chi}$ the coupling strength, $m_{\phi}$ is the mediator mass, and we define $m_{\chi}$ as the dark matter mass. There is no analytical form for the differential scattering cross-section due to a Yukawa potential, but by using the Born-approximation (\citealt{Ibe10}), valid when the scattering potential can be treated as a small perturbation, the differential cross-section of the DM-DM interactions results

\begin{equation}\label{sigmam}
\frac{{\rm{d}}\sigma}{{\rm{d}}\Omega}=\frac{\alpha_{\chi}^{2}}{m_{\chi}^{2}(m_{\phi}^{2}/m_{\chi}^{2}+v^{2}\sin^{2}(\theta/2))^{2}},
\end{equation}

\noindent which gives the following total cross-section,

\begin{equation}\label{sigma_total}
\sigma \equiv \int \frac{{\rm{d}}\sigma}{{\rm{d}}\Omega}{\rm{d}}\Omega = \frac{4\pi \alpha_{\chi}^{2}}{m_{\chi}^{2}(m_{\phi}^{2}/m_{\chi}^{2}+v^{2})}.
\end{equation}

In this velocity-dependent model, the scattering is anisotropic. This is because the velocity dependence results from a term in the scattering cross-section that depends on the exchanged momenta, which in turn depends on both the collision velocity and the scattering angle. For anisotropic scattering it is useful to consider the momentum transfer cross section,

\begin{equation}\label{sigmat}
\sigma_{T}/m_{\chi}\equiv 2\int(1-|\cos\theta|)\frac{{\rm{d}}\sigma}{{\rm{d}}\Omega}{\rm{d}}\Omega,
\end{equation}

\noindent for which interactions that lead to a large amount of momentum transfer contribute more, while those that transfer little momentum are down-weighted. \citet{Kahlhoefer15} shows that the momentum transfer cross section needs to be weighted by the scattering angle, in order to avoid overestimating the momentum transfer due to scattering with $\theta>\pi/2$, as in these cases the particles, which we assume to be identical, could be relabelled in such a way that they had scattered with $\theta<\pi/2$.

While \citet{Robertson17} (see also \citealt{Banerjee20}) implemented anisotropic scattering following eq.~(\ref{sigmam}), others (e.g. \citealt{Vogelsberger12,Zavala13,Vogelsberger16,Zeng21}) have instead simulated the scattering as isotropic but with a modified cross-section (e.g. eq~\ref{sigmat}), so that the effects of DM scattering closely follow the correct modelling of the particles interactions.

Fig.~\ref{SIDM_models} shows the momentum transfer cross sections of the SIDM models adopted in this work. The figure highlights three velocity-dependent models (dark blue, light blue and orange lines) where $\sigma_{T}/m_{\chi}$ reaches 100, 60 and 20 cm$^{2}$/g on $10^{9}~\rm{M}_{\odot}$ dwarf galaxies. Additionally three constant cross section models are considered $\sigma_{T}/m_{\chi}=10$ and 1 cm$^{2}$/g (black dashed and dotted lines) and $\sigma_{T}/m_{\chi}=0$ (CDM, not shown). While the bottom x-axis shows the relative velocity between DM particles, the top x-axis indicates the typical halo mass that hosts circular orbits of such velocities. 

The resulting $\sigma/m_{\chi}$ depends on the DM particles velocity, the DM particle mass, $m_{\chi}$, the mediator mass, $m_{\phi}$, and coupling strength, $\alpha$ of the interaction. These parameters have been adjusted so that the rate of scattering is important in dwarf DM haloes while being negligible in more massive (e.g. $>10^{11.5}\Msun$) haloes. This was done in order to avoid the destruction of satellite haloes in the simulations from excessive interactions between the DM particles from satellites and the host (\citealt{Nadler20}). Additionally, the velocity-dependent models are in agreement with the strong observational constraints from cluster-size haloes (see e.g.~\citealt{MiraldaEscude02,Randall08,Harvey15,Kim17,Wittman18,Harvey19,Sagunski21}). The models highlighted in dark blue line in Fig.~\ref{SIDM_models} has been fitted so that it matches the recent estimates of $\sigma/m_{\chi}$ on dwarf galaxy scales done by \citet{Correa21}. Note that although the model with $\sigma/m_{\chi}$=10 cm$^2$/g has been ruled out by observations of galaxy clusters, it will be used as a control model.

\begin{table*}
\begin{center}
\begin{tabular}{lrrrccccc}
\hline
\multicolumn{1}{c}{} & \multicolumn{3}{c}{\uline{SIDM parameters}} & \multicolumn{3}{c}{Cross Section} & \multicolumn{1}{c}{\uline{DM interaction}} \\
\cline{5-7}
Simulation & $m_{\chi}$ & $m_{\phi}$ & $\alpha$ & $\sigma_{T}/m_{\chi}$(10 km/s) & $\sigma_{T}/m_{\chi}$(50 km/s) & $\sigma_{T}/m_{\chi}$(100 km/s) &\\
Name & [GeV] & [MeV] &  & [cm$^{2}$/g]  & [cm$^{2}$/g] &  [cm$^{2}$/g] & \\
\hline\hline
CDM & / & / & / & 0 & 0 & 0 & No interaction\\
SigmaConstant1 & / & / & / & 1 & 1 & 1 & Isotropic\\
SigmaConstant10 & / & / & / & 10 & 10 & 10 & Isotropic\\
SigmaVel20 & 3.056 & 0.309 & $1.23\times 10^{-5}$ & 20 & 5 & 1 & Anisotropic\\
SigmaVel60 & 3.855 & 0.356 & $1.02\times 10^{-5}$& 60 & 12 & 2.5 & Anisotropic\\
SigmaVel100 & 4.236 & 0.350 & $4.96\times 10^{-6}$ & 100 & 19 & 3 & Anisotropic\\
\hline
\end{tabular}
\end{center}
\caption{SIDM models analysed in this work. Form left to right: Simulation name, SIDM parameters for each model (dark matter mass, $m_{\chi}$, mediator mass, $m_{\phi}$, and coupling strength, $\alpha$), momentum-transfer cross section at a relative velocity between DM particles of 10 km/s, 50 km/ and 100 km/s, DM type of interaction.}
\label{Table_models}
\end{table*}

\subsection{Simulations}

The simulations analysed in this paper are part of the TangoSIDM project, a simulation suite project that models cosmological simulations of structure formation in a $\Lambda$SIDM universe. TangoSIDM consists on a set of DM-only and hydrodynamical cosmological simulations of 25 Mpc on a side. These simulations have been produced using the SWIFT\footnote{https://swift.dur.ac.uk} code (\citealt{Schaller16,Schaller18}), that has been enhanced to include new DM physics modules (Sec.~\ref{SIDM_implementation}). SWIFT is an open-source, fast and accurate gravity and hydrodynamics solver that was specifically designed to be efficient on many core systems with several levels of parallelisation including vectorisation. It uses state-of-the-art algorithms to solve the equations of hydrodynamics and a modern gravity solver.

The analysis in this work focuses on six DM-only simulations of (25 Mpc)$^{3}$ that follows the evolution of 752$^{3}$ DM particles, reaching a spatial resolution of 650 pc and a mass resolution of $1.44\times 10^{6}~\rm{M}_{\odot}$. We use a comoving softening of 1.66 kpc at early times, which freezes at a maximum physical value of 650 pc at $z = 2.8$. The starting redshift of these simulations is $z=127$. The initial conditions were calculated using second-order Lagrangian perturbation theory with the method of \citet{Jenkins10,Jenkins13}. The adopted cosmological parameters are $\Omega_{\rm{m}}=0.307$, $\Omega_{\Lambda}=0.693$, $h=0.6777$, $\sigma_{8}=0.8288$ and $n_{S}=0.9611$.

Table~\ref{Table_models} highlights the SIDM model parameters adopted in this work. Each simulation from the TangoSIDM suite includes a different DM model, while three simulations have a constant scattering cross section, $\sigma/m_{\chi}$, of 10 cm$^{2}$/g (named SigmaConstant10), 1 cm$^2$/g (SigmaConstant1) and 0 cm$^2$/g (CDM), the other cosmological boxes have a $\sigma/m_{\chi}$ that depends on the particles velocity, as indicated by eq.~(\ref{sigma_total}). Table~\ref{Table_models} shows the $\sigma_{T}/m_{\chi}$ for the velocity-dependent models at DM particles velocities of 10, 50 and 100 km/s. The model that reaches $\sigma_{T}/m_{\chi}=100$ cm$^{2}$/g at 10 km/s is called SigmaVel100, similarly, the models reaching 60 and 20 cm$^{2}$/g at 10 km/s are called SigmaVel60, and SigmaVel20, respectively. The table indicates which models assume either isotropic or anisotropic scatter, and it also includes the values of the $m_{\chi}$, $m_{\phi}$ and $\alpha$ parameters that describe the velocity-dependent cross sections.

Halo catalogues and merger trees were generated using the VELOCIraptor halo finder (\citealt{Elahi11,Elahi19,Canas19}). VELOCIraptor uses a 3D-friends of friends (FOF) algorithm (\citealt{Davis85}) to identify field haloes, and subsequently applies a 6D-FOF algorithm to separate virialised structures and identify sub-haloes of the parent haloes (\citealt{Elahi19}). To link haloes through time, we use the halo merger tree code TreeFrog (\citealt{Elahi19b}), developed to work on the outputs of VELOCIraptor. Throughout this work, virial halo masses ($M_{200c}$) are defined as all matter within the virial radius $R_{200c}$, for which the mean internal density is 200 times the critical density. In each FOF halo, the `central' halo is the halo closest to the center (minimum of the potential), which is nearly always the most massive. The remaining haloes within the FOF halo are its satellites, also called subhaloes. For satellites, we do not use $M_{200c}$ for their mass definition, instead we use $M_{\rm{peak}}$ defined as the $M_{200c}$ mass that the satellite had before being accreted by a central more massive halo, and becoming a satellite. VELOCIraptor provides virial masses and radii for subhaloes, as well as for the main haloes. Additionally it calculates the concentration parameter ($c_{200c}$), defined as the ratio between $R_{200c}$ and the scale radius, $r_{s}$ (radius at which the logarithmic density slope is -2). The particle mass resolution of the simulations is sufficient to resolve (sub-)haloes down to ${\sim}10^{9}~\rm{M}_{\odot}$ with $10^{3}$ particles. 

\subsection{Self-interacting dark matter implementation}\label{SIDM_implementation}

We have modelled the interaction between DM simulation particles following a stochastic approach, where two DM particles $a$ and $b$ have a probability of interaction, $P_{ab}$, that depends on $\sigma/m_{\chi}$, as well as on the distance ($\delta {\bf{r}}_{ab}$) and relative velocity between them ($|{\bf{v}}_{a}-{\bf{v}}_{b}|$) as follows,

\begin{equation}\label{Probability}
P_{ab} = m_{b}(\sigma/m_{\chi})|{\bf{v}}_{a}-{\bf{v}}_{b}|g_{ab}(\delta {\bf{r}}_{ab})\Delta t,
\end{equation}

\noindent where 

\begin{equation}\label{double_kernel}
g_{ab} (\delta {\bf{r}}_{ab})= N\int_{0}^{{\rm{max}}(h_{a},h_{b})}d^{3}{\bf{r}}'W(|{\bf{r}}'|,h_{a})W(|\delta {\bf{r}}_{ab}+{\bf{r}}'|,h_{b}),
\end{equation}

\noindent with $W$ the particles kernel and $N$ a normalization factor. The derivation and full terms of the DM particles probability is further detailed in Appendix \ref{Appendix_SIDM_derivation}. For the simulations where the cross section is velocity-dependent, the term $\sigma/m_{\chi}$ in eq.~(\ref{Probability}) is given by eq.~(\ref{sigma_total}) and therefore it depends on the particles' relative velocity. In the simulations with constant cross section, the term $\sigma/m_{\chi}$ is constant and set to the value assumed in the simulation (1 for SigmaConstant1, or 10 for SigmaConstant10).

In eq.~(\ref{double_kernel}), the parameters $h_{a}$ and $h_{b}$ are the particles' search radii. The search radius, which encloses a region where a DM particle has the probability of interacting with its neighbours, is not constant, instead it follows the smoothing length of the DM particles kernel using an approach similar to SPH (e.g.~\citealt{Price12}). It is therefore adjusted according to the local DM density, allowing to better track the centre of objects. 

We compare $P_{ab}$ with a random number (that ranges between 0 and 1), if $P_{ab}$ is larger, the particles $a$ and $b$ are to scatter. Then, given the particles velocities ${\bf{v}}_{a}$ and ${\bf{v}}_{b}$, we move to the centre of momentum frame where the velocities result ${\bf{v'}}_{a}$ and ${\bf{v'}}_{b}=-{\bf{v'}}_{a}$. We use the direction of ${\bf{v'}}_{a}$ to define the $z$-axis, from which the polar scattering angle $\theta$ is measured. Given the two angles, $\theta$ and azimuthal $\phi$, that determine the unit vector ${\bf{\hat{e}}}$, the post-scatter velocities are 

\begin{eqnarray}
{\bf{\tilde{v}}}_{a}={\bf{V}}-w~{\bf{\hat{e}}},\\
{\bf{\tilde{v}}}_{b}={\bf{V}}+w~{\bf{\hat{e}}},
\end{eqnarray}

\noindent where ${\bf{V}}=({\bf{v}}_{a}+{\bf{v}}_{b})/2$, $w=|{\bf{v}}_{a}-{\bf{v}}_{b}|/2$. 

In the simulations, SigmaVel100, SigmaVel60 and SigmaVel20, where the cross section is velocity-dependent, the interactions are modelled as elastic and anisotropic collisions. We assume that the scattering potential follows the Yukawa potential (introduced in Section \ref{Scattering_cross_section_Section}), that produces an azimuthally-symmetric differential cross section (eq.~\ref{sigmam}), with a total cross-section as indicated in eq.~(\ref{sigma_total}). We determine the polar angle of scattering, $\theta$, following the probability density function,

\begin{equation}
p(\theta)=\frac{2\pi \sin\theta}{\sigma}\frac{\rm{d}\sigma}{\rm{d}\Omega}.
\end{equation}

\noindent where ${\rm{d}}\sigma/{\rm{d}}\Omega$ is given by eq.~(\ref{sigmam}) and $\sigma$ given by eq.~(\ref{sigma_total}). Integrating $p(\theta)$, we obtain the cumulative distribution function

\begin{equation}\label{Ptheta}
P(\theta)=\int_{0}^{\theta}p(\theta')\rm{d}\theta',
\end{equation}

\noindent of the probability that a particle scatters by an angle less than $\theta$. We draw a random variable, $X$, with a uniform distribution in the interval [0, 1], and calculate $\theta$, so that $P(\theta)=X$. 

In the simulations SigmaConstant1 and SigmaConstant10, where the cross section is constant, $\frac{\rm{d}\sigma}{\rm{d}\Omega}=\frac{\sigma}{4\pi}$. In this case the particles collisions are isotropic, where $\theta = \arccos(1-2X)$, and $X$ and $\phi$ are drawn from uniform distributions in the interval [0, 1] and [0, $2\pi$], respectively. 

SWIFT uses a KDK leapfrog time-stepping scheme,

\begin{eqnarray}\nonumber
v_{i+1/2} &=& v_{i}+a_{i}\Delta t/2,\\\nonumber
x_{i+1} &=& x_{i}+v_{i+1/2}\Delta t,\\\nonumber
v_{i+1} &=& v_{i+1/2}+a_{i+1}\Delta t/2,
\end{eqnarray}

\noindent where $x_{i}$ and $v_{i}$ are the positions and velocities at time step $i$, $a_{i}=a(x_{i})$ is the acceleration, or second derivative of $x$, at step $i$, $\Delta t$ is the size of each time step, and $x_{i+1}$, $v_{i+1}$ and $a_{i+1}$ correspond to positions, velocities and accelerations at step $i+1$. The SIDM scattering implementation modifies this scheme by inserting an extra kick driven by the collision between two particles. The extra kick modifies the initial particles' velocity, $v_{i+1/2}$ into $\tilde{v}_{i+1/2,\rm{SIDM}}$, here $v_{i+1/2}$ corresponds to the particles velocity after kick 1.

\begin{eqnarray}\nonumber
K(\Delta t/2) &:& v_{i+1/2}=v_{i}+a_{i}(x_{i})\Delta t/2,\\\nonumber
D(\Delta t) &:& x_{i+1}=x_{i}+v_{i+1/2}\Delta t,\\\nonumber
S(\Delta t) &:& {\tilde{v}_{i+1/2,\rm{SIDM}}}={\rm{Scatter}}(v_{i+1/2},x_{i+1},\Delta t),\\\nonumber
D(\Delta t/2) &:& \tilde{x}_{i+1}=x_{i+1}-v_{i+1/2}\Delta t/2+\tilde{v}_{i+1/2,\rm{SIDM}}\Delta t/2,\\\nonumber
K(\Delta t/2) &:& v_{i+1}=\tilde{v}_{i+1/2,\rm{SIDM}}+a_{i+1}(\tilde{x}_{i+1})\Delta t/2.
\end{eqnarray}

After the extra kick is introduced, the particles involved are drifted backwards half a step and then drifted forward half a step with the new velocities. SIDM kicks are assumed to be instantaneous, particles velocities are modified and also their positions. It is important to note that if an activate DM particle in the time-step $i$ kicks an inactive neighbour, the inactive particle is awakened for the following time-step and drifted accordingly.

The SIDM scattering is implemented on a particle-pair by particle-pair basis. The higher the cross section, the larger the probability of particles scattering, the larger the scattering events for the same particle in a single time-step. To conserve energy it is important that the scattering events are dealt in an appropriate way. Since the momentum kick from one scattering event alters the velocities of the particles for any future scattering event, we cannot allow a particle to scatter twice (or more) in a single time-step with the same initial velocity. Therefore we follow \citet{Vogelsberger12} and choose the individual particles time-step $\Delta t_{i}$ small enough by requiring that 

\begin{equation}\label{criterion}
\Delta t_{i}<\kappa\times[\rho_{a} \left\langle\sigma/m_{\chi}\right\rangle(v_{a}) \sigma_{v,a}]^{-1},
\end{equation}

\noindent where $\kappa=10^{-2}$, $\rho_{a}$ is the density of the DM particle $a$, $\left\langle\sigma/m_{\chi}\right\rangle$ is the average total cross section of the particle $a$ moving with velocity $v_{a}$ relative to its neighbours, and $\sigma_{v,a}$ is the velocity dispersion at the position of particle $a$. When comparing $\Delta t$ with the gravity time-step criterion $\Delta t_{\rm{grav}}$ ($\Delta t_{\rm{grav}}\propto \sqrt{\epsilon/|\boldsymbol{\rm{a}}|}$, with $\epsilon$ softening and {\bf{a}} gravitational acceleration),  we find that for the SigmaConstant1 model $\Delta t$ is always larger than $\Delta t_{\rm{grav}}$, except for the inner regions (< 3-5 kpc) of haloes more massive than $10^{11}$ M$_{\odot}$, where $\Delta t<\Delta t_{\rm{grav}}$ by up to a factor of 5. In the SigmaConstant10 model the same occurs, except that $\Delta t$ is a factor of 5 to 10 smaller than $\Delta t_{\rm{grav}}$ in the inner regions of haloes more massive than $10^{10}$ M$_{\odot}$. In the SigmaVel models, $\Delta t$ rapidly decreases in the inner regions (due to the increase of $\sigma_{T}/m_{\chi}$), reaching up to a factor of 100 lower values than $\Delta t_{\rm{grav}}$. This results in the SigmaVel models being computationally more expensive than the SigmaConstant1 and SigmaConstant10, and also CDM models. However, we find it necessary to implement eq.~(\ref{criterion}) in order to accurately model the scattering events.

Further details of the model, including validation tests and comparisons with previous simulation efforts, are included in Appendix \ref{Appendix_SIDM_derivation} and \ref{Model_validation_2_sec}. 

\begin{figure*} 
	\includegraphics[angle=0,width=0.49\textwidth]{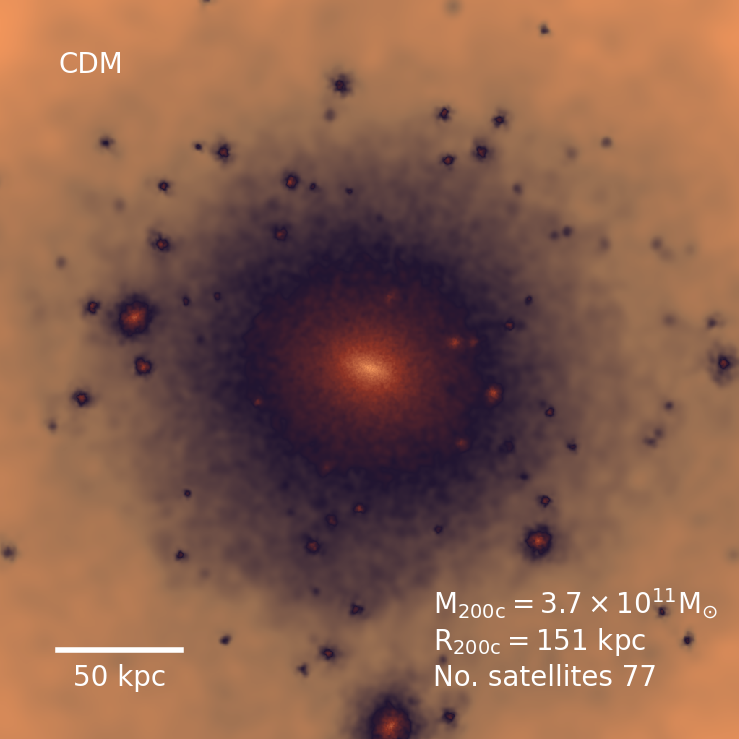}
	\includegraphics[angle=0,width=0.49\textwidth]{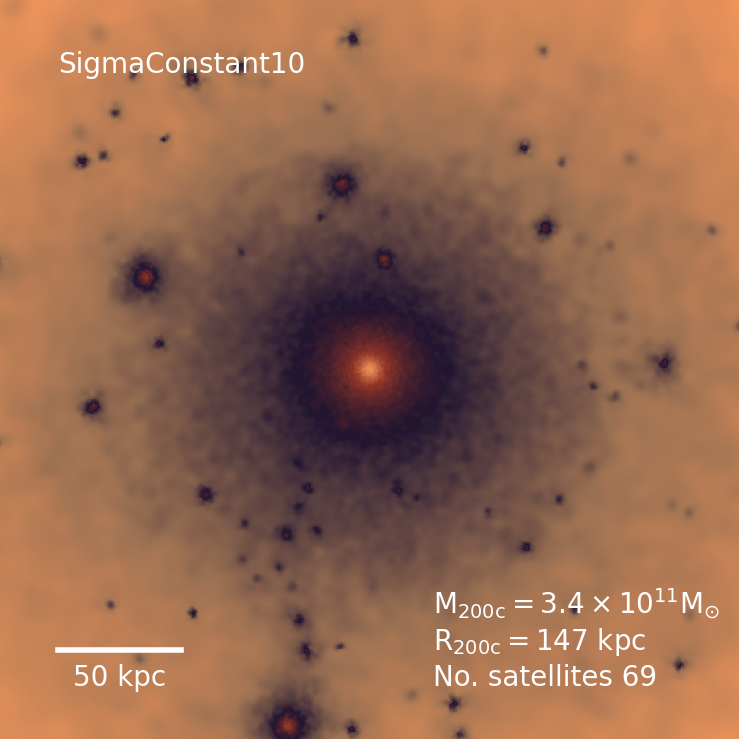}\\
	\hspace{0.01cm}
	\includegraphics[angle=0,width=0.49\textwidth]{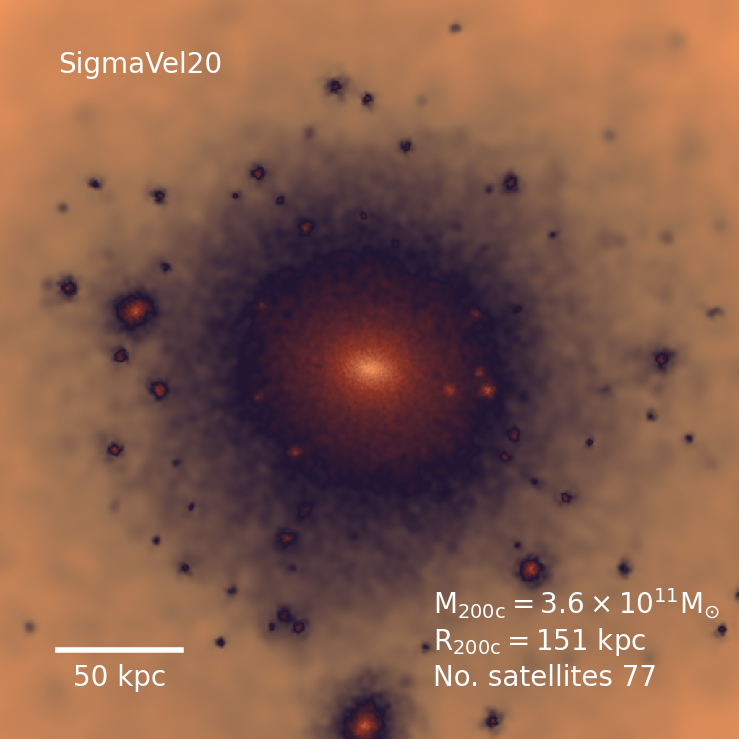}
	\includegraphics[angle=0,width=0.49\textwidth]{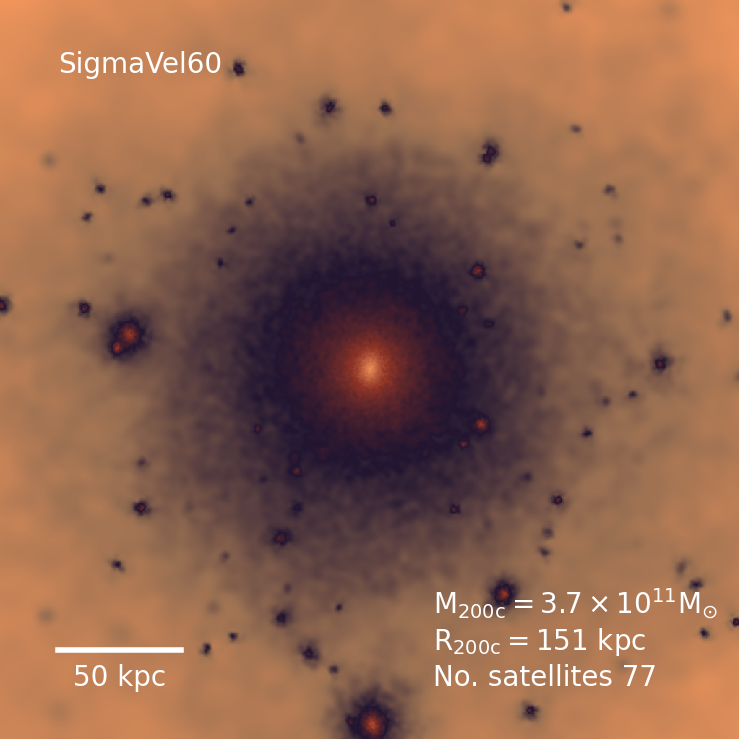}
	\caption{Density projections of the same central halo for some of the different models listed in Table~\ref{Table_models}. The central halo has a mass of $3.7\times 10^{11}~\rm{M}_{\odot}$ (CDM case, top-left panel), note, however, that this can vary depending on the SIDM model. The projection cube has a side length of 300 kpc. The SigmaConstant10 model (top-right panel), that has a constant cross section of 10 cm$^{2}$/g, not only largely destroys the surrounding low-mass subhaloes, but also thermalises the central halo, changing its elongated shape (as seen from the top-left panel) to spherical. The velocity-dependent model, SigmaVel20 (bottom-left panel), cannot be easily distinguished from the CDM case. The SigmaVel60 (bottom-right panel) shows a spherical-looking central halo but without subhalo disruption.}
	\label{SIDM_haloes}
\end{figure*}

\section{Results}

This section presents the first results of the TangoSIDM simulations. We analyse the evolution of satellites and the impact of the DM particles interactions in their internal structure.

\subsection{Subhalo population}

\begin{figure} 
    \centering
	\includegraphics[angle=0,width=0.45\textwidth]{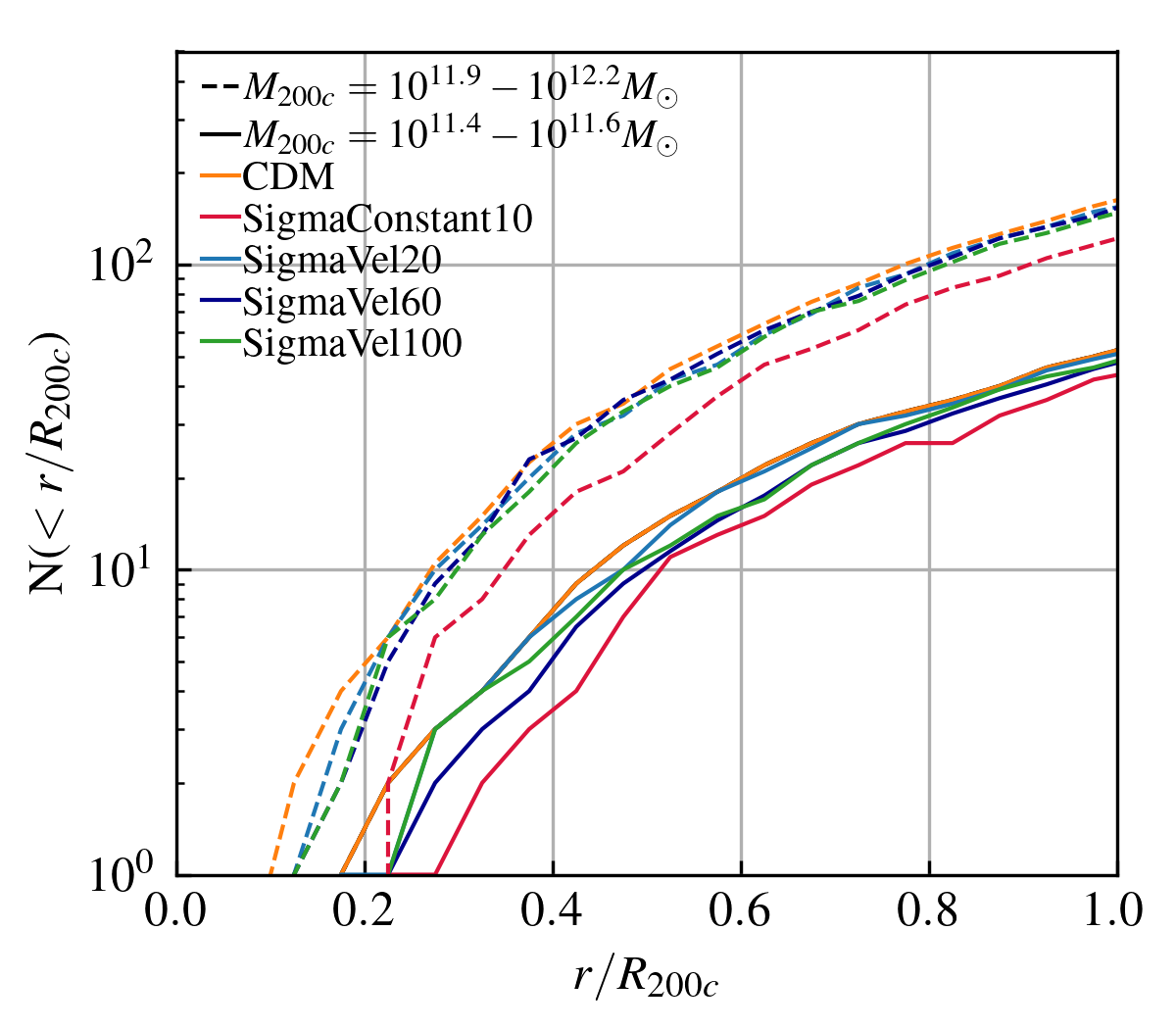}
	\caption{The average cumulative number of satellite haloes more massive than $10^{8}~\rm{M}_{\odot}$, as a function of the distance from the halo center (normalized by the virial radius, $R_{200c}$). Colour lines show the number of satellites in the CDM (orange line), SigmaConstant10 (red line), SigmaVel20 (light blue line), SigmaVel60 (dark blue line) and SigmaVel190 (green line) models. Dashed (solid) lines highlight the number of satellites around central haloes with $M_{200c}$ masses between $10^{11.9}$ and $10^{12.1}~\rm{M}_{\odot}$ ($10^{11.4}-10^{11.6}~\rm{M}_{\odot}$). There is no significant satellite disruption in the $10^{12}~\rm{M}_{\odot}$ haloes from the SigmaVel models relative to CDM. SigmaConstant10 shows lower number of satellites in both, the $10^{11.5}$ and $10^{12}~\rm{M}_{\odot}$ central haloes. In the $10^{11.5}~\rm{M}_{\odot}$ haloes, the SigmaVel60 and SigmaVel100 models have a lower number of satellites than SigmaVel20 and CDM, but a larger number relative to SigmaConstant10.}
	\label{SubhaloMassFunction}
\end{figure}

The simulations from the TangoSIDM project are all run from the same initial conditions, therefore it is possible the match the $z=0$ central haloes between simulations and compare their respective subhalo population. Fig.~\ref{SIDM_haloes} shows density projections of the same central halo for the different models listed in Table~\ref{Table_models}. The projection cubes have a side length of 300 kpc. In the CDM simulation (top-left panel), the central halo has a $M_{200c}$ mass of $3.7\times 10^{11}~\rm{M}_{\odot}$, a virial radius of 151 kpc and 77 satellites more massive than $10^8~\rm{M}_{\odot}$ that reside within $R_{200c}$. The SigmaConstant10 model (top-right panel) depicts a more spherical looking halo, of lower mass and fewer number of satellites. As expected, the frequent DM particle collisions in this simulation isotropise the particles orbit and produce a more spherical configuration (\citealt{MiraldaEscude02,Peter13,Vogelsberger12}). 

The visual impression from  Fig.~\ref{SIDM_haloes} shows that while the halo in the SigmaConstant10 model is quite spherical, it is elliptical in the SigmaVel20 model (bottom-left panel), looking similar to CDM (top-left panel). This is because at the scale of $10^{11}~\rm{M}_{\odot}$, SigmaVel20 is characterized by a momentum-transfer cross section of ${\sim}1$ cm$^2$/g (see Fig.~\ref{SIDM_models}), and therefore the rate of interactions is lower than in SigmaConstant10. In SigmaVel60 (bottom-right panel), $\sigma_{T}/m_{\chi}$ reaches 3-4 cm$^2$/g, and that seems to be a sufficient increase in $\sigma_{T}/m_{\chi}$ relative to the SigmaVel20 model to modify the halo's shape and make it slightly more spherical. 

SIDM interactions not only isotropise the DM particles' orbit, but also enhance the disruption of subhaloes by tidal stripping from the host (\citealt{Vogelsberger12,Nadler20}). The density projection from Fig.~\ref{SIDM_haloes} shows a larger number of destroyed satellites in the SigmaConstant10 model, relative to the SigmaVel models. Note that in the former the cross section has no velocity dependence, the rate of particle scattering is independent of the halo mass. This causes a larger number of DM particle interactions between the host and the satellite haloes. Although this model is ruled out by observational constraints, it still serves as a control study. It is interesting to also note that while SigmaVel60 model depicts a rather spherical-looking central halo, it does not seem to largely disrupt its subhalo population.

To better understand the rate of subhalo disruption in the different SIDM models, Fig.~\ref{SubhaloMassFunction} shows the cumulative number of satellite haloes around central haloes. We calculate the cumulative number of satellites around each central halo in radial bins of $0.05\times R_{200c}$, and then estimate the median. We select satellites more massive than $10^{8}~\rm{M}_{\odot}$ that orbit central haloes with virial masses between $10^{11.4}-10^{11.6}~\rm{M}_{\odot}$ (solid lines), and between $10^{11.9}$ and $10^{12.1}~\rm{M}_{\odot}$ (dashed lines). We do not find significant satellite disruption in the ${\sim}10^{12}~\rm{M}_{\odot}$ haloes from the SigmaVel models relative to CDM. As expected, SigmaConstant10 shows lower number of satellites in both, the ${\sim}10^{11.5}$ and ${\sim}10^{12}~\rm{M}_{\odot}$ central haloes. Interestingly, in the $10^{11.5}~\rm{M}_{\odot}$ haloes, the SigmaVel60 and SigmaVel100 models have a lower number of satellites than SigmaVel20 and CDM, but a larger number relative to SigmaConstant10. This indicates that even thought the cross section is velocity-dependent, there is still an impact of the particles collisions on the subhalo destruction.

\begin{figure*} 
	\includegraphics[angle=0,width=\textwidth]{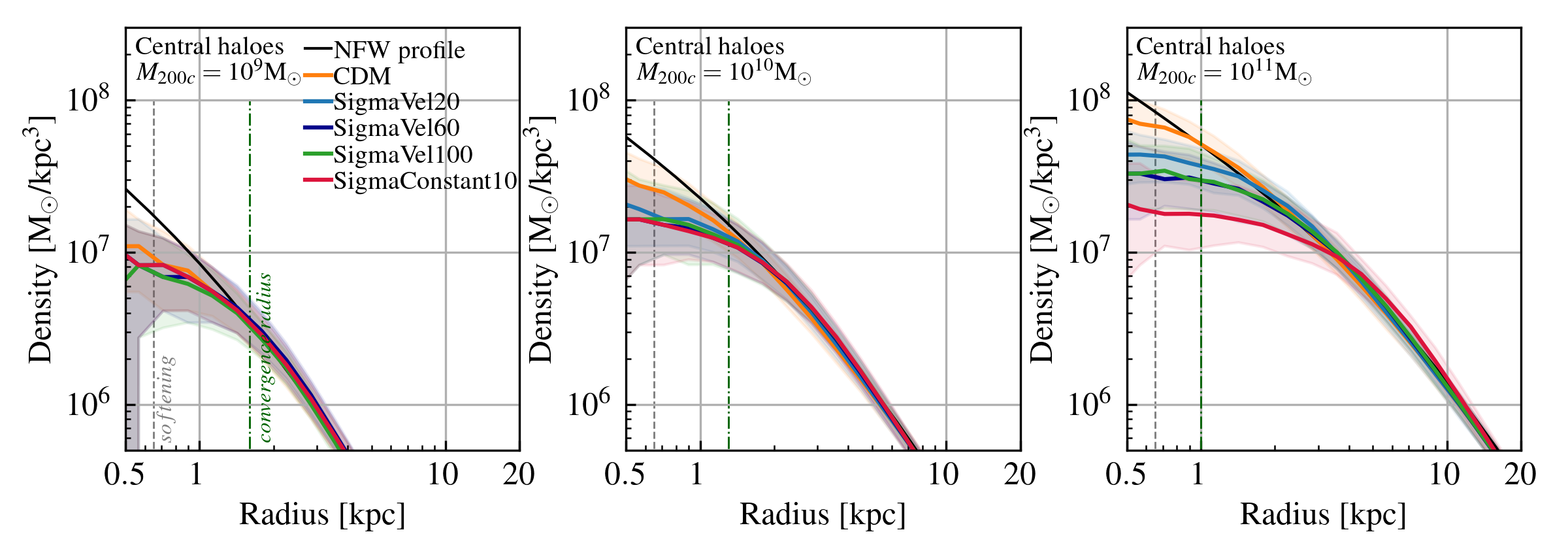}\\
	\vspace{-0.2cm}
	\includegraphics[angle=0,width=\textwidth]{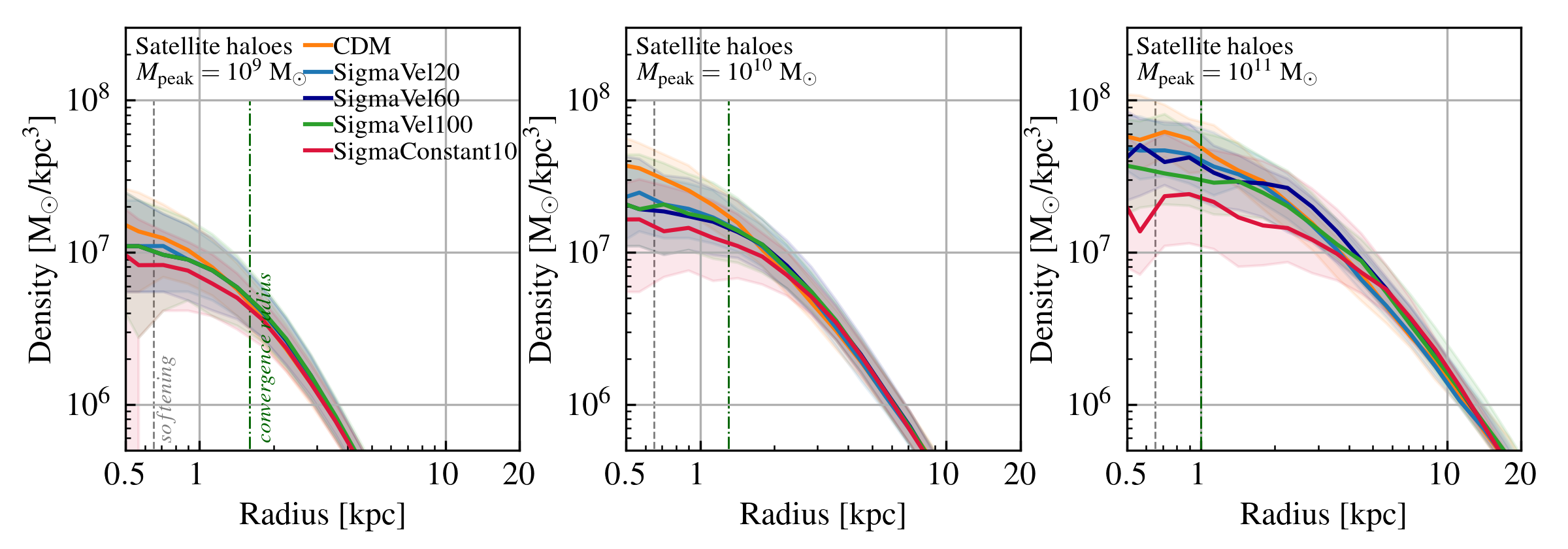}
	\caption{Density profile of central (top panels) and satellite (bottom panels) haloes from the CDM (orange lines), SigmaConstant10 (red lines), SigmaVel20 (light blue lines), SigmaVel60 (dark blue lines) and SigmaVel100 (green lines) models. The panels show the density profiles of $10^{9}~\rm{M}_{\odot}$ (left-panels), $10^{10}~\rm{M}_{\odot}$ (middle-panels) and  $10^{11}~\rm{M}_{\odot}$ (right-panels) haloes. The solid lines highlight the median values and the shaded regions the 16-84th percentiles. The black solid line in the top panels corresponds to the NFW profile, the green dashed-dotted lines indicate the convergence radius (see text for definition) and the grey dashed lines the softening scale. The difference in the profiles between same-mass haloes highlights the impact of dark matter particle interactions in the haloes densities.}
	\label{Density_Profiles}
\end{figure*}

\subsection{Density profiles}\label{Density_profiles_Section}

In this section we explore how the different rate of DM particles interactions shape the density profiles of central and satellite haloes. We select haloes with virial masses around $10^{9}~\rm{M}_{\odot}$, $10^{10}~\rm{M}_{\odot}$, and $10^{11}~\rm{M}_{\odot}$ (within 0.2 dex) and calculate the median densities in logarithmic radial bins of 0.2 dex. Fig.~\ref{Density_Profiles} shows the density profiles of central (top panels) and satellite (bottom panels) haloes from the CDM (orange lines), SigmaConstant10 (red lines), SigmaVel20 (light blue lines), SigmaVel60 (dark blue lines) and SigmaVel100 (green lines) models. The left panel includes $10^{9}~\rm{M}_{\odot}$ haloes, the middle panel $10^{10}~\rm{M}_{\odot}$ haloes, and the right panel $10^{11}~\rm{M}_{\odot}$ haloes. 

For comparison we include the NFW density profile in the top panels (solid black lines), which we estimated using the concentration-mass relation from \citet{Correa15c}. We also include the softening (grey dashed lines) and a convergence radius (green dashed-dotted lines). The original convergence criterion derived by \citet{Power03}, defined a convergence radius, $R_{P03}$, as the minimum radius where the mean density converges at the 10 per cent level relative to a simulation of higher resolution. However, because our SIDM models produce a larger variation in the haloes' internal density relative to the CDM simulation, we follow \citet{Schaller15} and relax \citet{Power03} convergence criterion requiring that the mean internal density converges at the 50 per cent level instead\footnote{Relaxing Power et al. (2003) criterion corresponds to calculating $R_{P03}$ as $0.15\leq \frac{\sqrt{200}}{8}\sqrt{\frac{4\pi\rho_{\rm{crit}}}{3m_{\rm{DM}}}}\frac{\sqrt{N(<R_{P03})}}{\ln N(<R_{P03})}R_{P03}^{3/2}$, where $N(<r)$ is the number of particles of mass, $m_{\rm{DM}}$, within radius $r$.}. $R_{P03}$ then results ${\sim}1.6$ kpc for $10^{9}~\rm{M}_{\odot}$ haloes, 1.3 and 1 kpc for $10^{10}~\rm{M}_{\odot}$ and $10^{11}~\rm{M}_{\odot}$ haloes, respectively.

The top panels of Fig.~\ref{Density_Profiles} show that while there is no visible core formation in the $10^{9}~\rm{M}_{\odot}$ haloes, higher mass haloes begin to exhibit a core. It can be seen from the top-right panel that $10^{11}~\rm{M}_{\odot}$ haloes in the SigmaConstant10 model form the largest core, reaching roughly a constant density of $2\times 10^{7}$M$_{\odot}$/kpc$^{3}$. In this constant cross-section model, core expansion maximizes in high-mass haloes. This is expected since the scattering rate of DM particles, $\Gamma(r)=\langle\sigma/m_{\chi}v_{\rm{pair}}\rangle(r)\rho(r)$, depends on the local density, so that DM particles in high-mass (${\sim}10^{11}~\rm{M}_{\odot}$) haloes, with typical central densities that reach $10^{8}~\rm{M}_{\odot}$/kpc$^{3}$, experience more frequent collisions, than lower mass haloes. Frequent DM particles collisions expel particles in wider out orbits, lower the central halo density, and increase the particles velocities, forming a `hot core'.

Core expansion is maximised at a time of around $t_{c}=25t_{0}$ (\citealt{Koda11}), with $t_{0}$ being

\begin{equation}
t_{0}^{-1} = \frac{\sigma_{T}/m_{\chi}}{2}\sqrt{\frac{GM_{200c}^{3}}{r_{s}^{7}}},
\end{equation}

\noindent where $r_{s}$ is the halo scale radius and $M_{200c}$ its mass. For a constant cross-section of $10$ cm$^{2}$/g, $t_{c}$ is around ${\sim}12.7$ Gyr for $10^{11}~\rm{M}_{\odot}$ haloes, ${\sim}18.6$ Gyr for $10^{10}~\rm{M}_{\odot}$ haloes and ${\sim}28.5$ Gyr for $10^{9}~\rm{M}_{\odot}$ haloes.

The top-right panel of Fig.~\ref{Density_Profiles} compares the $z=0$ density profile of $10^{11}~\rm{M}_{\odot}$ haloes from different SIDM models. It can be seen that the SigmaVel models do not include haloes with central cores as large as the SigmaConstant10 model. This is because the velocity-dependent models have lower cross-sections than 10 cm$^{2}$/g at this mass scale. Therefore these haloes are still in the process of core expansion.

The bottom panels of Fig.~\ref{Density_Profiles} compare the $z=0$ density profile of satellite haloes. At fixed radius satellite haloes exhibit higher densities than central haloes of the same mass. Because of this, satellite haloes are expected to host larger rates of DM particles interactions. Interesting to note that the bottom panels show a larger scatter around the median densities profiles of satellites, than in the same-mass centrals.

\begin{figure*} 
	\includegraphics[angle=0,width=\textwidth]{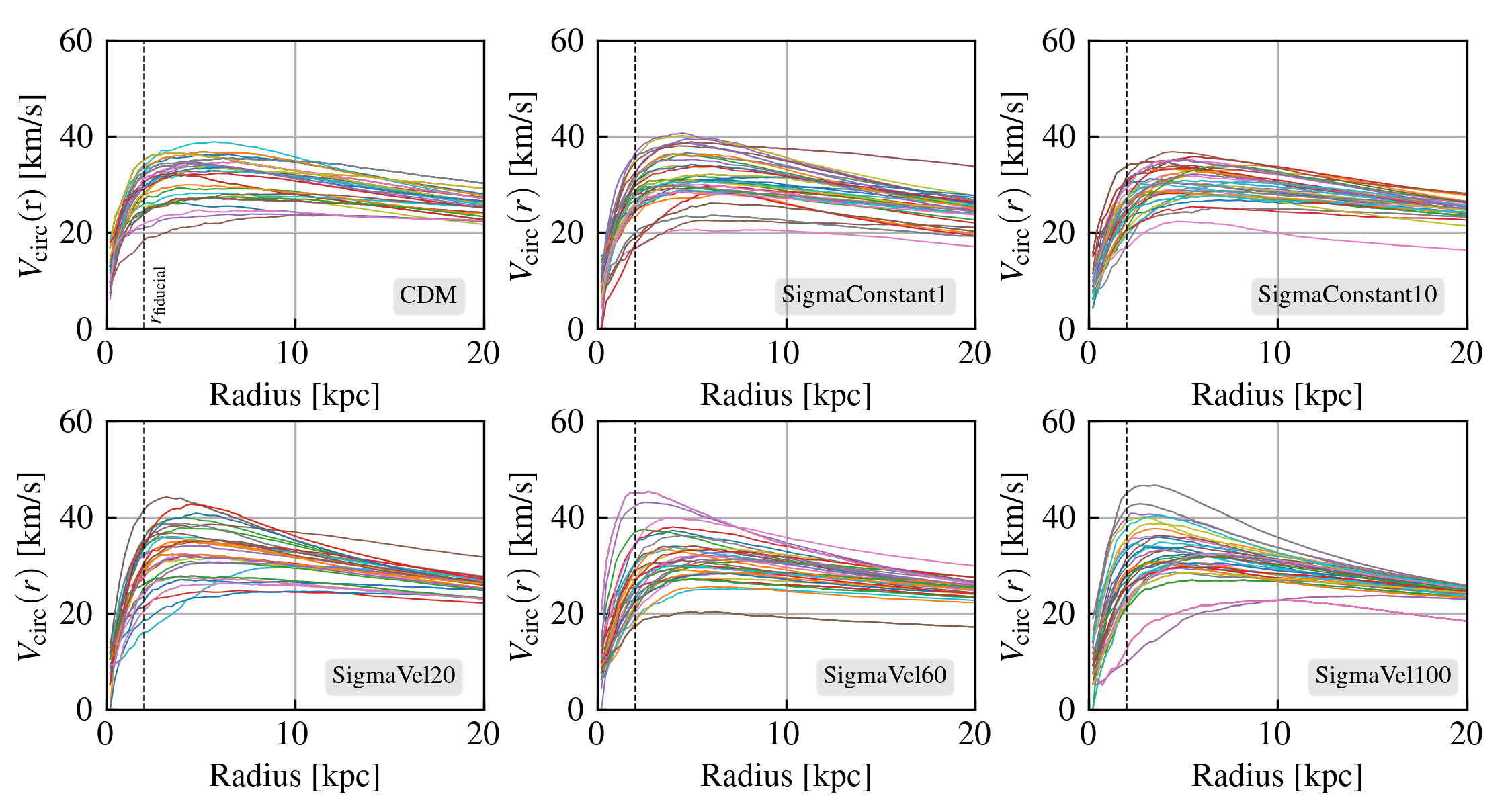}
	\vspace{-0.5cm}
	\caption{Circular velocity profiles of satellite haloes selected randomly in the mass range $10^{9.5-9.6}~\rm{M}_{\odot}$. The various colour lines show the profiles on the individual haloes, and each panel shows the `spread' in the haloes circular velocities from the CDM (top-left), SigmaConstant1 (top-middle), SigmaConstant10 (top-right), SigmaVel20 (bottom-left), SigmaVel60 (bottom-middle) and SigmaVel100 (bottom-right) models.}
	\label{RotationCurves}
\end{figure*}

\begin{figure} 
	\begin{center}
	\includegraphics[angle=0,width=0.43\textwidth]{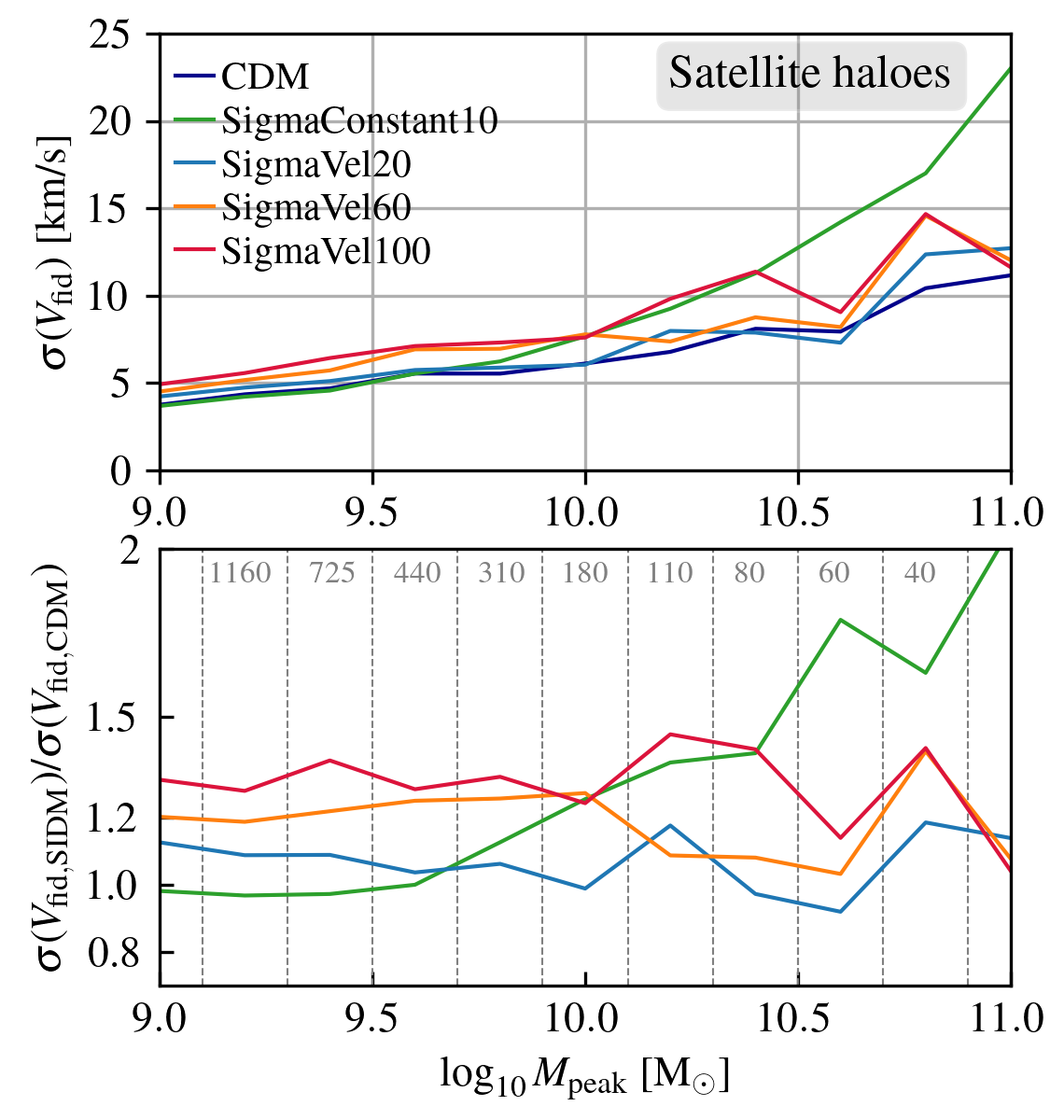}
	\caption{Scatter of the satellite haloes circular velocity as a function of halo mass. The curves correspond to the standard deviation of $V_{\rm{circ}}(r_{\rm{fid}})=V_{\rm{fid}}$ (circular velocity at the fiducial radius, defined in eq.~\ref{r_fiducial}). The various colour lines show the scatter of $V_{\rm{fid}}$ for satellite haloes in the CDM (dark blue line), SigmaConstant10 (green line), SigmaVel20 (light-blue line), SigmaVel60 (orange line) and SigmaVel100 (red line) models. The top panel shows the scatter as a function of halo mass, whereas the bottom shows the ratio between scatter from the SIDM models (SigmaConstant10, SigmaVel20, SigmaVel60 and SigmaVel100) and the CDM model. Each line highlights the respective model following the legends from the top panel. The grey numbers in the bottom panel indicates the umber of satellites (averaged over all simulations).}
	\label{Vfid_Scatter}
	\end{center}
\end{figure}

\subsection{Subhalo rotation curves}

Observations of rotation curves of dwarf galaxies reveal significant diversity in their shapes (\citealt{Oman15}). While many dwarf galaxies' rotation curves rise slowly toward the galaxies' outskirts, indicative of cored DM density profiles, others rise rapidly, indicative of cuspy profiles. This diversity has not only been observed in dwarf galaxies from the field, but also in the derived DM density profiles from the local spherical dwarfs, and ultra-faint dwarfs, that are satellites of the MW (see e.g. \citealt{Hayashi22, Hayashi21}).

Dwarf galaxies are dark-matter dominated systems, with a low contribution of baryons in their total mass. Although it has been shown that in hydrodynamical simulations DM halos of dwarf galaxies can be significantly modified by baryonic processes (e.g., \citealt{Governato10,Governato12,DiCintio14,Tollet16,Read16,Santos18}), this is very specific to the subgrid model adopted in the simulation (\citealt{BenitezLlambay19,Dutton20}). Throughout this work we relate the dwarf galaxies DM density with our results from pure DM-only simulations, under the assumption that, in the absence of SIDM, these profiles would be cuspy (as it was found in \citealt{Bose19}). More detailed analysis on the impact of baryons + SIDM will be addressed in a companion study (Correa et al. in prep.).

In this section we turn our focus to the circular velocity profiles of the smallest haloes from our sample. We select 30 random satellites haloes in the mass range of $10^{9.5-9.6}~\rm{M}_{\odot}$ from the CDM and SIDM simulations, and calculate their spherical circular velocity curves, $V^{2}_{\rm{circ}}(r)=GM({<}r)/r$, where $r$ is the 3D radius and $M({<}r)$ is the total mass enclosed within such radius. The goal of this section is to first produce a visual representation of the velocity profiles from low-mass haloes from the various models. In order to understand whether models with a velocity-dependent cross section can produce a diverse sample of rotation curves in same-mass haloes. Throughout this work we focus in the evolution of satellite haloes, but in Appendix~\ref{Diversity_Centrals} we extend the analysis for centrals.

Fig.~\ref{RotationCurves} shows the circular velocity profiles of 30 randomly selected satellite haloes from the CDM (top-left), SigmaConstant1 (top-middle), SigmaConstant10 (top-right), SigmaVel20 (bottom-left), SigmaVel60 (bottom-middle) and SigmaVel100 (bottom-right) models, in the mass range of $10^{9.5-9.6}~\rm{M}_{\odot}$. The various colour lines correspond to the profiles of the individual haloes. This allows for a comparison of the spread in the haloes circular velocities between the models, as well as for a visual inspection on the shapes. The black dashed line in the panels highlights a 2 kpc fiducial radius. Note that this radius corresponds is larger than the convergence radius of $10^{9}~\rm{M}_{\odot}$ haloes. We find that while the SigmaConstant10 model exhibits cored profiles with no large scatter in the circular velocities at 2 kpc, the SigmaVel100 model shows the highest spread, counting with both very cuspy and very cored profiles.

For a better comparison of the haloes rotational curves between the models, we define a fiducial radius, $r_{\rm{fid}}$, as

\begin{equation}\label{r_fiducial}
r_{\rm{fid}}=2\times (M/10^{9}~{\rm{M}}_{\odot})^{0.2},
\end{equation}

\noindent where $M=M_{200c}$ for central haloes and $M=M_{\rm{peak}}$ for satellites. We calculated $r_{\rm{fid}}$ by estimating the radius (as a function of halo mass) at which the circular velocity was maximum. We did this assuming the NFW density profile and the concentration-mass relation from \citet{Correa15c}. In this manner the normalization and slope of the relation were chosen so that $r_{\rm{fid}}=2$ kpc for $10^{9}~\rm{M}_{\odot}$ haloes, and it reaches 5 kpc for $10^{11}~\rm{M}_{\odot}$ haloes. $V_{\rm{circ}}(r_{\rm{fid}})$ largely coincides with the maximum circular velocity of the cuspy haloes in the SIDM models. 

We next define $V_{\rm{circ}}(r_{\rm{fid}})=V_{\rm{fid}}$, and quantify the scatter around $V_{\rm{fid}}$ for the different models as a function of halo mass. The top panel of Fig.~\ref{Vfid_Scatter} shows the standard deviation in $V_{\rm{fid}}$ from satellite haloes as a function of halo mass. The various colour lines show the scatter for satellite haloes in the CDM and SIDM models as indicated in the legend. We find that the mean DM circular velocity (at $r_{\rm{fid}}=2.6$ kpc) from $10^{9.5}~\rm{M}_{\odot}$ CDM satellite haloes is ${\sim}29$ km/s with a $1\sigma$ scatter of 5.5 km/s. The scatter increases to 5.8 km/s, 7 km/s and 7.2 km/s in the SigmaVel20, SigmaVel60 and SigmaVel100 models, respectively. The bottom panel of Fig.~\ref{Vfid_Scatter} shows the ratio between the scatter in the SIDM models and the CDM model for better comparison. It can be seen that the SigmaVel60 and SigmaVel100 models display the largest scatter in the satellite mass range $10^9-10^{10}~\rm{M}_{\odot}$, it increases by a factor of 1.2 and 1.3 relative to CDM. SigmaConstant10 shows a somewhat similar scatter as CDM in $10^{9}~\rm{M}_{\odot}$ haloes, but the scatter increases with halo mass. As a result SigmaConstant10 produces up to a factor of 2 larger scatter than CDM in $10^{11}~\rm{M}_{\odot}$ satellite haloes. Interestingly, the figure shows that in the velocity-dependent cross-section models, the higher $\sigma_{T}/m_{\chi}$ at dwarf galaxies scales, the larger the scatter in $V_{\rm{fid}}$. This, however, is found when analysing the rotation curves of satellites. To highlight the statistical significance of this scatter, the bottom panel of Fig.~\ref{Vfid_Scatter} indicates the average number of haloes in the various mass bins.

The increase in the $1\sigma$ scatter around $V_{\rm{fid}}$ from $10^9-10^{10}~\rm{M}_{\odot}$ satellites is not significant when comparing CDM with SigmaVel100. However, what largely increases is the number of satellites that have $V_{\rm{fid}}$ that are $\pm 2\sigma$ away from the $V_{\rm{fid}}$ CDM average. In the SigmaConstant1 and SigmaConstant10 models 4 and $6\%$, respectively, of the satellite population in the $10^{9.5-9.6}~\rm{M}_{\odot}$ mass range, have $V_{\rm{fid}}$ that are $\pm 2\sigma$ away from the CDM average. The number of $2\sigma$ outliers increases to $12$, $22$ and $24\%$ in the SigmaVel20, SigmaVel60 and SigmaVel100 models, respectively. We continue with an analysis of the distribution of $V_{\rm{fid}}$ outliers in the following subsection.

\begin{figure*} 
	\includegraphics[angle=0,width=\textwidth]{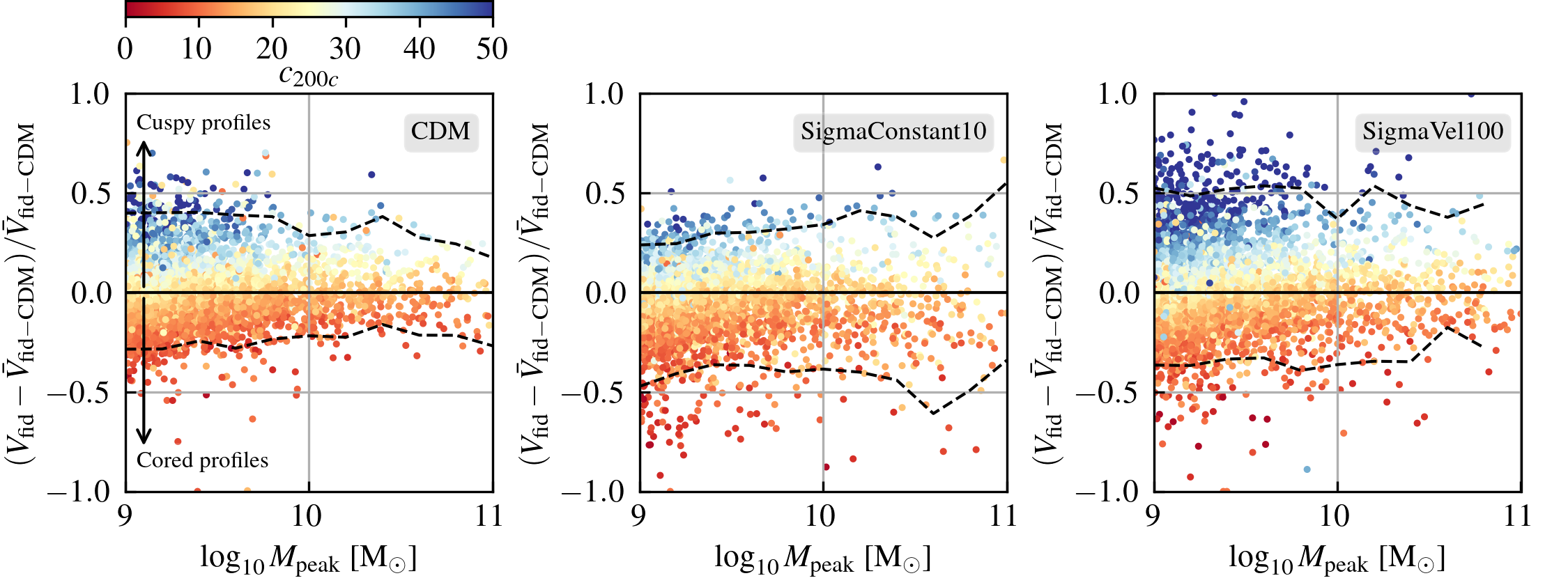}
	\caption{Circular velocities at the fiducial radius, $V_{\rm{fid}}$, relative to the median $\bar{V}_{\rm{fid}}$ from the CDM simulation. Each dot corresponds to a satellite halo, with a mass indicated by the x-axis, and with a concentration highlighted by the colour bar on the top of the figure. The panels show the ratio, $(V_{\rm{fid}}-\bar{V}_{\rm{fid-CDM}})/\bar{V}_{\rm{fid-CDM}}$, for the CDM (left), SigmaConstant10 (middle) and SigmaVel100 simulation (right). The figure indicates that if $V_{\rm{fid}}>\bar{V}_{\rm{fid-CDM}}$, a halo density profile is cuspy, whereas if $V_{\rm{fid}}<\bar{V}_{\rm{fid-CDM}}$, it is more cored. The color bar highlights that the scatter of the haloes circular velocities strongly correlates with the halo concentrations, with more concentrated haloes having cuspier density profiles. This correlation not only appears in the CDM simulation, but also in the SIDM models. While the middle panel does not show a large number of very cuspy satellite haloes from the $10^9-10^{11}~\rm{M}_{\odot}$ mass range, the right panel shows a large spread in the velocity ratio. The dashed black lines in the panels highlight the 97 and 3 percentiles of the distribution.}
	\label{M200_ratio_models}
\end{figure*}

\subsubsection{Diversity}\label{Diversity_section}

In this section we further assess the diversity of the rotation curves from the satellite population. We separate the CDM halo sample in halo mass bins of 0.1 dex, and calculate the median circular velocities at the fiducial radius (defined in eq.~\ref{r_fiducial}) at each mass bin. We refer to this median CDM fiducial velocity as $\bar{V}_{\rm{fid-CDM}}$. Finally, for each individual halo $i$ from the different simulations we calculate the ratio, $(V_{{\rm{fid}},i}-\bar{V}_{\rm{fid-CDM}})/\bar{V}_{\rm{fid-CDM}}$, where $\bar{V}_{\rm{fid-CDM}}$ is the median CDM $V_{\rm{fid}}$ from the mass bin the halo $i$ is, and $V_{{\rm{fid}},i}$ is the circular velocity at the fiducial radius of halo $i$.

Fig.~\ref{M200_ratio_models} shows the ratio, $(V_{\rm{fid}}-\bar{V}_{\rm{fid-CDM}})/\bar{V}_{\rm{fid-CDM}}$, as a function of halo mass for the CDM (left-panel), SigmaConstant10 (middle-panel) and SigmaVel100 (right-panel) models. Each dot in the figure is an individual satellite halo. Dots are coloured according to the color-bar at the top of the figure that indicates the haloes concentration. As expected, the scatter of the haloes circular velocities strongly correlates with the haloes concentration, so that more concentrated haloes have cuspier density profiles. This correlation not only appears in the CDM simulation, but also in the SIDM models. The dashed black lines in the panels of the figure highlight the 97 and 3 percentiles of the distribution for each mass bin.

The middle panel of Fig.~\ref{M200_ratio_models} shows that the SigmaConstant10 model has only very few cuspy satellites (i.e. with velocity ratios larger than 0.3), but a large number of satellites with very low densities in the inner regions (i.e. with velocity ratios lower than -0.5). We believe that some of these satellites have large cores due to the SIDM interactions, but others have low densities due to the excessive tidal disruption that there is in this model between the satellites and their hosts. From the right-panel we can see that satelllites in the SigmaVel100 model have become very cuspy over the $10^9-10^{10}~\rm{M}_{\odot}$ halo mass range, with velocity ratios larger than 0.5. These satellites are potentially in gravothermal core collapse. This figure clearly shows that a SIDM velocity-dependent model is able to increase the scatter in the rotation curves from low-mass satellite haloes in dark matter-only simulations. Interestingly, the increased diversity of rotation curves is not exclusive to satellite haloes, an increased scatter of $(V_{\rm{fid}}-\bar{V}_{\rm{fid-CDM}})/\bar{V}_{\rm{fid-CDM}}$ for central haloes is also found in the SigmaVel100 model (see Appendix~\ref{Diversity_Centrals}).

\begin{figure*} 
	\includegraphics[angle=0,width=\textwidth]{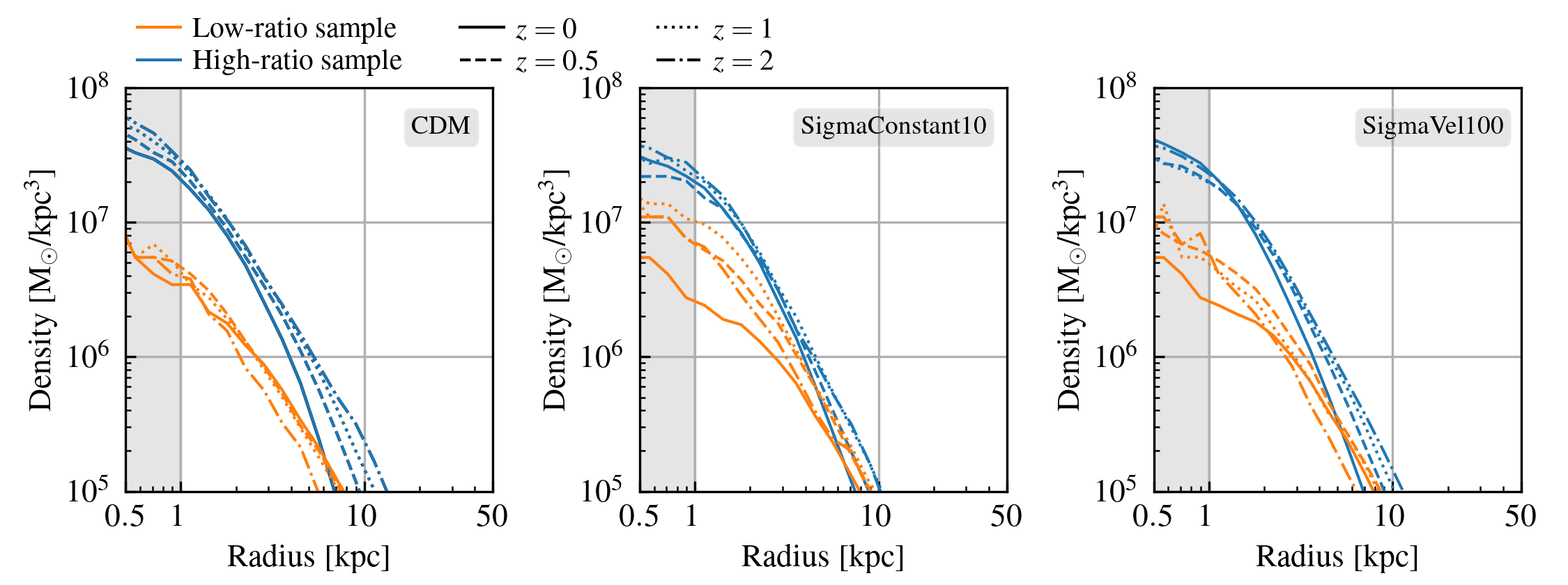}
	\caption{Satellite haloes' median density profiles from the CDM (left-panel), SigmaConstant10 (middle-panel) and SigmaVel100 (right-panel) simulations. The blue lines, labelled `high-ratio' sample, corresponds to satellite haloes in the mass range $10^{9}-10^{9.5}~\rm{M}_{\odot}$ that have velocity ratios, $(V_{\rm{fid}}-\bar{V}_{\rm{fid-CDM}})/\bar{V}_{\rm{fid-CDM}}$, larger than 0.3. Similarly, the orange lines, labelled `low-ratio' sample, corresponds to satellite haloes that have velocity ratios lower than -0.3. The different line types indicate different redshifts, with solid corresponding to median density profiles at redshift 0, dashed at redshift 0.5, dotted at redshift 1 and dashed-dotted at redshift 2. In this manner, each panel shows the evolution in density of haloes that at $z=0$ are satellites, have masses between $10^{9}$ and $10^{9.5}~\rm{M}_{\odot}$, and have either high or low velocity ratios. In the panels, the grey shaded area indicates the radial regime where the density profiles are below the convergence radius.}
	\label{SIDM_density_evolution}
\end{figure*}

\begin{figure*} 
	\includegraphics[angle=0,width=\textwidth]{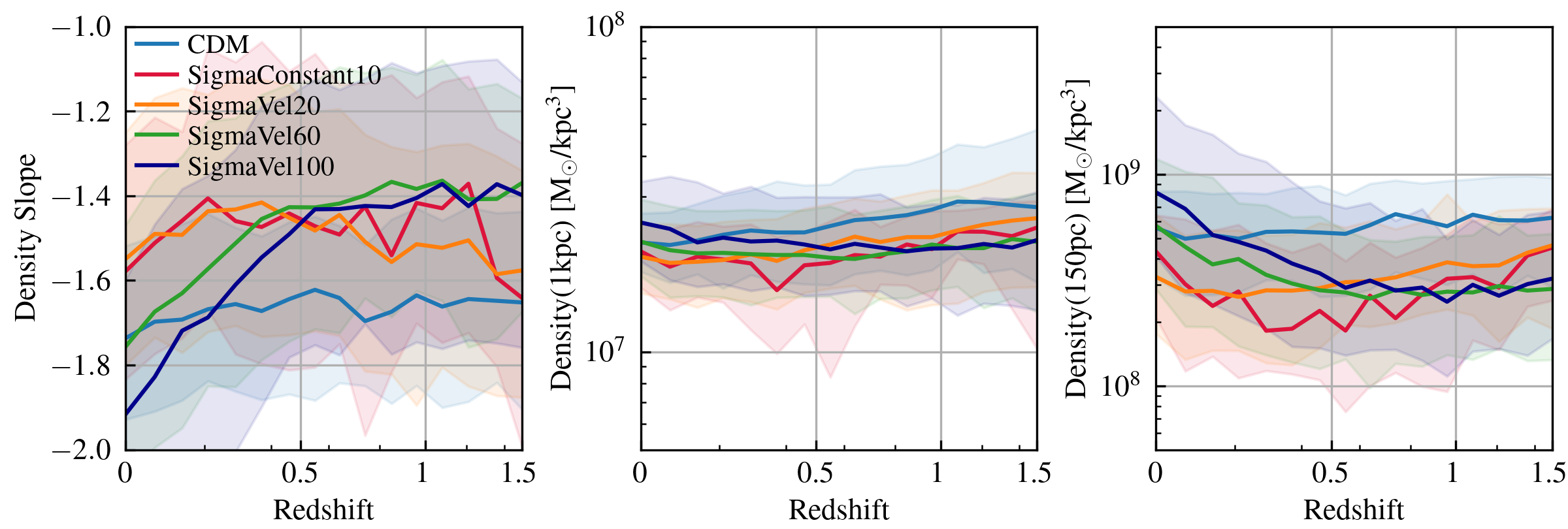}
	\caption{Evolution in the density slope (left panel), and in density at 1kpc (middle panel) and at 150 pc (right panel) of the high-ratio sample of $10^{9}-10^{9.5}~\rm{M}_{\odot}$ satellites. As indicated in the legends, each line shows the median evolution for the CDM, SigmaConstant10, SigmaVel20, SigmaVel60 and SigmaVel100 models. In the panels, the shaded areas highlight the 16-84\% percentiles.}
	\label{SIDM_density_evolution_2}
\end{figure*}

\subsection{Gravothermal core-collapse}

In this section we analyse the evolution of $z=0$ satellite haloes that can potentially be in gravothermal core collapse. We select satellites with masses between $10^{9}$ and $10^{9.5}~\rm{M}_{\odot}$ and follow their evolution throughout the simulations outputs. We create two subsamples, a high-velocity ratio sample that has $(V_{\rm{fid}}-\bar{V}_{\rm{fid-CDM}})/\bar{V}_{\rm{fid-CDM}}>0.3$, and a low-velocity ratio sample where $(V_{\rm{fid}}-\bar{V}_{\rm{fid-CDM}})/\bar{V}_{\rm{fid-CDM}}<-0.3$, and investigate the evolution of the median density profiles of the samples.

Fig.~\ref{SIDM_density_evolution} shows the satellite haloes' median density profiles from the CDM (left-panel), SigmaConstant10 (middle-panel) and SigmaVel100 (right-panel) simulations, where the blue lines correspond to the high-ratio sample and the orange lines to the low-ratio sample. The median densities at redshift 0 are shown in solid lines, at redshift 0.5 in dashed lines, at redshift 1 in dotted lines and at redshift 2 in dashed-dotted lines. In the panels, the grey shaded area indicates the radial region where the density profiles are below the convergence radius, and therefore we warn the reader that at these radii our results are not numerically resolved.

The left panel of the figure compares the evolution of CDM satellites haloes with low and high velocity ratios. We find that while the low-ratio sample does not significantly evolve in density in the 0-2 redshift range, the high-ratio sample does. High-ratio satellites experience significant mass loss, as can be seen from the decrease in density at the ${\sim}10$ kpc radius. The middle and right panels show that the low-ratio sample largely decrease their central densities during the redshift range 0-0.5. Since this feature is absent in the CDM sample, we find it to be produced by the SIDM interactions.

The evolution of the high-ratio satellite sample from the SigmaConstant10 and SigmaVel100 models shows that these satellites can potentially be in gravothermal core collapse. This can be seen by following the density evolution at ${\sim}1$ kpc radius in the middle and right panels. The median profiles show that the density of the sample decreases from $z=2$ to $z=0.5$, and then it begins to raise again. 

To better analyse this evolution we plot the evolution in the density at 1 kpc and at 150 pc, and the evolution of the logarithmic density slope in Fig.~\ref{SIDM_density_evolution_2}. For this figure we select satellite haloes with masses between $10^{9}$ and $10^{9.5}~\rm{M}_{\odot}$ that have $(V_{\rm{fid}}-\bar{V}_{\rm{fid-CDM}})/\bar{V}_{\rm{fid-CDM}}>0.3$, and plot the median values as a function of redshift. To calculate the slope in the density profile, for each individual halo from the sample, we fit the $\log_{10}\rho(r)-\log10(r)$ relation (with $r$ between 1-3 kpc) using a linear function. The left panel of Fig.~\ref{SIDM_density_evolution_2} shows the evolution of the logarithmic density slope of haloes from the CDM (blue line), SigmaConstant10 (red line), SigmaVel20 (orange), SigmaVel60 (green line) and SigmaVel100 (dark blue line) models. The middle panel of the figure shows the median density at 1kpc, and the right panel shows an extrapolation of the density at 150 pc using the linear fitting that estimated the density slope.

It can be seen from the left panel of Fig.~\ref{SIDM_density_evolution_2} that the high-ratio sample from the CDM simulation forms a steep density profile of roughly constant slope with decreasing redshift. The median steepness of the $z{=}0$ CDM profiles seems to match the slope of the density profiles from the SigmaVel60 model. SigmaVel60, however, shows a significant change in the evolution of the logarithmic density slope, being around -1.4 at $z{=}1.5$ and decreasing till -1.8 at $z{=}0$. SigmaVel100 shows a yet larger evolution in the haloes density, changing its median slope from $\sim -1.4$ at $z{=}0.5$ to -2 and $z{=}0$. This evolution indicates a contraction of the core and an increase in central density, that tends to mostly occur in the redshift range 0-0.5. The SigmaVel100 model also shows an important evolution in the haloes density, with the median slope decreasing from -1.4 at $z{=}0.5$ to -1.9 at $z{=}0$.

The middle panels of Fig.~\ref{SIDM_density_evolution_2} show the evolution in density at 1~kpc. From this panel it cannot be seen a significant raise in density from the SigmaVel60 or SigmaVel100 samples. But we find that the extrapolation of the shape of the profiles indicates that the satellites in the SigmaVel60 and SigmaVel100 model should largely increase in density in the central regions (${\sim}150$ pc). In a future work (Correa et al. in prep), using a higher-resolution set-up, we will further explore the evolution of satellites (as well as centrals), and analyse under which conditions (e.g. orbits, concentration, rate of mass gain/loss, environment) haloes undergo gravothermal core collapse.

\section{Discussion}

\subsection{Caveats}

In this work we have presented the first results of the TangoSIDM dark matter-only simulation suite, a set of cosmological simulations with different SIDM models. One limitation of this study is the resolution of the cosmological boxes. While we aim to explore and understand the internal evolution of low-mass haloes, we are limited to a minimum spatial resolution of 1 kpc (see Sec.~\ref{Density_profiles_Section} for a discussion on convergence radius). We have therefore analysed the density and rotation profiles on scales at or larger than 1 kpc, and we have shown that halo density profiles typically converge on scales down to half the softening (Fig.~\ref{Model_validation_2}, Sec.~\ref{Numerical_convergence}).

The lack of baryonic physics is an important effect that can impact our results. For galaxies in low-mass haloes, supernova feedback is able to alter the internal DM distribution of haloes (\citealt{Read05,Mashchenko08,Pontzen12, GarrisonKimmel14,DiCintio14,Tollet16}). The energy injection from supernovae produces gas outflows, leading to fluctuations of the nearby baryonic mass, that in turn modifies total gravitational potential within the inner DM halo. This causes a radial expansion of the orbits of the inner DM particles, and a result the formation of a core. While it has been shown that the SN-induced core formation is model-dependent (\citealt{BenitezLlambay19,Dutton20}), the process of gravothermal core collapse is potentially not. \citet{Burger22} has recently shown that for SIDM cross sections of at least 10 cm$^2$/g, the formation of cuspy central DM densities from gravothermal collapse occurs, irrespective of the star formation density threshold that controls the star formation burstiness of the galaxy, and hence the rate of SN explosions. This indicates that SIDM velocity-dependent models should still produce cuspy haloes and an increased diversity in the rotational curves, relative to CDM, even if baryonic feedback is included. In future work, however, we plan to analyse the impact of baryonic physics on the SIDM halo evolution.

\subsection{Gravothermal core-collapse in a cosmological set-up}

Several works have produced zoom-in simulations to study the phenomenology of SIDM on galaxy scales (i.e. \citealt{Vogelsberger12,Zavala13,Vogelsberger16,Robles19,Vogelsberger19,Zavala19,Nadler20,Bhattacharyya21,Sameie20,Shen21,Silverman22,Burger22}). However, their results can potentially depend on their specific set of initial conditions. This is because the rate of mass accretion, merger history, local environment, as well as other factors such as dynamical friction, tidal stripping and ram pressure, can alter the formation history of a galaxy and inner structure of a DM halo. Is it therefore important to study the SIDM effects in a cosmological set-up.

In this work we have shown that even in SIDM models where the cross section reaches 100 cm$^2$/g in $10^9~\rm{M}_{\odot}$ haloes, not all low-mass haloes enter in gravothermal core collapse, only a fraction. Merger history (e.g. \citealt{Colin02,Dave01}) and the impact of local environment can prevent a halo to enter in core collapse. This further supports the idea that the diversity in the rotation curves of dwarf galaxies can be a signature of velocity-dependent SIDM.




\section{Conclusions}

In this work we have presented the first results from the `Tantalizing models of Self-Interacting Dark Matter` project. A cosmological simulation suite project that aims to investigate the impact of SIDM on galaxies and DM haloes evolution. We have analysed DM-only cosmological simulations and compared the classical CDM model with SIDM models where the DM particles scattering cross section, $\sigma_{T}/m_{\chi}$ is constant, with $\sigma_{T}/m_{\chi}=1$ and 10 cm$^2$/g (called SigmaConstant1 and SigmaConstant10, respectively), or velocity-dependent, where $\sigma_{T}/m_{\chi}$ is lower than 10 cm$^2$/g in MW-mass haloes but reaches 100, 60 or 20 cm$^2$/g in $10^9~\rm{M}_{\odot}$ haloes (these models are refereed as SigmaVel100, SigmaVel60 and SigmaVel20, respectively), see Fig.~\ref{SIDM_models}.

Our SIDM implementation accurately models core formation in central haloes (Fig.~\ref{Density_Profiles}), it reaches numerical convergence (Fig.~\ref{Model_validation_2}) and it produces density profiles in agreement with what has been reported by  previous studies (Fig.~\ref{Model_validation_3}).

We have shown that a typical central halo of $10^{11.5}~\rm{M}_{\odot}$ changes morphology when we assume different SIDM models (Fig.~\ref{SIDM_haloes}). While it follows an elliptical elongated shape in the CDM scenario, in SIDM frequent DM particle collisions isotropise the particles orbit, making it more spherical. This however, depends on the cross section of the model, since it controls the rate of DM particles interactions. While the SigmaVel60 and SigmaConstant10 produce a spherical looking halo, SigmaVel20 does not. We have found that the largest subhalo destruction is produced in the SigmaConstant10 model, in contrast to the velocity-dependent models (Fig.~\ref{SubhaloMassFunction}).

We have focused on the evolution of satellites, and shown that the velocity-dependent $\sigma/m_{\chi}$ models produce a large diversity in the circular velocities of satellites haloes relative to CDM (Fig.~\ref{RotationCurves}). The scatter of the circular velocities at a fiducial radius increases with increasing cross sections, with the SigmaVel100 model reaching a factor of 1.3 larger scatter than CDM (Fig.~\ref{Vfid_Scatter}). We have further illustrated the increased diversity in rotation curves from the SigmaVel100 model, by calculating the deviation of the circular velocity at the fiducial radius, relative to the median CDM value. The increasing number of cuspy and cored haloes is shown in Fig.~\ref{M200_ratio_models}, where we have also compared with the SigmaConstant10 model.

The large variation in the haloes internal structure is driven by DM particles collisions, causing in some haloes the formation of extended cores, whereas in others gravothermal core collapse. Fig.~\ref{SIDM_density_evolution} shows the evolution in density of cuspy and core satellite haloes. We have found that very cuspy haloes are undergoing gravothermal core collapse. These haloes are changing the shape of their density distribution, by becoming steeper with decreasing redshift (Fig.~\ref{SIDM_density_evolution_2}).

An important motivation for this study is to understand whether SIDM can solve the so-called `cusp-core/diversity' problem of CDM. Our velocity-dependent SIDM models are able to produce DM haloes that are either cuspy or display a core, without the need of invoking a bursty star-forming galaxy. The models from the TangoSIDM project, therefore, offer a promising explanation for the diversity in the density and velocity profiles of observed dwarf galaxies.

\section*{Acknowledgements}

CC acknowledges the support of the Dutch Research Council (NWO Veni 192.020). NAM and CW have received funding from the European Research Council (ERC) under the European Union's Horizon 2020 research and innovation programme (Grant agreement No. 864035). The research in this paper made use of the SWIFT open-source simulation code (http://www.swiftsim.com, \citealt{Schaller18}) version 0.9.0. The TangoSIDM simulation suite have been produced using the DECI resource Mahti based in Finland at CSC, Finnish IT Center for Science, with support from the PRACE aisbl., project ID 17DECI0030-TangoSIDM. The TangoSIDM simulations design and analysis has been carried using the Dutch national e-infrastructure, Snellius, with the support of SURF Cooperative, project ID EINF-180-TangoSIDM. We acknowledge various public python packages that have greatly benefited this work: \verb|scipy| (\citealt{vanderWalt11}), \verb|numpy| (\citealt{vanderWalt11}), \verb|matplotlib| (\citealt{Hunter07}). This work has also benefited from the python analysis pipeline SwiftsimIO (\citealt{Borrow20}), and the Swift color map collection\footnote{https://github.com/JBorrow/swiftascmaps}.

\section*{Data availability}

The data supporting the plots within this article are available on reasonable request to the corresponding author.

\bibliography{biblio}

\begin{thebibliography}{}
\makeatletter
\relax
\def\mn@urlcharsother{\let\do\@makeother \do\$\do\&\do\#\do\^\do\_\do\%\do\~}
\def\mn@doi{\begingroup\mn@urlcharsother \@ifnextchar [ {\mn@doi@}
  {\mn@doi@[]}}
\def\mn@doi@[#1]#2{\def\@tempa{#1}\ifx\@tempa\@empty \href
  {http://dx.doi.org/#2} {doi:#2}\else \href {http://dx.doi.org/#2} {#1}\fi
  \endgroup}
\def\mn@eprint#1#2{\mn@eprint@#1:#2::\@nil}
\def\mn@eprint@arXiv#1{\href {http://arxiv.org/abs/#1} {{\tt arXiv:#1}}}
\def\mn@eprint@dblp#1{\href {http://dblp.uni-trier.de/rec/bibtex/#1.xml}
  {dblp:#1}}
\def\mn@eprint@#1:#2:#3:#4\@nil{\def\@tempa {#1}\def\@tempb {#2}\def\@tempc
  {#3}\ifx \@tempc \@empty \let \@tempc \@tempb \let \@tempb \@tempa \fi \ifx
  \@tempb \@empty \def\@tempb {arXiv}\fi \@ifundefined
  {mn@eprint@\@tempb}{\@tempb:\@tempc}{\expandafter \expandafter \csname
  mn@eprint@\@tempb\endcsname \expandafter{\@tempc}}}

\bibitem[\protect\citeauthoryear{{Ahn} \& {Shapiro}}{{Ahn} \&
  {Shapiro}}{2005}]{Ahn05}
{Ahn} K.,  {Shapiro} P.~R.,  2005, \mn@doi [\mnras]
  {10.1111/j.1365-2966.2005.09492.x}, \href
  {https://ui.adsabs.harvard.edu/abs/2005MNRAS.363.1092A} {363, 1092}

\bibitem[\protect\citeauthoryear{{Andrade}, {Fuson}, {Gad-Nasr}, {Kong},
  {Minor}, {Roberts}  \& {Kaplinghat}}{{Andrade} et~al.}{2022}]{Andrade22}
{Andrade} K.~E.,  {Fuson} J.,  {Gad-Nasr} S.,  {Kong} D.,  {Minor} Q.,
  {Roberts} M.~G.,   {Kaplinghat} M.,  2022, \mn@doi [\mnras]
  {10.1093/mnras/stab3241}, \href
  {https://ui.adsabs.harvard.edu/abs/2022MNRAS.510...54A} {510, 54}

\bibitem[\protect\citeauthoryear{{Applebaum}, {Brooks}, {Christensen},
  {Munshi}, {Quinn}, {Shen}  \& {Tremmel}}{{Applebaum}
  et~al.}{2021}]{Applebaum21}
{Applebaum} E.,  {Brooks} A.~M.,  {Christensen} C.~R.,  {Munshi} F.,  {Quinn}
  T.~R.,  {Shen} S.,   {Tremmel} M.,  2021, \mn@doi [\apj]
  {10.3847/1538-4357/abcafa}, \href
  {https://ui.adsabs.harvard.edu/abs/2021ApJ...906...96A} {906, 96}

\bibitem[\protect\citeauthoryear{{Arkani-Hamed}, {Finkbeiner}, {Slatyer}  \&
  {Weiner}}{{Arkani-Hamed} et~al.}{2009}]{ArkaniHamed09}
{Arkani-Hamed} N.,  {Finkbeiner} D.~P.,  {Slatyer} T.~R.,   {Weiner} N.,  2009,
  \mn@doi [\prd] {10.1103/PhysRevD.79.015014}, \href
  {https://ui.adsabs.harvard.edu/abs/2009PhRvD..79a5014A} {79, 015014}

\bibitem[\protect\citeauthoryear{{Balberg}, {Shapiro}  \& {Inagaki}}{{Balberg}
  et~al.}{2002}]{Balberg02}
{Balberg} S.,  {Shapiro} S.~L.,   {Inagaki} S.,  2002, \mn@doi [\apj]
  {10.1086/339038}, \href
  {https://ui.adsabs.harvard.edu/abs/2002ApJ...568..475B} {568, 475}

\bibitem[\protect\citeauthoryear{{Banerjee}, {Adhikari}, {Dalal}, {More}  \&
  {Kravtsov}}{{Banerjee} et~al.}{2020}]{Banerjee20}
{Banerjee} A.,  {Adhikari} S.,  {Dalal} N.,  {More} S.,   {Kravtsov} A.,  2020,
  \mn@doi [\jcap] {10.1088/1475-7516/2020/02/024}, \href
  {https://ui.adsabs.harvard.edu/abs/2020JCAP...02..024B} {2020, 024}

\bibitem[\protect\citeauthoryear{{Ben{\'\i}tez-Llambay}, {Frenk}, {Ludlow}  \&
  {Navarro}}{{Ben{\'\i}tez-Llambay} et~al.}{2019}]{BenitezLlambay19}
{Ben{\'\i}tez-Llambay} A.,  {Frenk} C.~S.,  {Ludlow} A.~D.,   {Navarro} J.~F.,
  2019, \mn@doi [\mnras] {10.1093/mnras/stz1890}, \href
  {https://ui.adsabs.harvard.edu/abs/2019MNRAS.488.2387B} {488, 2387}

\bibitem[\protect\citeauthoryear{{Bhattacharyya}, {Adhikari}, {Banerjee},
  {More}, {Kumar}, {Nadler}  \& {Chatterjee}}{{Bhattacharyya}
  et~al.}{2021}]{Bhattacharyya21}
{Bhattacharyya} S.,  {Adhikari} S.,  {Banerjee} A.,  {More} S.,  {Kumar} A.,
  {Nadler} E.~O.,   {Chatterjee} S.,  2021, arXiv e-prints, \href
  {https://ui.adsabs.harvard.edu/abs/2021arXiv210608292B} {p. arXiv:2106.08292}

\bibitem[\protect\citeauthoryear{{Boddy}, {Feng}, {Kaplinghat}, {Shadmi}  \&
  {Tait}}{{Boddy} et~al.}{2014}]{Boddy14}
{Boddy} K.~K.,  {Feng} J.~L.,  {Kaplinghat} M.,  {Shadmi} Y.,   {Tait} T.
  M.~P.,  2014, \mn@doi [\prd] {10.1103/PhysRevD.90.095016}, \href
  {https://ui.adsabs.harvard.edu/abs/2014PhRvD..90i5016B} {90, 095016}

\bibitem[\protect\citeauthoryear{{Borrow} \& {Borrisov}}{{Borrow} \&
  {Borrisov}}{2020}]{Borrow20}
{Borrow} J.,  {Borrisov} A.,  2020, \mn@doi [The Journal of Open Source
  Software] {10.21105/joss.02430}, \href
  {https://ui.adsabs.harvard.edu/abs/2020JOSS....5.2430B} {5, 2430}

\bibitem[\protect\citeauthoryear{{Borrow}, {Schaller}, {Bower}  \&
  {Schaye}}{{Borrow} et~al.}{2022}]{Borrow22}
{Borrow} J.,  {Schaller} M.,  {Bower} R.~G.,   {Schaye} J.,  2022, \mn@doi
  [\mnras] {10.1093/mnras/stab3166}, \href
  {https://ui.adsabs.harvard.edu/abs/2022MNRAS.511.2367B} {511, 2367}

\bibitem[\protect\citeauthoryear{{Bose} et~al.,}{{Bose} et~al.}{2019}]{Bose19}
{Bose} S.,  et~al., 2019, \mn@doi [\mnras] {10.1093/mnras/stz1168}, \href
  {https://ui.adsabs.harvard.edu/abs/2019MNRAS.486.4790B} {486, 4790}

\bibitem[\protect\citeauthoryear{{Boylan-Kolchin}, {Bullock}  \&
  {Kaplinghat}}{{Boylan-Kolchin} et~al.}{2011}]{BoylanKolchin11}
{Boylan-Kolchin} M.,  {Bullock} J.~S.,   {Kaplinghat} M.,  2011, \mn@doi
  [\mnras] {10.1111/j.1745-3933.2011.01074.x}, \href
  {https://ui.adsabs.harvard.edu/abs/2011MNRAS.415L..40B} {415, L40}

\bibitem[\protect\citeauthoryear{{Boylan-Kolchin}, {Bullock}  \&
  {Kaplinghat}}{{Boylan-Kolchin} et~al.}{2012}]{BoylanKolchin12}
{Boylan-Kolchin} M.,  {Bullock} J.~S.,   {Kaplinghat} M.,  2012, \mn@doi
  [\mnras] {10.1111/j.1365-2966.2012.20695.x}, \href
  {https://ui.adsabs.harvard.edu/abs/2012MNRAS.422.1203B} {422, 1203}

\bibitem[\protect\citeauthoryear{{Buckley} \& {Fox}}{{Buckley} \&
  {Fox}}{2010}]{Buckley10}
{Buckley} M.~R.,  {Fox} P.~J.,  2010, \mn@doi [\prd]
  {10.1103/PhysRevD.81.083522}, \href
  {https://ui.adsabs.harvard.edu/abs/2010PhRvD..81h3522B} {81, 083522}

\bibitem[\protect\citeauthoryear{{Burger}, {Zavala}, {Sales}, {Vogelsberger},
  {Marinacci}  \& {Torrey}}{{Burger} et~al.}{2022}]{Burger22}
{Burger} J.~D.,  {Zavala} J.,  {Sales} L.~V.,  {Vogelsberger} M.,  {Marinacci}
  F.,   {Torrey} P.,  2022, \mn@doi [\mnras] {10.1093/mnras/stac994}, \href
  {https://ui.adsabs.harvard.edu/abs/2022MNRAS.tmp..969B} {}

\bibitem[\protect\citeauthoryear{{Ca{\~n}as}, {Elahi}, {Welker}, {del P Lagos},
  {Power}, {Dubois}  \& {Pichon}}{{Ca{\~n}as} et~al.}{2019}]{Canas19}
{Ca{\~n}as} R.,  {Elahi} P.~J.,  {Welker} C.,  {del P Lagos} C.,  {Power} C.,
  {Dubois} Y.,   {Pichon} C.,  2019, \mn@doi [\mnras] {10.1093/mnras/sty2725},
  \href {https://ui.adsabs.harvard.edu/abs/2019MNRAS.482.2039C} {482, 2039}

\bibitem[\protect\citeauthoryear{{Carton Zeng}, {Peter}, {Du}, {Benson}, {Kim},
  {Jiang}, {Cyr-Racine}  \& {Vogelsberger}}{{Carton Zeng}
  et~al.}{2021}]{Zeng21}
{Carton Zeng} Z.,  {Peter} A. H.~G.,  {Du} X.,  {Benson} A.,  {Kim} S.,
  {Jiang} F.,  {Cyr-Racine} F.-Y.,   {Vogelsberger} M.,  2021, arXiv e-prints,
  \href {https://ui.adsabs.harvard.edu/abs/2021arXiv211000259C} {p.
  arXiv:2110.00259}

\bibitem[\protect\citeauthoryear{{Col{\'\i}n}, {Avila-Reese}, {Valenzuela}  \&
  {Firmani}}{{Col{\'\i}n} et~al.}{2002}]{Colin02}
{Col{\'\i}n} P.,  {Avila-Reese} V.,  {Valenzuela} O.,   {Firmani} C.,  2002,
  \mn@doi [\apj] {10.1086/344259}, \href
  {https://ui.adsabs.harvard.edu/abs/2002ApJ...581..777C} {581, 777}

\bibitem[\protect\citeauthoryear{{Correa}}{{Correa}}{2021}]{Correa21}
{Correa} C.~A.,  2021, \mn@doi [\mnras] {10.1093/mnras/stab506}, \href
  {https://ui.adsabs.harvard.edu/abs/2021MNRAS.503..920C} {503, 920}

\bibitem[\protect\citeauthoryear{{Correa}, {Wyithe}, {Schaye}  \&
  {Duffy}}{{Correa} et~al.}{2015}]{Correa15c}
{Correa} C.~A.,  {Wyithe} J. S.~B.,  {Schaye} J.,   {Duffy} A.~R.,  2015,
  \mn@doi [\mnras] {10.1093/mnras/stv697}, \href
  {https://ui.adsabs.harvard.edu/abs/2015MNRAS.450.1521C} {450, 1521}

\bibitem[\protect\citeauthoryear{{Dav{\'e}}, {Spergel}, {Steinhardt}  \&
  {Wandelt}}{{Dav{\'e}} et~al.}{2001}]{Dave01}
{Dav{\'e}} R.,  {Spergel} D.~N.,  {Steinhardt} P.~J.,   {Wandelt} B.~D.,  2001,
  \mn@doi [\apj] {10.1086/318417}, \href
  {https://ui.adsabs.harvard.edu/abs/2001ApJ...547..574D} {547, 574}

\bibitem[\protect\citeauthoryear{{Davis}, {Efstathiou}, {Frenk}  \&
  {White}}{{Davis} et~al.}{1985}]{Davis85}
{Davis} M.,  {Efstathiou} G.,  {Frenk} C.~S.,   {White} S.~D.~M.,  1985,
  \mn@doi [\apj] {10.1086/163168}, \href
  {https://ui.adsabs.harvard.edu/abs/1985ApJ...292..371D} {292, 371}

\bibitem[\protect\citeauthoryear{{Dawson} et~al.,}{{Dawson}
  et~al.}{2013}]{Dawson13}
{Dawson} W.,  et~al., 2013, in American Astronomical Society Meeting Abstracts
  \#221. p. 125.04

\bibitem[\protect\citeauthoryear{{Di Cintio}, {Brook}, {Macci{\`o}}, {Stinson},
  {Knebe}, {Dutton}  \& {Wadsley}}{{Di Cintio} et~al.}{2014}]{DiCintio14}
{Di Cintio} A.,  {Brook} C.~B.,  {Macci{\`o}} A.~V.,  {Stinson} G.~S.,  {Knebe}
  A.,  {Dutton} A.~A.,   {Wadsley} J.,  2014, \mn@doi [\mnras]
  {10.1093/mnras/stt1891}, \href
  {https://ui.adsabs.harvard.edu/abs/2014MNRAS.437..415D} {437, 415}

\bibitem[\protect\citeauthoryear{{Dooley}, {Peter}, {Vogelsberger}, {Zavala}
  \& {Frebel}}{{Dooley} et~al.}{2016}]{Dooley16}
{Dooley} G.~A.,  {Peter} A. H.~G.,  {Vogelsberger} M.,  {Zavala} J.,   {Frebel}
  A.,  2016, \mn@doi [\mnras] {10.1093/mnras/stw1309}, \href
  {https://ui.adsabs.harvard.edu/abs/2016MNRAS.461..710D} {461, 710}

\bibitem[\protect\citeauthoryear{{Dutton}, {Buck}, {Macci{\`o}}, {Dixon},
  {Blank}  \& {Obreja}}{{Dutton} et~al.}{2020}]{Dutton20}
{Dutton} A.~A.,  {Buck} T.,  {Macci{\`o}} A.~V.,  {Dixon} K.~L.,  {Blank} M.,
  {Obreja} A.,  2020, \mn@doi [\mnras] {10.1093/mnras/staa3028}, \href
  {https://ui.adsabs.harvard.edu/abs/2020MNRAS.499.2648D} {499, 2648}

\bibitem[\protect\citeauthoryear{{Ebisu}, {Ishiyama}  \& {Hayashi}}{{Ebisu}
  et~al.}{2022}]{Ebisu22}
{Ebisu} T.,  {Ishiyama} T.,   {Hayashi} K.,  2022, \mn@doi [\prd]
  {10.1103/PhysRevD.105.023016}, \href
  {https://ui.adsabs.harvard.edu/abs/2022PhRvD.105b3016E} {105, 023016}

\bibitem[\protect\citeauthoryear{{Elahi}, {Thacker}  \& {Widrow}}{{Elahi}
  et~al.}{2011}]{Elahi11}
{Elahi} P.~J.,  {Thacker} R.~J.,   {Widrow} L.~M.,  2011, \mn@doi [\mnras]
  {10.1111/j.1365-2966.2011.19485.x}, \href
  {https://ui.adsabs.harvard.edu/abs/2011MNRAS.418..320E} {418, 320}

\bibitem[\protect\citeauthoryear{{Elahi}, {Ca{\~n}as}, {Poulton}, {Tobar},
  {Willis}, {Lagos}, {Power}  \& {Robotham}}{{Elahi} et~al.}{2019a}]{Elahi19}
{Elahi} P.~J.,  {Ca{\~n}as} R.,  {Poulton} R. J.~J.,  {Tobar} R.~J.,  {Willis}
  J.~S.,  {Lagos} C. d.~P.,  {Power} C.,   {Robotham} A. S.~G.,  2019a, \mn@doi
  [\pasa] {10.1017/pasa.2019.12}, \href
  {https://ui.adsabs.harvard.edu/abs/2019PASA...36...21E} {36, e021}

\bibitem[\protect\citeauthoryear{{Elahi}, {Poulton}, {Tobar}, {Ca{\~n}as},
  {Lagos}, {Power}  \& {Robotham}}{{Elahi} et~al.}{2019b}]{Elahi19b}
{Elahi} P.~J.,  {Poulton} R. J.~J.,  {Tobar} R.~J.,  {Ca{\~n}as} R.,  {Lagos}
  C. d.~P.,  {Power} C.,   {Robotham} A. S.~G.,  2019b, \mn@doi [\pasa]
  {10.1017/pasa.2019.18}, \href
  {https://ui.adsabs.harvard.edu/abs/2019PASA...36...28E} {36, e028}

\bibitem[\protect\citeauthoryear{{Elbert}, {Bullock}, {Garrison-Kimmel},
  {Rocha}, {O{\~n}orbe}  \& {Peter}}{{Elbert} et~al.}{2015}]{Elbert15}
{Elbert} O.~D.,  {Bullock} J.~S.,  {Garrison-Kimmel} S.,  {Rocha} M.,
  {O{\~n}orbe} J.,   {Peter} A. H.~G.,  2015, \mn@doi [\mnras]
  {10.1093/mnras/stv1470}, \href
  {https://ui.adsabs.harvard.edu/abs/2015MNRAS.453...29E} {453, 29}

\bibitem[\protect\citeauthoryear{{Engler} et~al.,}{{Engler}
  et~al.}{2021}]{Engler21}
{Engler} C.,  et~al., 2021, \mn@doi [\mnras] {10.1093/mnras/stab2437}, \href
  {https://ui.adsabs.harvard.edu/abs/2021MNRAS.507.4211E} {507, 4211}

\bibitem[\protect\citeauthoryear{{Essig}, {McDermott}, {Yu}  \&
  {Zhong}}{{Essig} et~al.}{2019}]{Essig19}
{Essig} R.,  {McDermott} S.~D.,  {Yu} H.-B.,   {Zhong} Y.-M.,  2019, \mn@doi
  [\prl] {10.1103/PhysRevLett.123.121102}, \href
  {https://ui.adsabs.harvard.edu/abs/2019PhRvL.123l1102E} {123, 121102}

\bibitem[\protect\citeauthoryear{{Fattahi} et~al.,}{{Fattahi}
  et~al.}{2016}]{Fattahi16}
{Fattahi} A.,  et~al., 2016, \mn@doi [\mnras] {10.1093/mnras/stv2970}, \href
  {https://ui.adsabs.harvard.edu/abs/2016MNRAS.457..844F} {457, 844}

\bibitem[\protect\citeauthoryear{{Feng}, {Kaplinghat}  \& {Yu}}{{Feng}
  et~al.}{2010}]{Feng10}
{Feng} J.~L.,  {Kaplinghat} M.,   {Yu} H.-B.,  2010, \mn@doi [\prd]
  {10.1103/PhysRevD.82.083525}, \href
  {https://ui.adsabs.harvard.edu/abs/2010PhRvD..82h3525F} {82, 083525}

\bibitem[\protect\citeauthoryear{{Ferrero}, {Abadi}, {Navarro}, {Sales}  \&
  {Gurovich}}{{Ferrero} et~al.}{2012}]{Ferrero12}
{Ferrero} I.,  {Abadi} M.~G.,  {Navarro} J.~F.,  {Sales} L.~V.,   {Gurovich}
  S.,  2012, \mn@doi [\mnras] {10.1111/j.1365-2966.2012.21623.x}, \href
  {https://ui.adsabs.harvard.edu/abs/2012MNRAS.425.2817F} {425, 2817}

\bibitem[\protect\citeauthoryear{{Fischer}, {Br{\"u}ggen}, {Schmidt-Hoberg},
  {Dolag}, {Kahlhoefer}, {Ragagnin}  \& {Robertson}}{{Fischer}
  et~al.}{2021}]{Fischer21}
{Fischer} M.~S.,  {Br{\"u}ggen} M.,  {Schmidt-Hoberg} K.,  {Dolag} K.,
  {Kahlhoefer} F.,  {Ragagnin} A.,   {Robertson} A.,  2021, \mn@doi [\mnras]
  {10.1093/mnras/stab1198}, \href
  {https://ui.adsabs.harvard.edu/abs/2021MNRAS.505..851F} {505, 851}

\bibitem[\protect\citeauthoryear{{Garrison-Kimmel}, {Boylan-Kolchin}, {Bullock}
   \& {Kirby}}{{Garrison-Kimmel} et~al.}{2014}]{GarrisonKimmel14}
{Garrison-Kimmel} S.,  {Boylan-Kolchin} M.,  {Bullock} J.~S.,   {Kirby} E.~N.,
  2014, \mn@doi [\mnras] {10.1093/mnras/stu1477}, \href
  {https://ui.adsabs.harvard.edu/abs/2014MNRAS.444..222G} {444, 222}

\bibitem[\protect\citeauthoryear{{Garrison-Kimmel} et~al.,}{{Garrison-Kimmel}
  et~al.}{2019}]{GarrisonKimmel19}
{Garrison-Kimmel} S.,  et~al., 2019, \mn@doi [\mnras] {10.1093/mnras/stz1317},
  \href {https://ui.adsabs.harvard.edu/abs/2019MNRAS.487.1380G} {487, 1380}

\bibitem[\protect\citeauthoryear{{Governato} et~al.,}{{Governato}
  et~al.}{2010}]{Governato10}
{Governato} F.,  et~al., 2010, \mn@doi [\nat] {10.1038/nature08640}, \href
  {https://ui.adsabs.harvard.edu/abs/2010Natur.463..203G} {463, 203}

\bibitem[\protect\citeauthoryear{{Governato} et~al.,}{{Governato}
  et~al.}{2012}]{Governato12}
{Governato} F.,  et~al., 2012, \mn@doi [\mnras]
  {10.1111/j.1365-2966.2012.20696.x}, \href
  {https://ui.adsabs.harvard.edu/abs/2012MNRAS.422.1231G} {422, 1231}

\bibitem[\protect\citeauthoryear{{Graus}, {Bullock}, {Boylan-Kolchin}  \&
  {Nierenberg}}{{Graus} et~al.}{2018}]{Graus18}
{Graus} A.~S.,  {Bullock} J.~S.,  {Boylan-Kolchin} M.,   {Nierenberg} A.~M.,
  2018, \mn@doi [\mnras] {10.1093/mnras/sty1924}, \href
  {https://ui.adsabs.harvard.edu/abs/2018MNRAS.480.1322G} {480, 1322}

\bibitem[\protect\citeauthoryear{{Harvey}, {Massey}, {Kitching}, {Taylor}  \&
  {Tittley}}{{Harvey} et~al.}{2015}]{Harvey15}
{Harvey} D.,  {Massey} R.,  {Kitching} T.,  {Taylor} A.,   {Tittley} E.,  2015,
  \mn@doi [Science] {10.1126/science.1261381}, \href
  {https://ui.adsabs.harvard.edu/abs/2015Sci...347.1462H} {347, 1462}

\bibitem[\protect\citeauthoryear{{Harvey}, {Robertson}, {Massey}  \&
  {McCarthy}}{{Harvey} et~al.}{2019}]{Harvey19}
{Harvey} D.,  {Robertson} A.,  {Massey} R.,   {McCarthy} I.~G.,  2019, \mn@doi
  [\mnras] {10.1093/mnras/stz1816}, \href
  {https://ui.adsabs.harvard.edu/abs/2019MNRAS.488.1572H} {488, 1572}

\bibitem[\protect\citeauthoryear{{Hayashi}, {Chiba}  \& {Ishiyama}}{{Hayashi}
  et~al.}{2020}]{Hayashi20}
{Hayashi} K.,  {Chiba} M.,   {Ishiyama} T.,  2020, \mn@doi [\apj]
  {10.3847/1538-4357/abbe0a}, \href
  {https://ui.adsabs.harvard.edu/abs/2020ApJ...904...45H} {904, 45}

\bibitem[\protect\citeauthoryear{{Hayashi}, {Ibe}, {Kobayashi}, {Nakayama}  \&
  {Shirai}}{{Hayashi} et~al.}{2021}]{Hayashi21}
{Hayashi} K.,  {Ibe} M.,  {Kobayashi} S.,  {Nakayama} Y.,   {Shirai} S.,  2021,
  \mn@doi [\prd] {10.1103/PhysRevD.103.023017}, \href
  {https://ui.adsabs.harvard.edu/abs/2021PhRvD.103b3017H} {103, 023017}

\bibitem[\protect\citeauthoryear{{Hayashi}, {Hirai}, {Chiba}  \&
  {Ishiyama}}{{Hayashi} et~al.}{2022}]{Hayashi22}
{Hayashi} K.,  {Hirai} Y.,  {Chiba} M.,   {Ishiyama} T.,  2022, arXiv e-prints,
  \href {https://ui.adsabs.harvard.edu/abs/2022arXiv220602821H} {p.
  arXiv:2206.02821}

\bibitem[\protect\citeauthoryear{{Hernquist}}{{Hernquist}}{1990}]{Hernquist90}
{Hernquist} L.,  1990, \mn@doi [\apj] {10.1086/168845}, \href
  {https://ui.adsabs.harvard.edu/abs/1990ApJ...356..359H} {356, 359}

\bibitem[\protect\citeauthoryear{{Homma} et~al.,}{{Homma}
  et~al.}{2019}]{Homma19}
{Homma} D.,  et~al., 2019, \mn@doi [\pasj] {10.1093/pasj/psz076}, \href
  {https://ui.adsabs.harvard.edu/abs/2019PASJ...71...94H} {71, 94}

\bibitem[\protect\citeauthoryear{{Hopkins} et~al.,}{{Hopkins}
  et~al.}{2018}]{Hopkins18}
{Hopkins} P.~F.,  et~al., 2018, \mn@doi [\mnras] {10.1093/mnras/sty1690}, \href
  {https://ui.adsabs.harvard.edu/abs/2018MNRAS.480..800H} {480, 800}

\bibitem[\protect\citeauthoryear{Hunter}{Hunter}{2007}]{Hunter07}
Hunter J.~D.,  2007, \mn@doi [Computing in Science \& Engineering]
  {10.1109/MCSE.2007.55}, 9, 90

\bibitem[\protect\citeauthoryear{{Ibe} \& {Yu}}{{Ibe} \& {Yu}}{2010}]{Ibe10}
{Ibe} M.,  {Yu} H.-B.,  2010, \mn@doi [Physics Letters B]
  {10.1016/j.physletb.2010.07.026}, \href
  {https://ui.adsabs.harvard.edu/abs/2010PhLB..692...70I} {692, 70}

\bibitem[\protect\citeauthoryear{{Jee}, {Hoekstra}, {Mahdavi}  \&
  {Babul}}{{Jee} et~al.}{2014}]{Jee14}
{Jee} M.~J.,  {Hoekstra} H.,  {Mahdavi} A.,   {Babul} A.,  2014, \mn@doi [\apj]
  {10.1088/0004-637X/783/2/78}, \href
  {https://ui.adsabs.harvard.edu/abs/2014ApJ...783...78J} {783, 78}

\bibitem[\protect\citeauthoryear{{Jenkins}}{{Jenkins}}{2010}]{Jenkins10}
{Jenkins} A.,  2010, \mn@doi [\mnras] {10.1111/j.1365-2966.2010.16259.x}, \href
  {https://ui.adsabs.harvard.edu/abs/2010MNRAS.403.1859J} {403, 1859}

\bibitem[\protect\citeauthoryear{{Jenkins}}{{Jenkins}}{2013}]{Jenkins13}
{Jenkins} A.,  2013, \mn@doi [\mnras] {10.1093/mnras/stt1154}, \href
  {https://ui.adsabs.harvard.edu/abs/2013MNRAS.434.2094J} {434, 2094}

\bibitem[\protect\citeauthoryear{{Jethwa}, {Erkal}  \& {Belokurov}}{{Jethwa}
  et~al.}{2018}]{Jethwa18}
{Jethwa} P.,  {Erkal} D.,   {Belokurov} V.,  2018, \mn@doi [\mnras]
  {10.1093/mnras/stx2330}, \href
  {https://ui.adsabs.harvard.edu/abs/2018MNRAS.473.2060J} {473, 2060}

\bibitem[\protect\citeauthoryear{{Kahlhoefer}, {Schmidt-Hoberg}, {Kummer}  \&
  {Sarkar}}{{Kahlhoefer} et~al.}{2015}]{Kahlhoefer15}
{Kahlhoefer} F.,  {Schmidt-Hoberg} K.,  {Kummer} J.,   {Sarkar} S.,  2015,
  \mn@doi [\mnras] {10.1093/mnrasl/slv088}, \href
  {https://ui.adsabs.harvard.edu/abs/2015MNRAS.452L..54K} {452, L54}

\bibitem[\protect\citeauthoryear{{Kahlhoefer}, {Kaplinghat}, {Slatyer}  \&
  {Wu}}{{Kahlhoefer} et~al.}{2019}]{Kahlhoefer19}
{Kahlhoefer} F.,  {Kaplinghat} M.,  {Slatyer} T.~R.,   {Wu} C.-L.,  2019,
  \mn@doi [\jcap] {10.1088/1475-7516/2019/12/010}, \href
  {https://ui.adsabs.harvard.edu/abs/2019JCAP...12..010K} {2019, 010}

\bibitem[\protect\citeauthoryear{{Kaplinghat}, {Tulin}  \& {Yu}}{{Kaplinghat}
  et~al.}{2016}]{Kaplinghat16}
{Kaplinghat} M.,  {Tulin} S.,   {Yu} H.-B.,  2016, \mn@doi [\prl]
  {10.1103/PhysRevLett.116.041302}, \href
  {https://ui.adsabs.harvard.edu/abs/2016PhRvL.116d1302K} {116, 041302}

\bibitem[\protect\citeauthoryear{{Kaplinghat}, {Valli}  \& {Yu}}{{Kaplinghat}
  et~al.}{2019}]{Kaplinghat19}
{Kaplinghat} M.,  {Valli} M.,   {Yu} H.-B.,  2019, \mn@doi [\mnras]
  {10.1093/mnras/stz2511}, \href
  {https://ui.adsabs.harvard.edu/abs/2019MNRAS.490..231K} {490, 231}

\bibitem[\protect\citeauthoryear{{Kelley}, {Bullock}, {Garrison-Kimmel},
  {Boylan-Kolchin}, {Pawlowski}  \& {Graus}}{{Kelley} et~al.}{2019}]{Kelley19}
{Kelley} T.,  {Bullock} J.~S.,  {Garrison-Kimmel} S.,  {Boylan-Kolchin} M.,
  {Pawlowski} M.~S.,   {Graus} A.~S.,  2019, \mn@doi [\mnras]
  {10.1093/mnras/stz1553}, \href
  {https://ui.adsabs.harvard.edu/abs/2019MNRAS.487.4409K} {487, 4409}

\bibitem[\protect\citeauthoryear{{Kim} \& {Peter}}{{Kim} \&
  {Peter}}{2021}]{Kim21}
{Kim} S.~Y.,  {Peter} A. H.~G.,  2021, arXiv e-prints, \href
  {https://ui.adsabs.harvard.edu/abs/2021arXiv210609050K} {p. arXiv:2106.09050}

\bibitem[\protect\citeauthoryear{{Kim}, {Peter}  \& {Wittman}}{{Kim}
  et~al.}{2017}]{Kim17}
{Kim} S.~Y.,  {Peter} A. H.~G.,   {Wittman} D.,  2017, \mn@doi [\mnras]
  {10.1093/mnras/stx896}, \href
  {https://ui.adsabs.harvard.edu/abs/2017MNRAS.469.1414K} {469, 1414}

\bibitem[\protect\citeauthoryear{{Kim}, {Peter}  \& {Hargis}}{{Kim}
  et~al.}{2018}]{Kim18}
{Kim} S.~Y.,  {Peter} A. H.~G.,   {Hargis} J.~R.,  2018, \mn@doi [\prl]
  {10.1103/PhysRevLett.121.211302}, \href
  {https://ui.adsabs.harvard.edu/abs/2018PhRvL.121u1302K} {121, 211302}

\bibitem[\protect\citeauthoryear{{Klypin}, {Kravtsov}, {Valenzuela}  \&
  {Prada}}{{Klypin} et~al.}{1999}]{Klypin99}
{Klypin} A.,  {Kravtsov} A.~V.,  {Valenzuela} O.,   {Prada} F.,  1999, \mn@doi
  [\apj] {10.1086/307643}, \href
  {https://ui.adsabs.harvard.edu/abs/1999ApJ...522...82K} {522, 82}

\bibitem[\protect\citeauthoryear{{Koda} \& {Shapiro}}{{Koda} \&
  {Shapiro}}{2011}]{Koda11}
{Koda} J.,  {Shapiro} P.~R.,  2011, \mn@doi [\mnras]
  {10.1111/j.1365-2966.2011.18684.x}, \href
  {https://ui.adsabs.harvard.edu/abs/2011MNRAS.415.1125K} {415, 1125}

\bibitem[\protect\citeauthoryear{{Kummer}, {Br{\"u}ggen}, {Dolag}, {Kahlhoefer}
   \& {Schmidt-Hoberg}}{{Kummer} et~al.}{2019}]{Kummer19}
{Kummer} J.,  {Br{\"u}ggen} M.,  {Dolag} K.,  {Kahlhoefer} F.,
  {Schmidt-Hoberg} K.,  2019, \mn@doi [\mnras] {10.1093/mnras/stz1261}, \href
  {https://ui.adsabs.harvard.edu/abs/2019MNRAS.487..354K} {487, 354}

\bibitem[\protect\citeauthoryear{{Lazar} et~al.,}{{Lazar}
  et~al.}{2020}]{Lazar20}
{Lazar} A.,  et~al., 2020, \mn@doi [\mnras] {10.1093/mnras/staa2101}, \href
  {https://ui.adsabs.harvard.edu/abs/2020MNRAS.497.2393L} {497, 2393}

\bibitem[\protect\citeauthoryear{{Mashchenko}, {Wadsley}  \&
  {Couchman}}{{Mashchenko} et~al.}{2008}]{Mashchenko08}
{Mashchenko} S.,  {Wadsley} J.,   {Couchman} H.~M.~P.,  2008, \mn@doi [Science]
  {10.1126/science.1148666}, \href
  {https://ui.adsabs.harvard.edu/abs/2008Sci...319..174M} {319, 174}

\bibitem[\protect\citeauthoryear{{Massey} et~al.,}{{Massey}
  et~al.}{2015}]{Massey15}
{Massey} R.,  et~al., 2015, \mn@doi [\mnras] {10.1093/mnras/stv467}, \href
  {https://ui.adsabs.harvard.edu/abs/2015MNRAS.449.3393M} {449, 3393}

\bibitem[\protect\citeauthoryear{{Miralda-Escud{\'e}}}{{Miralda-Escud{\'e}}}{2002}]{MiraldaEscude02}
{Miralda-Escud{\'e}} J.,  2002, \mn@doi [\apj] {10.1086/324138}, \href
  {https://ui.adsabs.harvard.edu/abs/2002ApJ...564...60M} {564, 60}

\bibitem[\protect\citeauthoryear{{Monaghan} \& {Lattanzio}}{{Monaghan} \&
  {Lattanzio}}{1985}]{Monaghan85}
{Monaghan} J.~J.,  {Lattanzio} J.~C.,  1985, \aap, \href
  {https://ui.adsabs.harvard.edu/abs/1985A&A...149..135M} {149, 135}

\bibitem[\protect\citeauthoryear{{Moore}}{{Moore}}{1994}]{Moore94}
{Moore} B.,  1994, \mn@doi [\nat] {10.1038/370629a0}, \href
  {https://ui.adsabs.harvard.edu/abs/1994Natur.370..629M} {370, 629}

\bibitem[\protect\citeauthoryear{{Moore}, {Quinn}, {Governato}, {Stadel}  \&
  {Lake}}{{Moore} et~al.}{1999}]{Moore99}
{Moore} B.,  {Quinn} T.,  {Governato} F.,  {Stadel} J.,   {Lake} G.,  1999,
  \mn@doi [\mnras] {10.1046/j.1365-8711.1999.03039.x}, \href
  {https://ui.adsabs.harvard.edu/abs/1999MNRAS.310.1147M} {310, 1147}

\bibitem[\protect\citeauthoryear{{Nadler}, {Banerjee}, {Adhikari}, {Mao}  \&
  {Wechsler}}{{Nadler} et~al.}{2020}]{Nadler20}
{Nadler} E.~O.,  {Banerjee} A.,  {Adhikari} S.,  {Mao} Y.-Y.,   {Wechsler}
  R.~H.,  2020, arXiv e-prints, \href
  {https://ui.adsabs.harvard.edu/abs/2020arXiv200108754N} {p. arXiv:2001.08754}

\bibitem[\protect\citeauthoryear{{Navarro}, {Frenk}  \& {White}}{{Navarro}
  et~al.}{1997}]{NFW97}
{Navarro} J.~F.,  {Frenk} C.~S.,   {White} S. D.~M.,  1997, \mn@doi [\apj]
  {10.1086/304888}, \href
  {https://ui.adsabs.harvard.edu/abs/1997ApJ...490..493N} {490, 493}

\bibitem[\protect\citeauthoryear{{Nishikawa}, {Boddy}  \&
  {Kaplinghat}}{{Nishikawa} et~al.}{2020}]{Nishikawa19}
{Nishikawa} H.,  {Boddy} K.~K.,   {Kaplinghat} M.,  2020, \mn@doi [\prd]
  {10.1103/PhysRevD.101.063009}, \href
  {https://ui.adsabs.harvard.edu/abs/2020PhRvD.101f3009N} {101, 063009}

\bibitem[\protect\citeauthoryear{{Oh}, {Brook}, {Governato}, {Brinks}, {Mayer},
  {de Blok}, {Brooks}  \& {Walter}}{{Oh} et~al.}{2011}]{Oh11}
{Oh} S.-H.,  {Brook} C.,  {Governato} F.,  {Brinks} E.,  {Mayer} L.,  {de Blok}
  W.~J.~G.,  {Brooks} A.,   {Walter} F.,  2011, \mn@doi [\aj]
  {10.1088/0004-6256/142/1/24}, \href
  {https://ui.adsabs.harvard.edu/abs/2011AJ....142...24O} {142, 24}

\bibitem[\protect\citeauthoryear{{Oman} et~al.,}{{Oman} et~al.}{2015}]{Oman15}
{Oman} K.~A.,  et~al., 2015, \mn@doi [\mnras] {10.1093/mnras/stv1504}, \href
  {https://ui.adsabs.harvard.edu/abs/2015MNRAS.452.3650O} {452, 3650}

\bibitem[\protect\citeauthoryear{{Papastergis}, {Giovanelli}, {Haynes}  \&
  {Shankar}}{{Papastergis} et~al.}{2015}]{Papastergis15}
{Papastergis} E.,  {Giovanelli} R.,  {Haynes} M.~P.,   {Shankar} F.,  2015,
  \mn@doi [\aap] {10.1051/0004-6361/201424909}, \href
  {https://ui.adsabs.harvard.edu/abs/2015A&A...574A.113P} {574, A113}

\bibitem[\protect\citeauthoryear{{Peter}, {Rocha}, {Bullock}  \&
  {Kaplinghat}}{{Peter} et~al.}{2013}]{Peter13}
{Peter} A. H.~G.,  {Rocha} M.,  {Bullock} J.~S.,   {Kaplinghat} M.,  2013,
  \mn@doi [\mnras] {10.1093/mnras/sts535}, \href
  {https://ui.adsabs.harvard.edu/abs/2013MNRAS.430..105P} {430, 105}

\bibitem[\protect\citeauthoryear{{Planck Collaboration} et~al.}{{Planck
  Collaboration} et~al.}{2020}]{Planck20}
{Planck Collaboration} et~al., 2020, \mn@doi [\aap]
  {10.1051/0004-6361/201833293}, \href
  {https://ui.adsabs.harvard.edu/abs/2020A&A...641A...2P} {641, A2}

\bibitem[\protect\citeauthoryear{{Pontzen} \& {Governato}}{{Pontzen} \&
  {Governato}}{2012}]{Pontzen12}
{Pontzen} A.,  {Governato} F.,  2012, \mn@doi [\mnras]
  {10.1111/j.1365-2966.2012.20571.x}, \href
  {https://ui.adsabs.harvard.edu/abs/2012MNRAS.421.3464P} {421, 3464}

\bibitem[\protect\citeauthoryear{{Pospelov}, {Ritz}  \& {Voloshin}}{{Pospelov}
  et~al.}{2008}]{Pospelov08}
{Pospelov} M.,  {Ritz} A.,   {Voloshin} M.,  2008, \mn@doi [\prd]
  {10.1103/PhysRevD.78.115012}, \href
  {https://ui.adsabs.harvard.edu/abs/2008PhRvD..78k5012P} {78, 115012}

\bibitem[\protect\citeauthoryear{{Power}, {Navarro}, {Jenkins}, {Frenk},
  {White}, {Springel}, {Stadel}  \& {Quinn}}{{Power} et~al.}{2003}]{Power03}
{Power} C.,  {Navarro} J.~F.,  {Jenkins} A.,  {Frenk} C.~S.,  {White} S.~D.~M.,
   {Springel} V.,  {Stadel} J.,   {Quinn} T.,  2003, \mn@doi [\mnras]
  {10.1046/j.1365-8711.2003.05925.x}, \href
  {https://ui.adsabs.harvard.edu/abs/2003MNRAS.338...14P} {338, 14}

\bibitem[\protect\citeauthoryear{{Price}}{{Price}}{2012}]{Price12}
{Price} D.~J.,  2012, \mn@doi [Journal of Computational Physics]
  {10.1016/j.jcp.2010.12.011}, \href
  {https://ui.adsabs.harvard.edu/abs/2012JCoPh.231..759P} {231, 759}

\bibitem[\protect\citeauthoryear{{Randall}, {Markevitch}, {Clowe}, {Gonzalez}
  \& {Brada{\v{c}}}}{{Randall} et~al.}{2008}]{Randall08}
{Randall} S.~W.,  {Markevitch} M.,  {Clowe} D.,  {Gonzalez} A.~H.,
  {Brada{\v{c}}} M.,  2008, \mn@doi [\apj] {10.1086/587859}, \href
  {https://ui.adsabs.harvard.edu/abs/2008ApJ...679.1173R} {679, 1173}

\bibitem[\protect\citeauthoryear{{Read} \& {Gilmore}}{{Read} \&
  {Gilmore}}{2005}]{Read05}
{Read} J.~I.,  {Gilmore} G.,  2005, \mn@doi [\mnras]
  {10.1111/j.1365-2966.2004.08424.x}, \href
  {https://ui.adsabs.harvard.edu/abs/2005MNRAS.356..107R} {356, 107}

\bibitem[\protect\citeauthoryear{{Read}, {Iorio}, {Agertz}  \&
  {Fraternali}}{{Read} et~al.}{2016}]{Read16}
{Read} J.~I.,  {Iorio} G.,  {Agertz} O.,   {Fraternali} F.,  2016, \mn@doi
  [\mnras] {10.1093/mnras/stw1876}, \href
  {https://ui.adsabs.harvard.edu/abs/2016MNRAS.462.3628R} {462, 3628}

\bibitem[\protect\citeauthoryear{{Read}, {Walker}  \& {Steger}}{{Read}
  et~al.}{2018}]{Read18}
{Read} J.~I.,  {Walker} M.~G.,   {Steger} P.,  2018, \mn@doi [\mnras]
  {10.1093/mnras/sty2286}, \href
  {https://ui.adsabs.harvard.edu/abs/2018MNRAS.481..860R} {481, 860}

\bibitem[\protect\citeauthoryear{{Read}, {Walker}  \& {Steger}}{{Read}
  et~al.}{2019}]{Read19}
{Read} J.~I.,  {Walker} M.~G.,   {Steger} P.,  2019, \mn@doi [\mnras]
  {10.1093/mnras/sty3404}, \href
  {https://ui.adsabs.harvard.edu/abs/2019MNRAS.484.1401R} {484, 1401}

\bibitem[\protect\citeauthoryear{{Relatores} et~al.,}{{Relatores}
  et~al.}{2019}]{Relatores19}
{Relatores} N.~C.,  et~al., 2019, \mn@doi [\apj] {10.3847/1538-4357/ab5305},
  \href {https://ui.adsabs.harvard.edu/abs/2019ApJ...887...94R} {887, 94}

\bibitem[\protect\citeauthoryear{{Ren}, {Kwa}, {Kaplinghat}  \& {Yu}}{{Ren}
  et~al.}{2019}]{Ren19}
{Ren} T.,  {Kwa} A.,  {Kaplinghat} M.,   {Yu} H.-B.,  2019, \mn@doi [Physical
  Review X] {10.1103/PhysRevX.9.031020}, \href
  {https://ui.adsabs.harvard.edu/abs/2019PhRvX...9c1020R} {9, 031020}

\bibitem[\protect\citeauthoryear{{Robertson}}{{Robertson}}{2017}]{Robertson17Thesis}
{Robertson} A.,  2017, PhD thesis, Durham University, UK

\bibitem[\protect\citeauthoryear{{Robertson}, {Massey}  \& {Eke}}{{Robertson}
  et~al.}{2017}]{Robertson17}
{Robertson} A.,  {Massey} R.,   {Eke} V.,  2017, \mn@doi [\mnras]
  {10.1093/mnras/stx463}, \href
  {https://ui.adsabs.harvard.edu/abs/2017MNRAS.467.4719R} {467, 4719}

\bibitem[\protect\citeauthoryear{{Robertson}, {Harvey}, {Massey}, {Eke},
  {McCarthy}, {Jauzac}, {Li}  \& {Schaye}}{{Robertson}
  et~al.}{2019}]{Robertson19}
{Robertson} A.,  {Harvey} D.,  {Massey} R.,  {Eke} V.,  {McCarthy} I.~G.,
  {Jauzac} M.,  {Li} B.,   {Schaye} J.,  2019, \mn@doi [\mnras]
  {10.1093/mnras/stz1815}, \href
  {https://ui.adsabs.harvard.edu/abs/2019MNRAS.488.3646R} {488, 3646}

\bibitem[\protect\citeauthoryear{{Robertson}, {Massey}, {Eke}, {Schaye}  \&
  {Theuns}}{{Robertson} et~al.}{2021}]{Robertson21}
{Robertson} A.,  {Massey} R.,  {Eke} V.,  {Schaye} J.,   {Theuns} T.,  2021,
  \mn@doi [\mnras] {10.1093/mnras/staa3954}, \href
  {https://ui.adsabs.harvard.edu/abs/2021MNRAS.501.4610R} {501, 4610}

\bibitem[\protect\citeauthoryear{{Robles}, {Kelley}, {Bullock}  \&
  {Kaplinghat}}{{Robles} et~al.}{2019}]{Robles19}
{Robles} V.~H.,  {Kelley} T.,  {Bullock} J.~S.,   {Kaplinghat} M.,  2019,
  \mn@doi [\mnras] {10.1093/mnras/stz2345}, \href
  {https://ui.adsabs.harvard.edu/abs/2019MNRAS.490.2117R} {490, 2117}

\bibitem[\protect\citeauthoryear{{Rocha}, {Peter}, {Bullock}, {Kaplinghat},
  {Garrison-Kimmel}, {O{\~n}orbe}  \& {Moustakas}}{{Rocha}
  et~al.}{2013}]{Rocha13}
{Rocha} M.,  {Peter} A. H.~G.,  {Bullock} J.~S.,  {Kaplinghat} M.,
  {Garrison-Kimmel} S.,  {O{\~n}orbe} J.,   {Moustakas} L.~A.,  2013, \mn@doi
  [\mnras] {10.1093/mnras/sts514}, \href
  {https://ui.adsabs.harvard.edu/abs/2013MNRAS.430...81R} {430, 81}

\bibitem[\protect\citeauthoryear{{Sagunski}, {Gad-Nasr}, {Colquhoun},
  {Robertson}  \& {Tulin}}{{Sagunski} et~al.}{2021}]{Sagunski21}
{Sagunski} L.,  {Gad-Nasr} S.,  {Colquhoun} B.,  {Robertson} A.,   {Tulin} S.,
  2021, \mn@doi [\jcap] {10.1088/1475-7516/2021/01/024}, \href
  {https://ui.adsabs.harvard.edu/abs/2021JCAP...01..024S} {2021, 024}

\bibitem[\protect\citeauthoryear{{Sameie}, {Yu}, {Sales}, {Vogelsberger}  \&
  {Zavala}}{{Sameie} et~al.}{2020}]{Sameie20}
{Sameie} O.,  {Yu} H.-B.,  {Sales} L.~V.,  {Vogelsberger} M.,   {Zavala} J.,
  2020, \mn@doi [\prl] {10.1103/PhysRevLett.124.141102}, \href
  {https://ui.adsabs.harvard.edu/abs/2020PhRvL.124n1102S} {124, 141102}

\bibitem[\protect\citeauthoryear{{Santos-Santos}, {Di Cintio}, {Brook},
  {Macci{\`o}}, {Dutton}  \& {Dom{\'\i}nguez-Tenreiro}}{{Santos-Santos}
  et~al.}{2018}]{Santos18}
{Santos-Santos} I.~M.,  {Di Cintio} A.,  {Brook} C.~B.,  {Macci{\`o}} A.,
  {Dutton} A.,   {Dom{\'\i}nguez-Tenreiro} R.,  2018, \mn@doi [\mnras]
  {10.1093/mnras/stx2660}, \href
  {https://ui.adsabs.harvard.edu/abs/2018MNRAS.473.4392S} {473, 4392}

\bibitem[\protect\citeauthoryear{{Santos-Santos} et~al.,}{{Santos-Santos}
  et~al.}{2020}]{Santos20}
{Santos-Santos} I. M.~E.,  et~al., 2020, \mn@doi [\mnras]
  {10.1093/mnras/staa1072}, \href
  {https://ui.adsabs.harvard.edu/abs/2020MNRAS.495...58S} {495, 58}

\bibitem[\protect\citeauthoryear{{Sawala} et~al.,}{{Sawala}
  et~al.}{2016}]{Sawala16}
{Sawala} T.,  et~al., 2016, \mn@doi [\mnras] {10.1093/mnras/stw145}, \href
  {https://ui.adsabs.harvard.edu/abs/2016MNRAS.457.1931S} {457, 1931}

\bibitem[\protect\citeauthoryear{{Schaller} et~al.,}{{Schaller}
  et~al.}{2015}]{Schaller15}
{Schaller} M.,  et~al., 2015, \mn@doi [\mnras] {10.1093/mnras/stv1067}, \href
  {https://ui.adsabs.harvard.edu/abs/2015MNRAS.451.1247S} {451, 1247}

\bibitem[\protect\citeauthoryear{{Schaller}, {Gonnet}, {Chalk}  \&
  {Draper}}{{Schaller} et~al.}{2016}]{Schaller16}
{Schaller} M.,  {Gonnet} P.,  {Chalk} A. B.~G.,   {Draper} P.~W.,  2016, arXiv
  e-prints, \href {https://ui.adsabs.harvard.edu/abs/2016arXiv160602738S} {p.
  arXiv:1606.02738}

\bibitem[\protect\citeauthoryear{{Schaller}, {Gonnet}, {Draper}, {Chalk},
  {Bower}, {Willis}  \& {Hausammann}}{{Schaller} et~al.}{2018}]{Schaller18}
{Schaller} M.,  {Gonnet} P.,  {Draper} P.~W.,  {Chalk} A. B.~G.,  {Bower}
  R.~G.,  {Willis} J.,   {Hausammann} L.,  2018, {SWIFT: SPH With
  Inter-dependent Fine-grained Tasking}, Astrophysics Source Code Library,
  record ascl:1805.020 (\mn@eprint {ascl} {1805.020})

\bibitem[\protect\citeauthoryear{{Shen}, {Hopkins}, {Necib}, {Jiang},
  {Boylan-Kolchin}  \& {Wetzel}}{{Shen} et~al.}{2021}]{Shen21}
{Shen} X.,  {Hopkins} P.~F.,  {Necib} L.,  {Jiang} F.,  {Boylan-Kolchin} M.,
  {Wetzel} A.,  2021, \mn@doi [\mnras] {10.1093/mnras/stab2042}, \href
  {https://ui.adsabs.harvard.edu/abs/2021MNRAS.506.4421S} {506, 4421}

\bibitem[\protect\citeauthoryear{{Silverman}, {Bullock}, {Kaplinghat}, {Robles}
   \& {Valli}}{{Silverman} et~al.}{2022}]{Silverman22}
{Silverman} M.,  {Bullock} J.~S.,  {Kaplinghat} M.,  {Robles} V.~H.,   {Valli}
  M.,  2022, arXiv e-prints, \href
  {https://ui.adsabs.harvard.edu/abs/2022arXiv220310104S} {p. arXiv:2203.10104}

\bibitem[\protect\citeauthoryear{{Spergel} \& {Steinhardt}}{{Spergel} \&
  {Steinhardt}}{2000}]{Spergel00}
{Spergel} D.~N.,  {Steinhardt} P.~J.,  2000, \mn@doi [\prl]
  {10.1103/PhysRevLett.84.3760}, \href
  {https://ui.adsabs.harvard.edu/abs/2000PhRvL..84.3760S} {84, 3760}

\bibitem[\protect\citeauthoryear{{Springel}}{{Springel}}{2005}]{Springel05}
{Springel} V.,  2005, \mn@doi [\mnras] {10.1111/j.1365-2966.2005.09655.x},
  \href {https://ui.adsabs.harvard.edu/abs/2005MNRAS.364.1105S} {364, 1105}

\bibitem[\protect\citeauthoryear{{Tollerud}, {Boylan-Kolchin}  \&
  {Bullock}}{{Tollerud} et~al.}{2014}]{Tollerud14}
{Tollerud} E.~J.,  {Boylan-Kolchin} M.,   {Bullock} J.~S.,  2014, \mn@doi
  [\mnras] {10.1093/mnras/stu474}, \href
  {https://ui.adsabs.harvard.edu/abs/2014MNRAS.440.3511T} {440, 3511}

\bibitem[\protect\citeauthoryear{{Tollet} et~al.,}{{Tollet}
  et~al.}{2016}]{Tollet16}
{Tollet} E.,  et~al., 2016, \mn@doi [\mnras] {10.1093/mnras/stv2856}, \href
  {https://ui.adsabs.harvard.edu/abs/2016MNRAS.456.3542T} {456, 3542}

\bibitem[\protect\citeauthoryear{{Torrealba} et~al.}{{Torrealba}
  et~al.}{2018}]{Torrealba18}
{Torrealba} G.,  et~al., 2018, \mn@doi [\mnras] {10.1093/mnras/sty170}, \href
  {https://ui.adsabs.harvard.edu/abs/2018MNRAS.475.5085T} {475, 5085}

\bibitem[\protect\citeauthoryear{{Tulin} \& {Yu}}{{Tulin} \&
  {Yu}}{2018}]{Tulin18}
{Tulin} S.,  {Yu} H.-B.,  2018, \mn@doi [\physrep]
  {10.1016/j.physrep.2017.11.004}, \href
  {https://ui.adsabs.harvard.edu/abs/2018PhR...730....1T} {730, 1}

\bibitem[\protect\citeauthoryear{{Turner}, {Lovell}, {Zavala}  \&
  {Vogelsberger}}{{Turner} et~al.}{2021}]{Turner21}
{Turner} H.~C.,  {Lovell} M.~R.,  {Zavala} J.,   {Vogelsberger} M.,  2021,
  \mn@doi [\mnras] {10.1093/mnras/stab1725}, \href
  {https://ui.adsabs.harvard.edu/abs/2021MNRAS.505.5327T} {505, 5327}

\bibitem[\protect\citeauthoryear{{Valli} \& {Yu}}{{Valli} \&
  {Yu}}{2018}]{Valli18}
{Valli} M.,  {Yu} H.-B.,  2018, \mn@doi [Nature Astronomy]
  {10.1038/s41550-018-0560-7}, \href
  {https://ui.adsabs.harvard.edu/abs/2018NatAs...2..907V} {2, 907}

\bibitem[\protect\citeauthoryear{{Vogelsberger}, {Zavala}  \&
  {Loeb}}{{Vogelsberger} et~al.}{2012}]{Vogelsberger12}
{Vogelsberger} M.,  {Zavala} J.,   {Loeb} A.,  2012, \mn@doi [\mnras]
  {10.1111/j.1365-2966.2012.21182.x}, \href
  {https://ui.adsabs.harvard.edu/abs/2012MNRAS.423.3740V} {423, 3740}

\bibitem[\protect\citeauthoryear{{Vogelsberger}, {Zavala}, {Cyr-Racine},
  {Pfrommer}, {Bringmann}  \& {Sigurdson}}{{Vogelsberger}
  et~al.}{2016}]{Vogelsberger16}
{Vogelsberger} M.,  {Zavala} J.,  {Cyr-Racine} F.-Y.,  {Pfrommer} C.,
  {Bringmann} T.,   {Sigurdson} K.,  2016, \mn@doi [\mnras]
  {10.1093/mnras/stw1076}, \href
  {https://ui.adsabs.harvard.edu/abs/2016MNRAS.460.1399V} {460, 1399}

\bibitem[\protect\citeauthoryear{{Vogelsberger}, {Zavala}, {Schutz}  \&
  {Slatyer}}{{Vogelsberger} et~al.}{2019}]{Vogelsberger19}
{Vogelsberger} M.,  {Zavala} J.,  {Schutz} K.,   {Slatyer} T.~R.,  2019,
  \mn@doi [\mnras] {10.1093/mnras/stz340}, \href
  {https://ui.adsabs.harvard.edu/abs/2019MNRAS.484.5437V} {484, 5437}

\bibitem[\protect\citeauthoryear{{Walker} \& {Pe{\~n}arrubia}}{{Walker} \&
  {Pe{\~n}arrubia}}{2011}]{Walker11}
{Walker} M.~G.,  {Pe{\~n}arrubia} J.,  2011, \mn@doi [\apj]
  {10.1088/0004-637X/742/1/20}, \href
  {https://ui.adsabs.harvard.edu/abs/2011ApJ...742...20W} {742, 20}

\bibitem[\protect\citeauthoryear{{Walker}, {McGaugh}, {Mateo}, {Olszewski}  \&
  {Kuzio de Naray}}{{Walker} et~al.}{2010}]{Walker10}
{Walker} M.~G.,  {McGaugh} S.~S.,  {Mateo} M.,  {Olszewski} E.~W.,   {Kuzio de
  Naray} R.,  2010, \mn@doi [\apjl] {10.1088/2041-8205/717/2/L87}, \href
  {https://ui.adsabs.harvard.edu/abs/2010ApJ...717L..87W} {717, L87}

\bibitem[\protect\citeauthoryear{{Wetzel}, {Hopkins}, {Kim},
  {Faucher-Gigu{\`e}re}, {Kere{\v{s}}}  \& {Quataert}}{{Wetzel}
  et~al.}{2016}]{Wetzel16}
{Wetzel} A.~R.,  {Hopkins} P.~F.,  {Kim} J.-h.,  {Faucher-Gigu{\`e}re} C.-A.,
  {Kere{\v{s}}} D.,   {Quataert} E.,  2016, \mn@doi [\apjl]
  {10.3847/2041-8205/827/2/L23}, \href
  {https://ui.adsabs.harvard.edu/abs/2016ApJ...827L..23W} {827, L23}

\bibitem[\protect\citeauthoryear{{Wittman}, {Golovich}  \& {Dawson}}{{Wittman}
  et~al.}{2018}]{Wittman18}
{Wittman} D.,  {Golovich} N.,   {Dawson} W.~A.,  2018, \mn@doi [\apj]
  {10.3847/1538-4357/aaee77}, \href
  {https://ui.adsabs.harvard.edu/abs/2018ApJ...869..104W} {869, 104}

\bibitem[\protect\citeauthoryear{{Yoshida}, {Springel}, {White}  \&
  {Tormen}}{{Yoshida} et~al.}{2000}]{Yoshida00}
{Yoshida} N.,  {Springel} V.,  {White} S. D.~M.,   {Tormen} G.,  2000, \mn@doi
  [\apjl] {10.1086/317306}, \href
  {https://ui.adsabs.harvard.edu/abs/2000ApJ...544L..87Y} {544, L87}

\bibitem[\protect\citeauthoryear{{Zavala}, {Vogelsberger}  \&
  {Walker}}{{Zavala} et~al.}{2013}]{Zavala13}
{Zavala} J.,  {Vogelsberger} M.,   {Walker} M.~G.,  2013, \mn@doi [\mnras]
  {10.1093/mnrasl/sls053}, \href
  {https://ui.adsabs.harvard.edu/abs/2013MNRAS.431L..20Z} {431, L20}

\bibitem[\protect\citeauthoryear{{Zavala}, {Lovell}, {Vogelsberger}  \&
  {Burger}}{{Zavala} et~al.}{2019}]{Zavala19}
{Zavala} J.,  {Lovell} M.~R.,  {Vogelsberger} M.,   {Burger} J.~D.,  2019,
  \mn@doi [\prd] {10.1103/PhysRevD.100.063007}, \href
  {https://ui.adsabs.harvard.edu/abs/2019PhRvD.100f3007Z} {100, 063007}

\bibitem[\protect\citeauthoryear{{de Blok} \& {McGaugh}}{{de Blok} \&
  {McGaugh}}{1997}]{deBlok97}
{de Blok} W.~J.~G.,  {McGaugh} S.~S.,  1997, \mn@doi [\mnras]
  {10.1093/mnras/290.3.533}, \href
  {https://ui.adsabs.harvard.edu/abs/1997MNRAS.290..533D} {290, 533}

\bibitem[\protect\citeauthoryear{{eBOSS Collaboration} et~al.}{{eBOSS
  Collaboration} et~al.}{2021}]{eBOSS2021}
{eBOSS Collaboration} et~al., 2021, \mn@doi [\prd]
  {10.1103/PhysRevD.103.083533}, \href
  {https://ui.adsabs.harvard.edu/abs/2021PhRvD.103h3533A} {103, 083533}

\bibitem[\protect\citeauthoryear{van~der Walt, Colbert  \& Varoquaux}{van~der
  Walt et~al.}{2011}]{vanderWalt11}
van~der Walt S.,  Colbert S.~C.,   Varoquaux G.,  2011, \mn@doi [Computing in
  Science & Engineering] {10.1109/mcse.2011.37}, 13, 22–30

\makeatother
\end{thebibliography}
\bibliographystyle{mnras}

\appendix
\section{SIDM implementation}\label{Appendix_SIDM_derivation}

\subsection{Scattering probability}\label{scatter_probability_sec}

In the simulations dark matter particles represent a patch of the DM field in phase space, with their distributions in physical-space and velocity-space defined by the distribution function $f({\bf{r}}, {\bf{v}}, t)$, with dM=$f({\bf{r}}, {\bf{v}}, t){\rm{d}}^{3}{\bf{r}}{\rm{d}}^{3}{\bf{v}}$ the mass of dark matter in the volume ${\rm{d}}^{3}{\bf{r}}$ centered on $\bf{r}$, with velocity in the velocity-space element $\rm{d}^{3}\bf{v}$ centered on ${\bf{v}}$.

In the presence of collisions, the distribution function evolves as

\begin{equation}\label{distribution_function}
\frac{Df({\bf{r}}, {\bf{v}},t)}{Dt} = \Gamma[f,\sigma]=\Gamma_{\rm{out}}-\Gamma_{\rm{in}},
\end{equation}

\noindent where the `out' term accounts for collisions in which a particle at position ${\bf{r}}$ and with velocity ${\bf{v}}$ scatters from another particle.

We take the ansatz that the evolution of the coarse-grained distribution function $\hat{f}$ (the distribution function averaged over several times the interparticle spacing) is a good representation of the evolution of the fine-grained distribution function $f$. Therefore the solution to $D\hat{f}({\bf{r}}, {\bf{v}},t)/Dt = \Gamma[\hat{f},\sigma]$ is the same as the solution for $f$. We next discretizate eq.~(\ref{distribution_function}) assuming $\hat{f}$ is

\begin{equation}
\hat{f}({\bf{r}}, {\bf{v}},t)=\sum_{i}m_{i}W(|{\bf{r}}-{\bf{r}}_{i}|;h_{i})\delta^{3}({\bf{v}}-{\bf{v}}_{i}),
\end{equation}

\noindent where a delta-function form for the velocity distribution is used since each simulation particle travels at only one speed. We treat each particle as being smoothed out in configuration space with a smoothing kernel $W$ with smoothing length $h_{i}$.

To calculate the scatter probability we follow \citet{Rocha13}, who wrote the particle-based discretization Boltzmann equation and integrated over the patch of phase space inhabited by a single particle of size $\delta {\bf{r}}_{p}\delta {\bf{v}}_{p}$ as follows

\begin{eqnarray}
&& \int_{\delta {\bf{r}}_{p}}d^{3}{\bf{r}}\int_{\delta {\bf{v}}_{p}}d^{3}{\bf{v}}\frac{D\hat{f}}{Dt} =\\\nonumber
&& \int_{\delta {\bf{r}}_{p}}d^{3}{\bf{r}}\int_{\delta {\bf{v}}_{p}}d^{3}{\bf{v}}\int d^{3}{\bf{v}}_{1}\int d\Omega\frac{d\sigma}{d\Omega}|{\bf{v}}-{\bf{v}}_{1}|\times\\
&& [\hat{f}({\bf{r}},{\bf{v}}',t)\hat{f}({\bf{r}},{\bf{v}}'_{1},t)-\hat{f}({\bf{r}},{\bf{v}},t)\hat{f}({\bf{r}},{\bf{v}}_{1},t)],
\end{eqnarray}

\noindent where a particle with initial velocity $v$ collides with a target particle of initial velocity $v_{1}$, the velocities after the collision are $v'$ and $v'_{1}$.

The `scattering in' part of the equation is

\begin{eqnarray}\nonumber
\Gamma(p) & = & \int_{\delta {\bf{r}}_{p}}d^{3}{\bf{r}}\int_{\delta {\bf{v}}_{p}}d^{3}{\bf{v}}\int d^{3}{\bf{v}}_{1}\int d\Omega\frac{d\sigma}{d\Omega}|{\bf{v}}-{\bf{v}}_{1}|\times\\
&& \hat{f}({\bf{r}},{\bf{v}},t)\hat{f}({\bf{r}},{\bf{v}}_{1},t),\\\nonumber
& = & \int_{\delta {\bf{r}}_{p}}d^{3}{\bf{r}}\int_{\delta {\bf{v}}_{p}}d^{3}{\bf{v}}\int d^{3}{\bf{v}}_{1}(\sigma/m)|{\bf{v}}-{\bf{v}}_{1}|m_{p}^{-1}\times \\\nonumber
&& \sum_{j}m_{j}W(|{\bf{r}}-{\bf{r}}_{j}|;h_{j})\delta^{3}({\bf{v}}-{\bf{v}}_{j})\times\\\label{Gamma_p_appendix}
&& \sum_{q}m_{q}W(|{\bf{r}}-{\bf{r}}_{q}|;h_{q})\delta^{3}({\bf{v}}-{\bf{v}}_{q})
\end{eqnarray}

\noindent where the $m_{p}^{-1}$ addition is to calculate the scattering probability for a single particle $j=p$. Doing the integration from eq.~(\ref{Gamma_p_appendix}) yields

\begin{eqnarray}\nonumber
\Gamma(p) & = & \int_{\delta {\bf{r}}_{p}}d^{3}{\bf{r}}\int d^{3}{\bf{v}}_{1}(\sigma/m)|{\bf{v}}_{p}-{\bf{v}}_{1}|\times\\\nonumber
&& \sum_{q}m_{q}W(|{\bf{r}}-{\bf{r}}_{p}|;h_{p})W(|{\bf{r}}-{\bf{r}}_{q}|;h_{q})\delta^{3}({\bf{v}}-{\bf{v}}_{q}),\\\nonumber
& = & \sum_{q}m_{q}(\sigma/m)|{\bf{v}}_{p}-{\bf{v}}_{p}|\times\\
&& \int_{\delta {\bf{r}}_{p}}d^{3}{\bf{r}} W(|{\bf{r}}-{\bf{r}}_{p}|;h_{p})W(|{\bf{r}}-{\bf{r}}_{q}|;h_{q}),\\
& = & \sum_{q}m_{q}(\sigma/m)|{\bf{v}}_{p}-{\bf{v}}_{p}|g_{pq}.
\end{eqnarray}

Using these last equations we can define the probability of particles $i$ and $j$ scattering as

\begin{eqnarray}\label{Probability_appendix}
P_{ij} = m_{j}(\sigma/m)|{\bf{v}}_{i}-{\bf{v}}_{j}| g_{ij}(\delta {\bf{r}}_{ij})\Delta t,
\end{eqnarray}

\noindent where
 
\begin{equation}\label{double_kernel_appendix}
g_{ij} (\delta {\bf{r}}_{ij})= N\int_{0}^{{\rm{max}}(h_{i},h_{j})}d^{3}{\bf{r}}'W(|{\bf{r}}'|,h_{i})W(|\delta {\bf{r}}_{ij}+{\bf{r}}'|,h_{j}),
\end{equation}

\noindent with $\delta {\bf{r}}_{ij}$ the distance between particles $i$ and $j$, and $N$ a normalization factor that requires $\int_{0}^{{\rm{max}}(h_{i},h_{j})}d^{3}{\bf{r}}'g_{ij} ({\bf{r}}')=1$ (suggested by \citealt{Dave01}). 

Other expressions to calculate the DM particles interactions have also been suggested. As an example, \citet{Robertson17} argued that the simplest way to estimate the scattering rate from the particles enclosed in the search region is for all neighbour particles to contribute equally to the probability of collision, independent of their location within the search region. They proposed that the probability of two particles, $i$ and $j$ (separated by a distance less than $h_{SI}$) of scattering within the next time step, $\Delta t$, is given by

\begin{equation}
P_{ij}=\frac{\sigma_{p}|{\bf{v}}_{i}-{\bf{v}}_{j}|\Delta t}{\frac{4}{3}\pi h_{SI}^{3}}.
\end{equation}

\begin{figure} 
	\includegraphics[angle=0,width=0.49\textwidth]{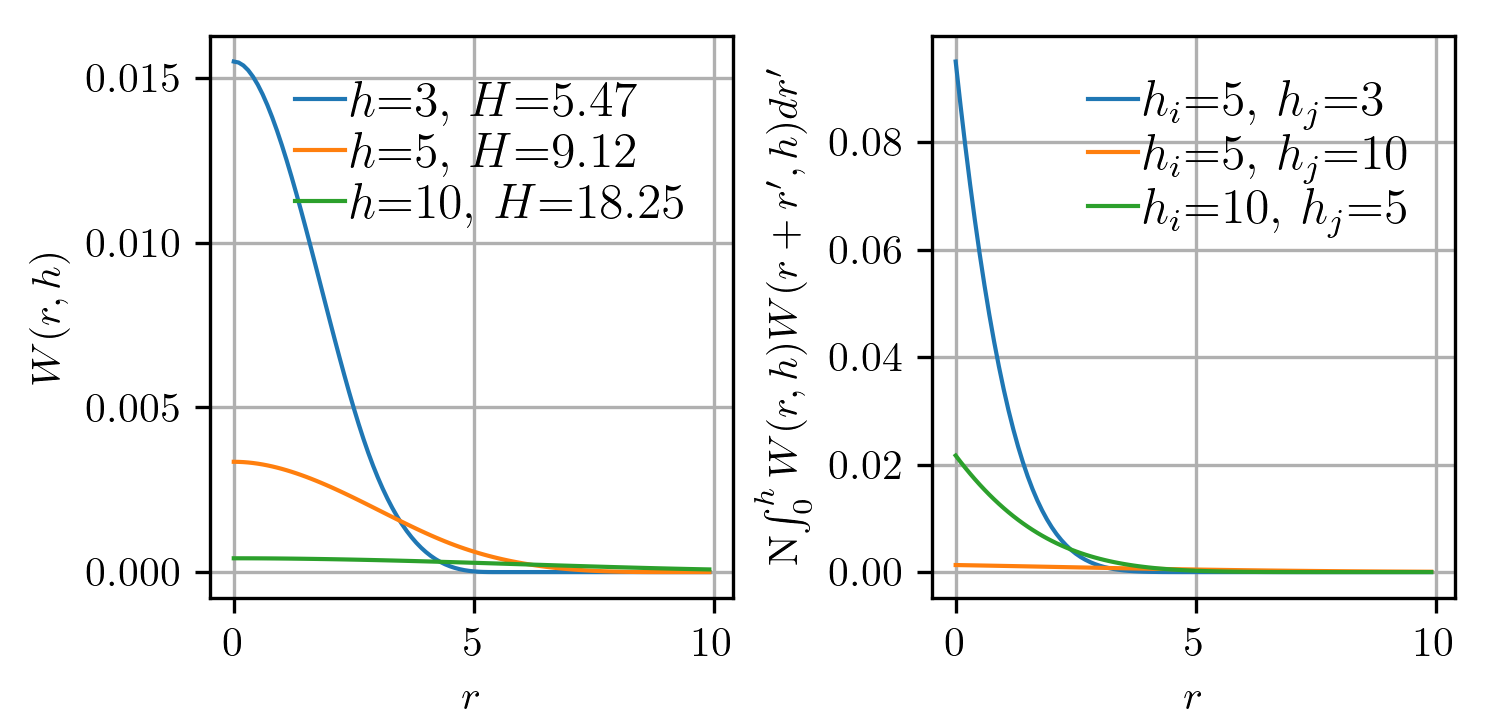}
	\caption{Cubic spline kernel (left) for particles with smoothing lengths $h$ = 3, 5 and 10 (and with kernel's radius $H=5.47$, 9.12, and 18.25, respectively). Convolution of kernels (right) from a pair of particles $i$ and $j$ with different smoothing lengths (see legend) as a function of separation distance. The right panel of the figure shows that particles that are closer to each other and whose kernels largely overlap have a higher probability of collision. This is because the probability $P_{ij}$ depends on the convolution of kernels (eq.~\ref{Probability_appendix}).}
	\label{convolution_kernels_example}
\end{figure}

The difference in our approach, relative to Robertson et al., is that the probability of particles colliding depends (1) on the particles distance and (2) on the particles kernel. Particles that are closer relative to each other have a higher probability of collision, and particles whose kernels largely overlap also have a higher probability of collision. This can be seen from Fig.~\ref{convolution_kernels_example}, that shows the cubic spline kernel for particles with smoothing lengths $h$ = 3, 5 and 10 (left panel), and the convolution of kernels as a function of separation distance of a pair of particles $i$ and $j$, that have different smoothing lengths (right panel).

An important feature of the SIDM implementation is how it selects the neighbouring particles for which the probability of scattering is calculated. We do it by defining the search radius as the DM particle smoothing length $h$. The smoothing length is not fixed, instead it is adapted according to the local DM density around the particles. The smoothing length of each individual particle, $h_{i}$, is calculated by requiring

\begin{equation}\label{smoothing_length_def}
\sum_{j}W(|{\bf{r}}_{j}-{\bf{r}}_{i}|;h_{i})=\left(\frac{\eta}{h_{i}}\right)^{3},
\end{equation}

\noindent when summing around its neighbours. In eq.~(\ref{smoothing_length_def}), $W$ is the kernel (defined in the following section) and $\eta$ a resolution parameter. This method follows the classical SPH formulation (see e.g. \citealt{Price12} for a review of the algorithm), and it has also been implemented in SWIFT to model the evolution of the gas particles (\citealt{Borrow22}).

\subsubsection{Integration of double kernel}

The kernel adopted in SWIFT to calculate the DM particles' density follows the spline kernel (\citealt{Monaghan85}) defined as

\begin{equation*}
W(x,h) = \frac{16}{\pi H} \begin{cases}
1/2 - 3x^{2}+3x^{3}  & x \le 1/2,\\
(1-x)^{3} & 1/2 <x \le 1,\\
0 & x > 1,
\end{cases}
\end{equation*}

\noindent where $H=\gamma h$ is the kernel's support radius and $\gamma=1.825742$. 

The numerical integral of the double kernel (eq.~\ref{double_kernel_appendix}) is computationally expensive, therefore we first do an analytical estimation of the function using the WolframAlpha online tool. 

Given two dark matter particles $i$ and $j$ separated by a distance $r$, we define the following variables, $H_{i}=\gamma h_{i}$, $H_{j}=\gamma h_{j}$, $x_{1}=H_{i}/2$, $x_{2}=H_{j}/2-r$, $x_{3}=H_{i}$, $x_{4}=H_{j}-r$, and find the following expression for $g_{ij}$,

\begin{eqnarray*}
g_{ij}(r) = \frac{(16/\pi)^{2}N}{H_{i}^{3}H_{j}^{3}} \times \hat{g}_{ij}(r),
\end{eqnarray*}

\noindent where $\hat{g}_{ij}(r)$ is

\begin{eqnarray*}
\hat{g}_{ij}(r) = \begin{cases}
W_{11}(x_{1})+W_{21}(x_{2})-W_{21}(x_{1})\\
+W_{22}(x_{3})-W_{22}(x_{2}), &\text{if}~x_{1} \le x_{2} \le x_{3} \le x_{4},\\
W_{11}(x_{1})+W_{21}(x_{2})-W_{21}(x_{1})\\
+W_{22}(x_{4})-W_{22}(x_{2}), &\text{if}~x_{1} \le x_{2} \le x_{4} \le x_{3},\\
W_{11}(x_{2})+W_{12}(x_{1})-W_{12}(x_{2})\\
+W_{22}(x_{3})-W_{22}(x_{1}),&\text{if}~x_{2} \le x_{1} \le x_{3} \le x_{4},\\
W_{11}(x_{2})+W_{12}(x_{4})-W_{12}(x_{2}), &\text{if}~x_{2} \le x_{4} \le x_{1} \le x_{3},\\
W_{11}(x_{1})+W_{21}(x_{3})-W_{21}(x_{1}), &\text{if}~x_{1} \le x_{3} \le x_{2} \le x_{4}.
\end{cases}
\end{eqnarray*}

\noindent Below we show the expression for $W_{11}$,

\begin{eqnarray}\nonumber
w_{11}(x) &=& 14 H_{j}^{3} [1-6(r/H_{j})^{2}+6(r/H_{J})^{3}] \\\nonumber
&& \times(5H_{i}^{3}-18H_{i}x^{2}+15x^{3})\\\nonumber
&& +45 H_{j}^{2}(r/H_{j})[-2+3r/H_{j}]x(7H_{i}^{3}-28H_{i}x^{2}+24x^{3})\\\nonumber
&& +9 H_{j}(-1+3r/H_{j})x^{2}(28H_{i}^{3}-120H_{i}x^{2}+105x^{3})\\
&& +105 (2H_{i}^{3}x^{3}-9H_{i}x^{5}+8x^{6}),\\
W_{11}(x) &=& \frac{0.00119048\times 4\pi x^{3}}{H_{j}^{3}H_{i}^{3}}w_{11}(x).
\end{eqnarray}

\subsection{Model validation}\label{Model_validation_1_sec}

In this section we test the scattering probability derived in Section~\ref{scatter_probability_sec}. To do so, we first generate a distribution of particles' positions and velocities that follow a Hernquist profile (\citealt{Hernquist90}), which is defined by its total mass, $M_{\rm{tot}}$, and a scale radius, $a$ (at which the enclosed mass is $M_{\rm{tot}}/4$), as follows

\begin{equation}\label{Hernquist_rho}
\rho(r)=\frac{M_{\rm{tot}}}{2\pi}\frac{a}{r(r+a)^{3}}.
\end{equation}

The 1-D velocity dispersion profile for the Hernquist halo follows from the Jeans equation as

\begin{eqnarray}\label{Hernquist_vel_disp}
\sigma_{1D}^{2} &=& \frac{GM_{\rm{tot}}}{12a}\left(\frac{12r(r+a)^{3}}{a^{4}}\ln\left(\frac{r+a}{r}\right) \right. \\\nonumber
&& \left. -\frac{r}{r+a}\left[25+52\left(\frac{r}{a}\right)+42\left(\frac{r}{a}\right)^{2}+12\left(\frac{r}{a}\right)^{3}\right]\right).
\end{eqnarray}

We next run simulations for an isolated halo that follows a Hernquist profile. The simulations are run for 1 Gyr, and count with different number of particles (and therefore different resolution), ranging from $64^3$, $128^3$ till $256^3$. For this test the algorithm determines the particles that collide and saves the effective kicks, but we disable the actual collisions and changes in the particles velocities, so that the halo maintains the same profile during its evolution. We calculate the scattering rate of the simulations by determining the location of all the collisions, and binning them in logarithmically-spaced radial bins. This is then divided by the averaged number of particles that reside in the same radial bins to get the scattering rate per particle.

We compare the scattering rate from the simulation output with the analytic solution. For an isolated halo, the number of scattering events as a function of radius can be calculated as

\begin{equation}\label{analytic}
\Gamma(r) = \rho(r)\left\langle(\sigma_{T}/m_{\chi})v_{\rm{pair}}\right\rangle(r),
\end{equation}

\noindent where $\rho(r)$ is the local DM density, and $\left\langle(\sigma_{T}/m_{\chi})v_{\rm{pair}}\right\rangle(r)$ is the averaged of the momentum transfer cross section times the relative velocity of DM particles. In the non-relativistic limit, the average of the cross section time the velocity can be calculated assuming a Maxwell-Boltzmann distribution function,

\begin{equation}\label{sigma_vel}
\left\langle(\sigma_{T}/m_{\chi})v_{\rm{pair}}\right\rangle(r) = \frac{1}{2\sigma^{3}_{v}(r)\sqrt{\pi}}\int(\sigma_{T}/m_{\chi})v^{3}e^{-v^{2}/4\sigma^{2}_{v}(r)}dv,
\end{equation}

\noindent where $\sigma_{v}(r)$ is the local velocity dispersion.

For a constant cross section $\left\langle(\sigma_{T}/m_{\chi})v_{\rm{pair}}\right\rangle(r)=(\sigma/m_{\chi})\left\langle v_{\rm{pair}}\right\rangle(r)=(\sigma/m_{\chi})/(4/\sqrt{\pi})\sigma_{1D}(r)$. In this case the scattering rate can be easily calculated from eqs.~(\ref{Hernquist_rho}) and (\ref{Hernquist_vel_disp}). In the velocity-dependent case, we calculate the integral (eq.~\ref{sigma_vel}) where $\sigma_{T}/m_{\chi}$ depends on $v$ according to eq.~(\ref{sigmat}).

\begin{figure} 
\begin{center}
	\includegraphics[angle=0,width=0.45\textwidth]{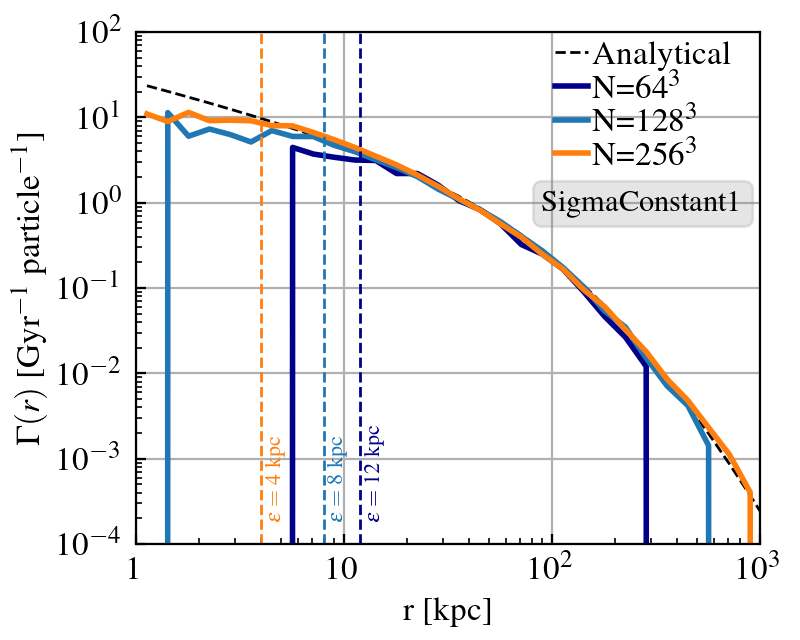}\\
	\includegraphics[angle=0,width=0.45\textwidth]{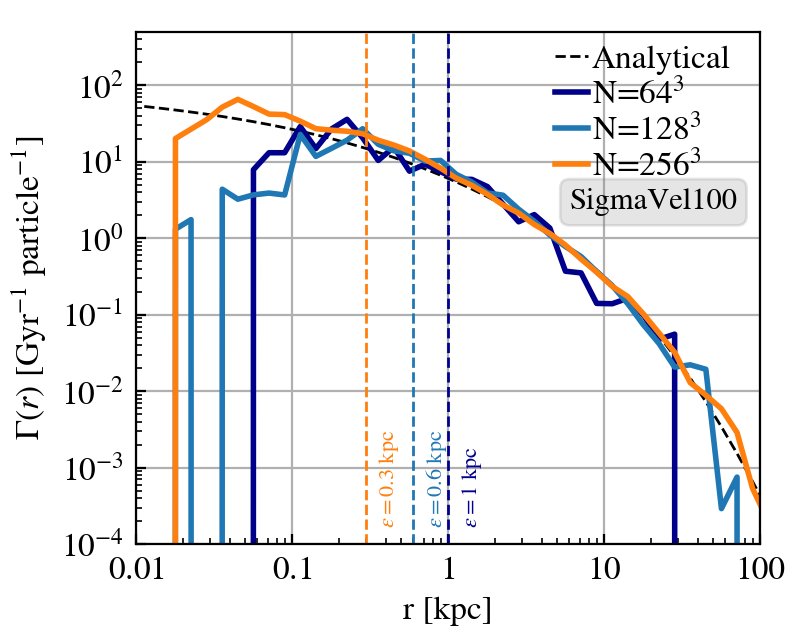}
	\caption{Comparison between the scatter rate profiles of Hernquist haloes obtained from the simulation outputs (coloured lines), and an analytic estimation (black dashed line). The top panel shows the scatter rate of a $10^{14}~\rm{M}_{\odot}$ halo, with a scale radius of 225 kpc, and a constant cross section of $\sigma/m_{\chi}=1$ cm$^{2}$/g, whereas the bottom panel shows the scatter rate of a $10^{10}~\rm{M}_{\odot}$ halo, with a scale radius of 25 kpc, and a velocity-dependent cross section that follows the SigmaVel100 model (see table~\ref{Table_models} for the model parameters). In the panels, the different simulations contain $64^3$ (orange lines), $128^3$ (light blue lines) and $256^3$ (dark blue lines) particles, and they are therefore increasing in resolution. The softening lengths (matching the simulations colors) are highlighted with dashed lines. The figure shows that for a constant cross section, or for a velocity-dependent cross section, the simulation outputs match the analytic estimation.}
   \end{center}
	\label{Model_validation_1}
\end{figure}

Fig.~\ref{Model_validation_1} shows a comparison between the scatter profiles of Hernquist haloes obtained from the simulation outputs, and the analytic estimation given by eq.~(\ref{analytic}). The top panel shows the scatter rate of a $10^{14}~\rm{M}_{\odot}$ halo, with a scale radius of 225 kpc, and a constant cross section of $\sigma/m_{\chi}=1$ cm$^{2}$/g, whereas the bottom panel shows the scatter rate of a $10^{10}~\rm{M}_{\odot}$ halo, with a scale radius of 25 kpc, and a velocity-dependent cross section that follows the SigmaVel100 model (see table~\ref{Table_models} for the model parameters). The panels show the numerical convergence in the simulations scattering rate, by comparing simulations with different number of particles, ranging from $64^3$ (orange lines), $128^3$ (light blue lines) till $256^3$ (dark blue lines) particles. These simulation different resolution as it is highlighted by the softening lengths (that match the simulations colors) with dashed lines. We conclude from Fig.~\ref{Model_validation_1} that the simulation outputs are able to reproduce the analytic estimates. The top panel shows that the simulations smoothly follow the analytic curve, whereas the bottom panel shows a some-what larger scatter around the correct answer. This is due to the fact that in this model the probability of scattering strongly depends on the particles relative velocity, instead of only on the particles positions. 

\begin{figure*} 
	\includegraphics[angle=0,width=\textwidth]{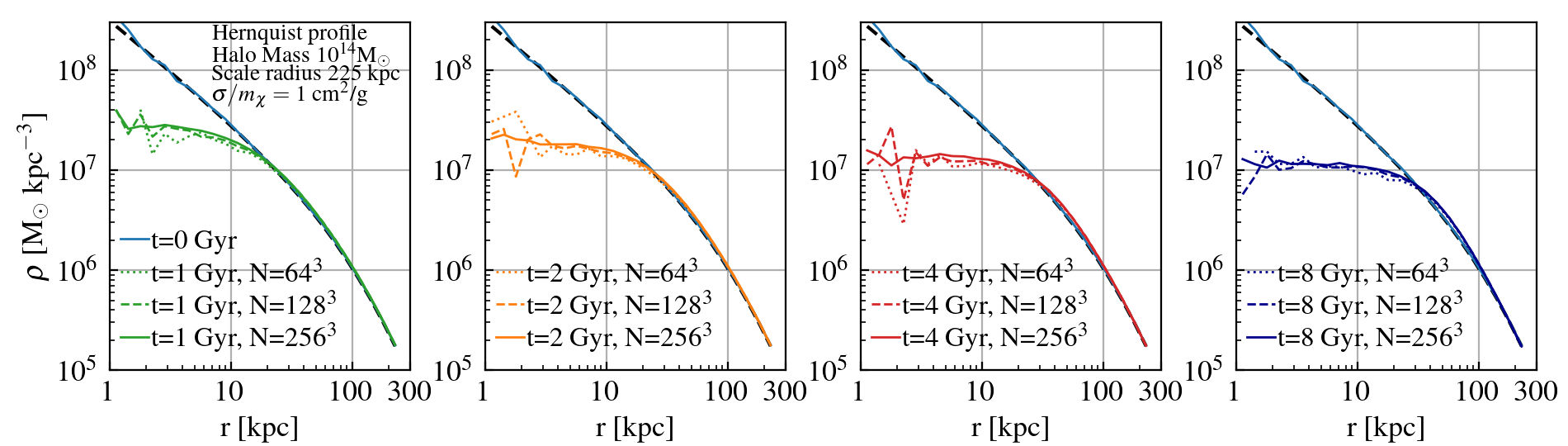}
	\caption{Analysis of the evolution of a $10^{14}~\rm{M}_{\odot}$ halo, with a scale radius of 225 kpc, and a constant cross section of $\sigma/m_{\chi}=1$ cm$^{2}$/g, produced with simulations that contain $64^3$ (shown as dotted lines), $128^3$ (shown as dashed lines) and $256^3$ (shown as solid lines) particles. From left to right, the panels show the evolution in the density profile of a Hernquist halo after 1, 2, 4 and 8 Gyrs of evolution. We find good convergence between the simulations with different resolution.}
	\label{Model_validation_2}
\end{figure*}

\begin{figure*} 
	\includegraphics[angle=0,width=0.75\textwidth]{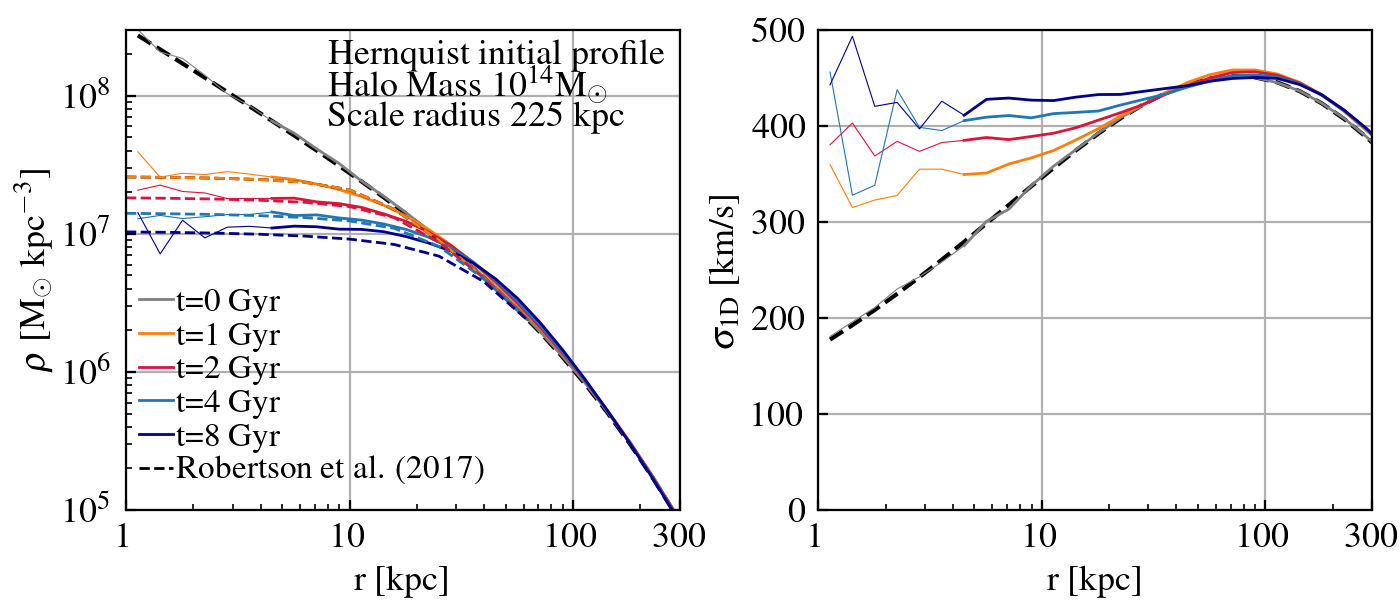}
	\caption{Evolution in the density profile (left panel) and velocity dispersion (right panel) of a $10^{14}~\rm{M}_{\odot}$ Hernquist halo. The figure shows the evolution of the halo after 1 (orange line), 2 (red line), 4 (light blue line) and 8 (dark blue line) Gyrs. The left panel compares the evolution obtained from the $256^3$ simulation, with the evolution reported by Robertson et al. (2017) (dashed lines). We find good agreement with Robertson et al. during the first 6 Gyrs of evolution. At later times our model does not produce such a large halo core at the point of maximum expansion as Robertson et al.}
	\label{Model_validation_3}
\end{figure*}

\section{Model validation}\label{Model_validation_2_sec}

\subsection{Numerical convergence}\label{Numerical_convergence}

In this section we analyse the numerical convergence of the simulations. We run simulations for an isolated halo that follows a Hernquist profile for 10 Gyr, and follow the evolution in the halo's density profile. Differently from section~\ref{Model_validation_1_sec}, we allow the effective collisions to modify the particles velocities. We model a $10^{14}~\rm{M}_{\odot}$ halo with a scale radius of 225 kpc, and assume a constant cross section of $\sigma/m_{\chi}=1$ cm$^{2}$/g. We run three simulation with different number of particles, ranging from $64^3$, $128^3$ till $256^3$, with gravitational softenings equal to 12, 8 and 4 kpc, respectively. Fig.~\ref{Model_validation_2} shows the evolution in the density profile of a Hernquist halo after 1, 2, 4  and 8 Gyrs of evolution. The simulation that contains $64^3$ particles is shown as dotted lines, the one that has $128^3$ particles is shown as dashed lines, and the simulation with $256^3$ particles is shown as solid lines. From the figure it can be seen that we achieve good convergence in the evolution of an isolated halo as shown by simulations with different resolution.

\subsection{Comparison with previous works}

We compare the evolution in density and velocity dispersion of the N=$256^{3}$ simulation, with the evolution reported by \citet{Robertson17Thesis}. As shown in the previous section, we follow the evolution of a $10^{14}~\rm{M}_{\odot}$ Hernquist halo with a scale radius of 225 kpc. We assume constant scattering cross section of $\sigma/m_{\chi}=1$ cm$^{2}$/g. Fig.~\ref{Model_validation_3} shows the evolution in density (left panel) and velocity dispersion (right panel) after 1 (orange line), 2 (red line), 4 (light blue line) and 8 (dark blue line) Gyrs. To compare with Robertson et al. we use the cored-Hernquist profile defined as

\begin{equation}
\rho(r) = \frac{M_{\rm{tot}}}{2\pi}\frac{a}{(r^\beta+r_{c}^\beta)^{1/\beta}}\frac{1}{(r+a)^3},
\end{equation}

\noindent where $r_{c}$ is the core-radius and $\beta$ a free parameters that controls the transition in density from constant core to $\rho\propto 1/r$. Robertson et al. model a Hernquist halo of same mass, scale radius and similar resolution. They fixed $\beta=4$, and obtained a core radius of 12, 17, 22 and 30 kpc, after 1, 2, 4 and 8 Gyrs respectively. 

The left panel of Fig.~\ref{Model_validation_3} compares the density between our model (solid lines) and the best-fit profile from Robertson et al. (dashed lines). We find good agreement with Robertson et al. during the first 6 Gyrs of evolution. At later times our model does not produce such a large halo core at the point of maximum expansion as Robertson et al. This is likely due to the different manner in which the probability of DM particles interaction is calculated (see Section~\ref{scatter_probability_sec}). Robertson et al. estimated $r_{c}$ running a $10^{14}~\rm{M}_{\odot}$ Hernquist halo in a $256^{3}$ simulation, with 2 kpc gravitational softening and assumed a fixed search radius (for the SIDM interactions) equal to the gravitational softening length.

\citet{Fischer21} derived a new approach to model frequent scattering based on an effective drag force, which they implemented into the N-body code GADGET-3 (an updated version of the N-body code GADGET-2, \citealt{Springel05}). Similar to this work, Fischer et al. calculate a DM particle drag force based on a kernel function representing the DM density distribution. In their comparison with Robertson et al., they also reported a smaller maximum core size, but overall a similar evolution.

\section{Diversity in the rotation curves from central haloes}\label{Diversity_Centrals}

\begin{figure*} 
	\includegraphics[angle=0,width=\textwidth]{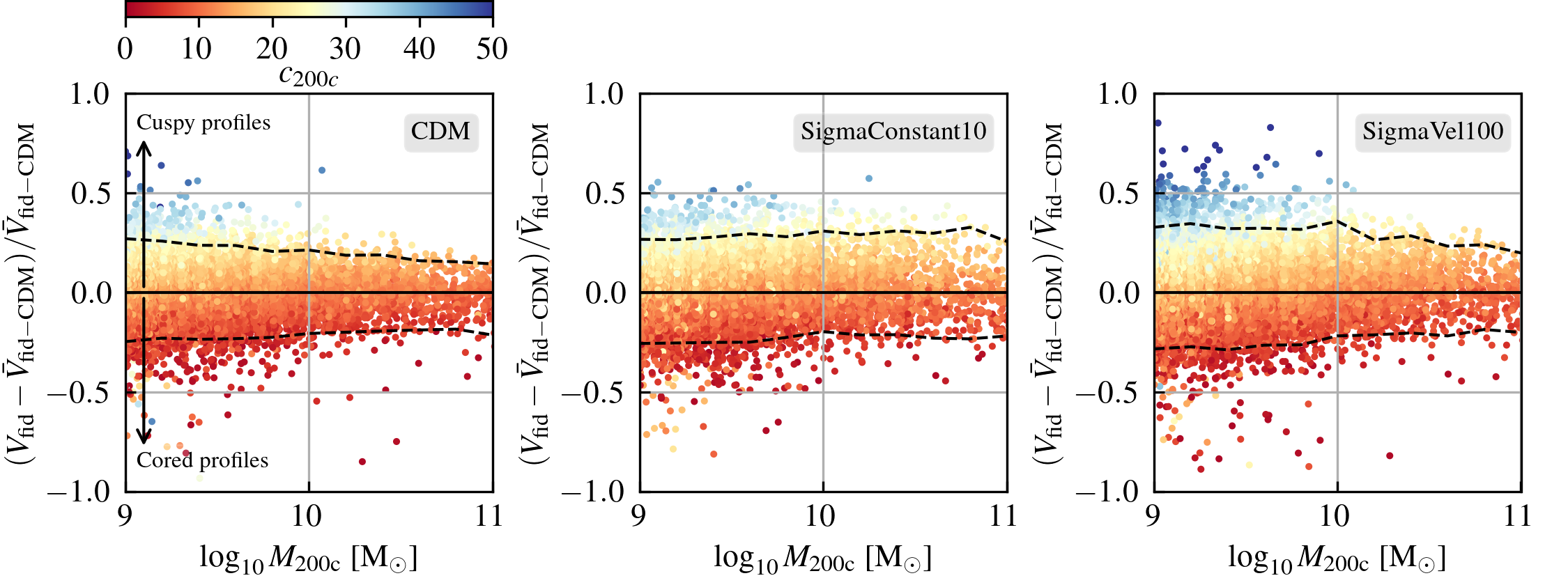}
	\caption{Same as Fig.~\ref{M200_ratio_models}, but for central haloes. Circular velocities at the fiducial radius, $V_{\rm{fid}}$, relative to the median $\bar{V}_{\rm{fid}}$ from the CDM simulation. Each dot corresponds to a central halo, with a mass indicated by the x-axis, and with a concentration highlighted by the colour bar on the top of the figure. The panels show the ratio, $(V_{\rm{fid}}-\bar{V}_{\rm{fid-CDM}})/\bar{V}_{\rm{fid-CDM}}$, for the CDM (left), SigmaConstant10 (middle) and SigmaVel100 simulation (right). The dashed black lines in the panels highlight the 97 and 3 percentiles of the distribution. The figure indicates that the SigmaVel100 model contains a larger scatter in the velocity ratios from central haloes, relative to CDM.}
	\label{M200_ratio_models_centrals}
\end{figure*}

This section extends the analysis presented in Section~\ref{Diversity_section} for the case of central haloes. Fig.~\ref{M200_ratio_models_centrals} shows the velocity ratio, $(V_{\rm{fid}}-\bar{V}_{\rm{fid-CDM}})/\bar{V}_{\rm{fid-CDM}}$, as a function of halo mass. As in Section~\ref{Diversity_section}, the fiducial velocity $V_{\rm{fid}}$ is calculated for each individual central halo $i$, and then related to $\bar{V}_{\rm{fid-CDM}}$, defined as the median CDM $V_{\rm{fid}}$ from the mass bin the halo $i$ is. Fig.~\ref{M200_ratio_models_centrals} depicts the relation for the CDM (left-panel), SigmaConstant10 (middle-panel) and SigmaVel100 (right-panel) models. By comparing the panels it can be seen that the SigmaVel100 model shows a larger number of outliers with $(V_{\rm{fid}}-\bar{V}_{\rm{fid-CDM}})/\bar{V}_{\rm{fid-CDM}}>0.3$ (as well as with $(V_{\rm{fid}}-\bar{V}_{\rm{fid-CDM}})/\bar{V}_{\rm{fid-CDM}}<-0.3$) relative to CDM. 

\end{document}